\documentclass[aps,prb,twocolumn,groupedaddress,showpacs]{revtex4}

\usepackage{graphicx} 
\begin{document}
\title{Orbital Landau level dependence of the fractional quantum Hall
effect in quasi-two dimensional electron layers: finite-thickness
effects} \author{Michael R. Peterson$^1$, Th. Jolicoeur$^2$ and S. Das
Sarma$^1$} \affiliation{$^1$Condensed Matter Theory Center, Department
of Physics, University of Maryland, College Park, MD 20742, USA}
\affiliation{$^2$Laboratoire de Physique Th\'eorique et Mod\`eles
Statistiques, Universit\'e Paris-Sud, 91405 Orsay Cedex, France}
\begin{abstract}

The fractional quantum Hall effect (FQHE) in the second orbital Landau
level at even-denominator filling factor 5/2 remains enigmatic and
motivates our work.  We theoretically consider the effect of the
quasi-2D nature of the experimental fractional quantum Hall system on
a number of FQH states (filling factors 1/3, 1/5, and 1/2) in the
lowest, second, and third orbital Landau levels (LLL, SLL, and TLL,
respectively) by calculating the wavefunction overlap, as a function
of quasi-2D layer thickness, between the exact ground state of a model
Hamiltonian and the consensus variational ansatz wavefunctions, i.e.,
the Laughlin wavefunction for 1/3 and 1/5 and the Moore-Read Pfaffian
wavefunction for 1/2.  Using large numerical overlap as a stability,
or FQHE robustness, criterion we find that the FQHE does not occur in
the TLL (for any quasi-2D layer thickness), is the most robust for
zero thickness (strict 2D limit) in the LLL for 1/3 and 1/5 and for
11/5 in the SLL, and is the most robust at finite-thickness (4-5
magnetic lengths) in the SLL for the mysterious even-denominator 5/2
state and the presumably more conventional 7/3 state.  We do not find
any FQHE at 1/2 in the LLL for any thickness for the quasi-2D models
considered in our work.  Furthermore, we examine the orbital effects
of a non-zero in-plane (parallel) magnetic field finding that its
application effectively reduces the quasi-2D layer thickness and,
therefore, could destroy the FQHE at 5/2 and 7/3, while enhancing the
FQHE at 11/5, in the SLL.  The in-plane field also enhances the LLL
FQHE states by making the quasi-2D system more purely 2D.  The
in-plane field effects could thus be qualitatively different in the
LLL and the SLL by virtue of magneto-orbital coupling through the
finite thickness effect.  Using exact diagonalization on the torus
geometry, we show the appearance of the characteristic threefold
topological degeneracy expected for the Pfaffian state. This signature
is enhanced by nonzero thickness corroborating our findings from
overlap calculations.  Our results have ramifications for
{\textit{wavefunction engineering}}, opening the possibility of
creating an optimal experimental system where the 5/2 FQHE state is
more likely described by the Moore-Read Pfaffian state with obvious
applications in the burgeoning field of fault-tolerant topological
quantum computing.

\end{abstract}


\maketitle

\section{Introduction}
\label{intro}

The fractional quantum Hall effect (FQHE), discovered in
1982\cite{fqhe}, is a quintessentially strongly correlated quantum
phenomenon where electrons in a two-dimensional (2D) electron system
condense into a strongly interacting incompressible quantum fluid
ground state\cite{laughlin,cftheory} with fractionally charged
quasiparticle excitations which obey
anyonic\cite{fracstats,fracstats1,fracstats2}, rather than ordinary
fermionic or bosonic, quantum statistics.  The FQHE occurs at low
temperatures in clean (high mobility) 2D semiconductor structures
under the influence of a strong external magnetic field applied normal
to the 2D plane of confinement of the electron layer.  The subject has
been studied extensively during the last 25 years, and
reviews\cite{qhe-persp,qhe-girvin,cfbook} can be found in the
literature.

In this article we provide a detailed numerical theoretical study of
{\textit{the orbital Landau level}} (LL) dependence of the FQHE,
emphasizing the relative importance of the {\textit{quasi}}-2D layer
width, i.e., the ``finite-thickness effect", of the electron system
transverse to the plane of confinement in the lowest three LLs.  In
the non-interacting 2D system, taken here to be confined in the
$x$-$y$ plane with the magnetic field $B$ along the $z$-direction
which is also the direction of confinement with a typical layer width
of $d$ (finite $d$ corresponds to the realistic quasi-2D system
studied in the laboratory and $d=0$ corresponds to the strictly 2D
idealized system often studied theoretically for convenience), the
application of the external magnetic field leads to the Landau
quantization of electronic energy levels given by
$E_n=(n+1/2)\hbar\omega_c$ where $n=0,1,2,3,\ldots$ is the orbital LL
index and $\omega_c=eB/mc$ is the cyclotron frequency defining the
equidistant energy level separation (i.e., the simple harmonic
oscillator spectrum) between the 2D LLs.  Each LL has a macroscopic
degeneracy given by $(2\pi l^2)^{-1}$ per unit area where
$l=\sqrt{\hbar c/eB}\equiv\sqrt{\hbar/m\omega_c}$ is the magnetic
length (which is used as the unit of length throughout).  For a given
2D electron density of $N_s$ (per unit area), one has a LL filling
factor $\nu=N_s/(2\pi l^2)^{-1}=2\pi l^2 N_s$ indicating the filling
of the macroscopically degenerate LLs in the 2D electron system.  If
$\nu<1$, only the lowest (orbital) Landau level (LLL), by convention
denoted as $n=0$, is fractionally occupied by electrons.  Our
discussion, so far, has neglected electron spin which is equivalent to
assuming the 2D system to be spin polarized (by a sufficiently strong
$B$ field, for example).  Including spin degeneracy in the picture
introduces a factor of two since each orbital LL state (i.e.,
$n=0,1,2,\ldots$) can be filled with both up and down spins.
Incorporating spin in this ``trivial'' manner (i.e., assuming each
orbital LL to be occupied sequentially by spin up and down electrons),
we get that $0<\nu<2$, $2<\nu<4$, $4<\nu<6$, and so on, correspond,
respectively, to spin up-down orbital LLs $n=0$ (LLL), $n=1$ (the
second Landau level, SLL), $n=2$ (the third Landau level, TLL).  Our
goal in this work is to theoretically investigate the FQHE in $n=$0,
1, and 2 and provide a critical \textit{comparative} study of the FQHE
in the LLL, SLL, and TLL, emphasizing the key role played by the
quasi-2D layer thickness parameter $d$ (or more precisely the
dimensionless parameter $d/l$) in determining the relative strength,
stability, or importance of various FQH states in different orbital
LLs.  We consider only completely spin-polarized (i.e., spinless) FQH
states in our calculations since the primary fractional states 1/2,
1/3, and 1/5 are universally thought to be spin-polarized.

The motivation of our work stems from the experimental observation
that FQH states are ubiquitous in the LLL (about 70 distinct FQH
states have been experimentally
observed\cite{qhe-persp,qhe-girvin,cfbook} in the $n=0$ LLL with
$0<\nu<2$), fairly rare in the SLL (less than 10 FQH states have been
observed\cite{willett,pan-52,fqhe-SLL,fqhe-SLL-1,fqhe-SLL-2,choi,pan}
in the $n=1$ SLL with $2<\nu<4$, and these are much ``weaker'' than
the corresponding LLL FQH states in the sense that the observation of
the SLL FQHE requires much lower temperatures and much higher sample
mobilities than in the corresponding LL situation), and essentially
non-existent in the TLL (no robust $n=2$, with $4<\nu<6$, TLL FQH
state has so far been convincingly observed
experimentally\cite{pan,fqhe-TLL}).  We establish definitively, in
this work, that even a \textit{qualitative} understanding of the
higher LL FQHE (i.e., for $n>0$) must necessarily include the finite
width effect (in the $z$-direction) of the quasi-2D electron layer.
We point out that the \textit{quantitative} role of the layer width in
the FQHE (even in the LLL) has long been theoretically
known\cite{macdonald,he,zds,ortalano-zhang-sds,park-jain,park-meskini-jain}.
What we show in the current work, through the detailed comparison of
theoretical numerical results in the $n=0,1,2$ LLs obtained on an
equal footing within the same model and approximation scheme, is that
the higher (i.e., $n=1,2$) LL FQHE has fundamentally different
qualitative dependence on the quasi-2D layer width parameter $d/l$
compared with the LLL ($n=0$) case.  {\textit{In particular, we find
that a finite value of $d/l$ is essential in establishing the FQHE in
higher LLs whereas in the LLL, finite $d/l$ only serves to
quantitatively weaken the FQHE!}}  The FQH states weaken (strengthen)
in $n=$0 (1) LLs as $d/l$ increases from the strictly 2D $d=0$ limit.
In the TLL ($n=2$), we do not find a stable FQHE at all, although our
limited numerical results show similar trends in $n=1$ and $n=2$ LLs.

The driving stimulus for studying the FQHE physics in the $n>0$
orbital LLs is, of course, to shed light on the enigmatic $\nu=5/2$
FQHE, originally observed\cite{willett} experimentally in 1987 and
subsequently confirmed and further studied experimentally
repeatedly\cite{pan-52,gammel-52,eisenstein-52} over the last two
decades.  The great fundamental significance of the 5/2 FQHE cannot be
overstated.  With the obvious exceptions of the original discoveries
of the (integer) quantum Hall effect itself\cite{iqhe} and the
subsequent 1/3 FQHE\cite{fqhe}, the 5/2 FQHE may arguably be the next
most important experimental discovery in the field.  It is the
\textit{only} known (so far) exception (for a single 2D layer system)
to the famous ``odd denominator'' rule for the FQHE, i.e., the FQHE
occurs at an odd denominator filling factor $\nu=p/q$ with $q$ an odd
integer (and $p$ either even or odd) with a concomitant quantization
of the Hall conductance into a fractionally quantized Hall
conductivity $\sigma_{xy}=(p/q)(e^2/h)$ and a zero (or a deep minimum)
in the longitudinal conductivity $\sigma_{xx}$ (and also in the
longitudinal resistivity
$\rho_{xx}=\sigma_{xx}[\sigma_{xx}+\sigma_{xy}]^{-2}$).  All other
(i.e., except for the 5/2 state in the SLL) observed FQHE states
(e.g., $\nu=1/3,2/5,3/7,\ldots$ in the LLL; $\nu=7/3, 8/3,
11/5,\ldots$ in the SLL) strictly obey the odd denominator rule, but
the $\nu=5/2$ state, with its well defined quantized Hall conductance
$\sigma_{xy}=(5/2)(e^2/h)$, stands in stark contrast to the odd
denominator rule.  We emphasize that the even denominator nature of
the 5/2 FQHE is not only a curious anomaly, it challenges our
understanding of the FQHE, as developed\cite{laughlin} in the Laughlin
$\nu=1/q$ (with $q$ an odd integer) wavefunction and further developed
in the Jain composite fermion theory based
wavefunctions\cite{cftheory,cfbook} for the $\nu=p/q$ type FQH states
(still with $q$ odd).  The odd integer restriction of the filling
factor denominator $q$ in the ``standard'' (i.e., Laughlin-Jain) FQHE
model is inescapable since it arises from the Pauli exclusion
principle for the electrons.  Any even-denominator FQHE must thus fall
outside the standard Laughlin-Jain FQHE paradigm, and must somehow
correspond to the condensation of bosons (which do not obey Pauli
principle and therefore allow for even denominator FQHE) in the
$\nu=5/2$ incompressible FQH liquid.

As an aside, it is worthwhile to mention that the 5/2 state is the
only observed even-denominator FQHE, as emphasized above, only for
\textit{single-layer} 2D systems.  In bilayer (or more generally,
multilayer) 2D systems, where experiments are carried out in two
parallel 2D layers separated by a barrier (i.e., a double quantum well
structure), even-denominator (e.g., $\nu=1/2$, 1/4) FQH states have
been
observed~\cite{shayegan-1992,eisenstein-1992,eisenstein-persp,macdonald-girvin-persp}
rather routinely.  These bilayer FQH states are theoretically
well-understood~\cite{he-xie-sds-prb-1993} to be strong-coupling
paired Laughlin states, which were postulated by
Halperin~\cite{halp-331} some time ago.  For example, the observed
$\nu=1/2$ bilayer FQH state~\cite{shayegan-1992,eisenstein-1992} has
been shown~\cite{he-xie-sds-prb-1993} to be the Halperin 331
state~\cite{halp-331}, where tightly bound pairs of electrons condense
into a bosonic Laughlin state which is allowed to describe
even-denominator fractions since the Pauli principle does not apply to
bosons.  Such strongly paired Halperin even-denominator states are
Abelian states, similar to ordinary Laughlin FQH states, in contrast
to the Moore-Read non-Abelian Pfaffian even-denominator
state~\cite{pfaff} (see below) which is thought to describe the weakly
paired BCS state underlying the 5/2 single-layer even-denominator FQH
state.  We also mention here that no single-layer (as opposed to
bilayer) $\nu=1/2$ FQHE has even been observed experimentally although
we know of no fundamental reason ruling out such paired (either
strong-coupling or weak-coupling) LLL states.

The leading theoretical candidate for the $\nu=5/2$ (and its
electron-hole counterpart at $\nu=7/2$) FQH state is the so-called
``Pfaffian'' wavefunction\cite{pfaff} of Moore and Read (MR), which
has been extensively studied over the last 17
years\cite{morf-overlap,rez-hald,rr,morf,scarola1,scarola,toke1,toke2,wojs1,moller,feiguin,mrref}.
In fact, there is no other proposed viable candidate ground state
wavefunction for the observed 5/2 FQH state, and as such, a consensus
has emerged that the Moore-Read Pfaffian wavefunction is the likely
description for the enigmatic even-denominator 5/2 FQHE.  In spite of
this consensus, arising primarily out of a lack of any other viable
alternative, there has been a minority viewpoint\cite{toke1,toke2}
questioning the validity of the MR Pfaffian in describing the 5/2 FQH
ground state.  In addition, a nagging issue of substantial importance
is why there is an incompressible FQH state at half filling in the SLL
(i.e., $\nu=2+1/2=5/2$), but not at $\nu=1/2$ in the LLL (or for that
matter, at half filling in the TLL, i.e. $\nu=4+1/2=9/2$) since the
Pfaffian wavefunction carries no LL index label and is presumably an
allowed variational description for the half-filled LL in any orbital
level $n=0,1,2,\ldots$.  Important early
work\cite{morf-overlap,rez-hald} showed that the Pfaffian wavefunction
is a rather fragile description of the $\nu=5/2$ FQH state, and slight
variations in the effective interaction between the carriers, as, for
example, could arise from the finite layer width or from changing the
orbital LL index could, in principle, affect the validity of the
Pfaffian wavefunction as a suitable description of the $\nu=5/2$
ground state.  The ``mundane'' details of the dependence on the
effective electron-electron interaction, rather than any deep
fundamental principle, is, in fact, the reason for the SLL 5/2 state
to be an incompressible FQHE state whereas the corresponding LLL 1/2
state is a compressible Fermi liquid state.  Thus, the orbital LL and
the finite layer thickness effects (and also perhaps the inter-Landau
level coupling, an effect we uncritically ignore in this work) on the
effective electron-electron interaction are the key in determining the
relative stability of various incompressible FQH states in different
orbital LLs.

In this work, we concentrate on three primary FQH states at filling
factors 1/2, 1/3, and 1/5 in three orbital LLs $n=0,1,2$, and
theoretically investigate the relative stability of the incompressible
quantized Hall states at these fillings by calculating, as a function
of the finite width parameter, the ground state wavefunction overlap
between the {\textit{exact}} (numerically exactly diagonalized)
few-particle ground state wavefunction with the corresponding
candidate variational wavefunction (i.e., Laughlin for $\nu=$1/3, 1/5
and MR Pfaffian for $\nu=1/2$) for the incompressible FQH state.  A
high (low) overlap provides a strong hint that the corresponding
realistic FQHE state is (is \textit{not}) described by the
corresponding variational state (i.e., Laughlin for 1/3, 1/5 and
Moore-Read for 1/2) in the appropriate orbital LL.  Such exact
diagonalization studies of small systems have been the main
theoretical tool in learning about the nature of incompressible FQH
states ever since the original discovery of the FQHE.  In particular,
the universal acceptance of the celebrated Laughlin wavefunction as
the appropriate description for the observed 1/3 FQH state in the LLL
is based almost entirely on the remarkably large (essentially unity)
overlap between the analytical Laughlin wavefunction and the exact
small-system numerical wavefunction.  Similarly, the Jain composite
fermion theory based variational wavefunctions are thought to be
excellent descriptions for the non-primary (i.e.,
$\nu=m/(2pm\pm1)$--with $m>1$ and $p$ an integer--such as $\nu=2/5$,
3/7, 4/9, etc.) LLL FQH states because of the good overlap between the
composite fermion wavefunctions and exact numerical wavefunctions for
small systems.  Even the MR Pfaffian wavefunction is accepted to be
the reasonable description for the experimental SLL 1/2 state (i.e.,
at $\nu=2+1/2=5/2$) based simply on the observation, made originally
by Morf\cite{morf-overlap} and followed up in other subsequent
theoretical publications\cite{rez-hald,scarola}, that the Pfaffian has
good, albeit {\textit{not}} spectacular, overlap ($\sim$0.8-0.9 for
the strict 2D system) with the exact small-system numerical
wavefunction.  Conversely, alternative possibilities for the 5/2 FQHE
state, such as the Halperin 331 state~\cite{halp-331} or the
Haldane-Rezayi spin-singlet hollow-core~\cite{hald-rez-ss} state, are
ruled out~\cite{eisenstein-persp,macdonald-girvin-persp} theoretically
essentially entirely on the basis of very poor calculated overlap of
these candidate states with exact small system numerical
wavefunctions.

Another measure of the stability--besides the wavefunction
overlap--that applies to the MR Pfaffian state, in particular, is the
appearance of the threefold topological degeneracy in the torus
geometry that is a signature of the Pfaffian state.  This degeneracy
is a direct signature of the non-Abelian nature of the state (and
therefore would not arise if the state is Abelian such as the Halperin
331 state).  The existence of this degeneracy for a physical
interaction, i.e., a modified Coulomb interaction, has not been
earlier reported in the literature.  Our finding of the appropriate MR
degeneracy in the 5/2 numerical state precisely where the overlap is
maximal is an important advance in theoretical understanding.

We choose the Laughlin (for 1/3 and 1/5) and the Pfaffian (for 1/2) as
our candidate variational ansatz because these are the \textit{only}
proposed incompressible FQH states at these primary fractional
fillings.  Our work will, therefore, miss out on discovering any
\textit{other} possible ground state wavefunctions (i.e., other than
Laughlin for 1/3 and 1/5 and Moore-Read for 1/2) describing the
experimental FQH state since we restrict our overlap calculations
entirely to the Laughlin or Moore-Read candidate states.  This
restriction is, however, not a serious drawback of our work since no
other candidate wavefunctions exist in the literature for the FQH
states of our interest.

Another incentive for our work comes from the fact that existing
numerical work\cite{macdonald-SLL,toke,reynolds,scarola-hll}, much of
it carried out in the idealized 2D limit, concludes that the 1/3 FQH
state in the SLL (i.e., the $\nu=2+1/3=7/3$ state or its electron-hole
counterpart $\nu=8/3$ state) is unlikely to be a simple Laughlin state
(although in Ref.~\onlinecite{toke} the deviation from the Laughlin
wavefunction stems from residual quasiparticle interactions) since the
SLL small-system exact diagonalization studies give only modest
wavefunction overlap between the Laughlin wavefunction and the exact
numerical ground state in the SLL.  As mentioned above, for the 5/2
FQH state, the calculated overlap\cite{morf-overlap,rez-hald,scarola}
between the MR Pfaffian and the exact diagonalization finite system
wavefunction is also quite modest, much below unity, in the idealized
2D limit.  Our results would shed light on whether the
finite-thickness, or width, of the quasi-2D layer enhances the overlap
between the Laughlin (or Moore-Read) state and the exact wavefunction
so that the Landau level dependent stability of the incompressible
FQHE can be discussed in terms of a systematic tuning of the
Hamiltonian through the variation in the electron-electron interaction
caused by changing the layer width parameter $d/l$.  We find that
indeed a finite 2D layer width, i.e., a true {\textit{quasi}}-2D
system, is necessary for stabilizing the FQHE in the SLL, at least
within the restricted wavefunction space (i.e., Laughlin and
Moore-Read) we investigate.  We note that increasing $d/l$ from zero
(i.e., the strict 2D ideal limit) is equivalent to softening the
interaction.

One rather direct consequence of the quasi-2D finite-thickness effect
on the FQHE is the \textit{orbital} coupling of an in-plane (i.e.,
parallel to the 2D layer) applied magnetic field, in addition to the
quantizing perpendicular magnetic field.  Having both parallel and
perpendicular magnetic fields is, of course, equivalent to having a
tilted magnetic field, which is often used in 2D physics to study spin
polarization effects -- the idea being that the parallel component of
the magnetic field couples only to the electron spin through the
modification of the Zeeman energy.  This is certainly true in the
idealized strictly 2D limit where a magnetic field in the 2D plane has
\textit{no} coupling to the orbital 2D dynamics of the electrons, but
does couple to the electron spin.  In a quasi-2D system, however, the
situation is more complex since an in-plane field could, in principle,
couple to the quasi-2D orbital dynamics of the electrons through the
finite-thickness of the electron layer.  One effect of the parallel
field would, for example, be to squeeze the electron layer in the
third direction, reducing its effective quasi-2D layer width.  We
consider such magneto-orbital coupling effects due to a parallel
applied field in our work within the context of our LL-dependent
finite-thickness studies, using a parabolic or simple harmonic
confinement model.

A compelling and timely reason for the detailed investigation of
higher LL FQHE is the recent interest in using the $\nu=5/2$ (and
possibly $\nu=12/5$, recently observed by Xia, et al in
Ref.~\onlinecite{fqhe-SLL-1}, and $\nu=13/5$) FQH states for
fault-tolerant topological quantum computation~\cite{tqc-1,tqc-2}
using the non-Abelian quasiparticles associated with the Pfaffian
state.  Since the non-Abelian nature of the quasiparticle excitations
is crucially tied to the specific form of the MR Pfaffian
wavefunction, it is important that we know whether the observed 5/2
state is really the Moore-Read state or not.  Recently, serious
questions have been raised~\cite{toke1,toke2} about whether the
Pfaffian is the appropriate description for the 5/2 state.  Our work,
investigating the detailed nature of the incompressibility as a
function of the quasi-2D layer width parameter in higher LLs, thus is
timely and necessary for further progress in the subject of
topological quantum computation.  There have also been several recent
experimental investigations of the SLL FQHE motivated by topological
quantum computation
considerations~\cite{willett-2008,dolev-2008,radu-2008}.

The rest of the paper is organized as follows: In
section~\ref{sec-hamil} we introduce the essential Hamiltonians we are
studying, namely, the electron-electron interaction Hamiltonian and
ones that give either the Laughlin or Pfaffian states as zero-energy
ground states.  In section~\ref{sec-ft} we describe the various models
used to characterize the quasi-2D nature of the experimental quantum
confinement, i.e. the ``finite-thickness'' models for various physical
systems.  The $f$- and $g$-functions (defined below) are then
considered in section~\ref{fg-funcs} as a first attempt to understand
the effective electron-electron interaction due to the quasi-2D nature
of the system.  Overlaps between the exactly diagonalized system for
the ``realistic'' quasi-2D system and the Laughlin (fillings 1/3 and
1/5) or Pfaffian (filling 1/2) wavefunction as a function of the
quasi-2D layer width are reported in section~\ref{overlaps1}.  The
appearance of the ground state threefold degeneracy signature of the
MR Pfaffian state on the torus--particularly for quasi-2D systems--is
investigated in Sec.~\ref{sec-topo}.  The effect of an in-plane
magnetic field {\textit{and}} the quasi-2D nature of the system are
considered in section~\ref{sec-ip}.  In section~\ref{contour-sec} we
discuss connections of our overlap results to previous work and, in
particular, whether or not the physics of the FQHE is adequately
captured by the first few pseudopotentials.  The finite thickness
effects on the excitation gaps are also briefly discussed in
section~\ref{sec-gap}.  Finally, conclusions are given in
section~\ref{conc} with some additional discussions.  A short letter
reporting some of our results has recently appeared in the
literature~\cite{Peterson08}.

\section{Model} 

\subsection{Hamiltonians: Coulomb, Laughlin, and Pfaffian}
\label{sec-hamil}

We consider spin polarized electrons entirely confined (i.e., 
no inter-LL coupling) to a Landau
level of index $n=0$ (LLL), $1$ (SLL), and $2$ (TLL) interacting
through a pair potential $V(r_{ij})$, where $r_{ij}=|\vec r_i - \vec
r_j|$ is the distance between two electrons (distance is measured in
units of magnetic length $l$).  Since the electrons are confined to a
single LL the kinetic energy is a constant, therefore, the Hamiltonian
for $N$ electrons is taken to be the interaction Hamiltonian
\begin{eqnarray}
\hat H = \sum_{i<j}^{N} V(r_{ij})\;.
\end{eqnarray}

\begin{figure}[t]
\begin{center}
\mbox{\includegraphics[width=5.5cm,angle=-90]{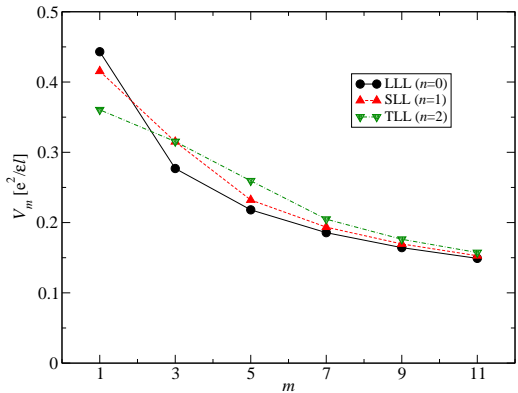}}
\end{center}
\caption{(Color online) $V_m$ as a function of $m$ ($m$ odd) for the pure 
Coulomb potential (zero-thickness), i.e., $V(k)=(e^2l/\epsilon) (1/k)$, in 
the LLL (solid circle), the SLL (upward triangle), 
and the TLL (downward triangle).  $V_m$ are given 
in units of $e^2/\epsilon l$.  The lines are a guide to the eye.}
\label{vms-coulomb}
\end{figure}

Haldane~\cite{exact-laughlin,haldane-qhe} showed how this Hamiltonian
can be parameterized by the relative angular momentum $m$ between two
electrons, through an expansion in the ``pseudopotential'' functions
$V_m^{(n)}$, which serve as a complete set of basis functions due to
angular momentum conservation:
\begin{eqnarray}
\label{hamil}
\hat H = \sum_{i<j}^{N} V(r_{ij}) = \sum_{m=1(\mbox{odd})}^{\infty} V^{(n)}_m \sum_{i<j}^{N} \hat P_m(m_{ij})
\end{eqnarray}
where $\hat P_m(m_{ij})$ is an operator that projects onto the states
of relative angular momentum $m_{ij}=m$.  Since we are considering
spin polarized fermions only odd pseudopotentials are relevant.  The
Haldane pseudopotentials $V_m^{(n)}$ for electrons confined to a LL
with index $n$, in the planar geometry (as opposed to the spherical
geometry), are written as
\begin{eqnarray}
V_m^{(n)}=\int_{0}^{\infty} dk k \left[L_n(k^2/2)\right]^2 L_m(k^2)e^{-k^2}V(k)
\label{hald-pp}
\end{eqnarray}
where $L_n(x)$ are Laguerre polynomials, and $V(k)$ is the Fourier
transform of the interaction potential $V(r)$.  To define our Fourier
transform convention we write
\begin{eqnarray}
V(k)&=&\frac{1}{2\pi}\int d^2 k e^{i\vec k\cdot\vec r}V(r)\nonumber\\
&=&\int_{0}^{\infty}dr r J_0(kr) V(r)\;.
\end{eqnarray}
This parameterization allows all the calculations to be done
entirely within the Hilbert space of the lowest Landau level, i.e.,
the information about higher LLs is completely contained within the
$V_m^{(n)}$.  Note that this simplification depends on our neglecting
all Landau level mixing effects, which, along with the assumption 
of complete spin-polarization, is a key and uncritical assumption 
for our theory.

In a purely 2D system the electron-electron interaction is the Coulomb
interaction $V(r)=(e^2/\epsilon l) (1/r)$, where $r$ is the distance
in the 2D plane between a pair of electrons, yielding a Fourier
transform of $V(k)=(e^2l/\epsilon)(1/k)$ where $k$ is units of $1/l$.
Figure~\ref{vms-coulomb} displays the first six pseudopotentials as a
function of $m$ for the LLL, the SLL, and the TLL.  Although the
differences between the pseudopotentials in different LLs are small
quantitatively, the qualitative changes in the system behavior can be
severe, as discussed below.  The main qualitative difference in the
pseudopotentials $V_m^{(n)}$ among the three LLs is that
$V_1^{(0)}>V_1^{(1)}>V_1^{(2)}$, but $V_2^{(0)}<V_2^{(1)}$,
$V_2^{(2)}$ and $V_3^{(0)}<V_3^{(1)}<V_3^{(2)}$.

We will denote the celebrated Laughlin~\cite{laughlin} wavefunction at
$\nu=1/q$ to be $|\Psi_L\rangle$.  It was shown by
Haldane~\cite{exact-laughlin} that $|\Psi_L\rangle$ at filling
$\nu=1/3$ is the exact zero-energy ground state for a ``hard-core''
Hamiltonian $\hat H^{(3)}_L$ where
\begin{eqnarray}
\hat H^{(3)}_L = (\mbox{const.})\sum_{i<j}^{N} \hat P_1(m_{ij})\;.
\end{eqnarray}
This can be obtained from the original Hamiltonian (Eq.~\ref{hamil})
by setting $V^{(n)}_1=\mbox{constant}$ and $V^{(n)}_m=0$ for all
$m\geq 3$; in other words, the Laughlin state at 1/3 avoids all
electron pairs with $m<3$ since $V_1/V_m=\infty$ for $m\geq 3$.
Further, the Laughlin state at $\nu=1/q$ (filling 1/5 for example) is
the exact zero-energy ground state for a Hamiltonian
\begin{eqnarray}
\hat H^{(q)}_L = (\mbox{const.})\sum_{m=1(\mbox{odd})}^{q-2}\sum_{i<j}^{N}\hat P_m(m_{ij})\;.
\label{laugh-H}
\end{eqnarray}
Generally, $\hat{H}^{(q)}_L$ is a Hamiltonian that penalizes two
electrons that have an angular momentum smaller than $m$.

The Pfaffian~\cite{pfaff} wavefunction $|\Psi_{Pf}\rangle$ is thought
to be the leading candidate for the FQHE at filling 1/2 in the SLL,
i.e., $\nu=5/2$, and is known~\cite{pfaff-exact} to be an exact
zero-energy ground state for a three-body Hamiltonian $\hat H_{Pf}$
which penalizes states where three electrons are in a relative angular
momentum state smaller than some value.  There is no known two-body
interaction Hamiltonian which has the Pfaffian state as the ground
eigenstate, so we should think of the Pfaffian as a variational ansatz
for our two-body interaction Hamiltonians.

\begin{figure}[t]
\begin{center}
\mbox{\includegraphics[width=5.5cm,angle=-90]{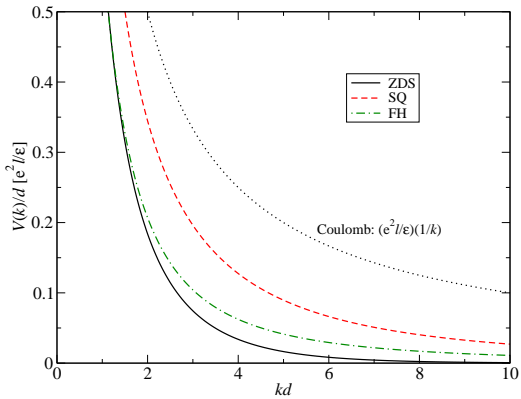}}
\end{center}
\caption{(Color online) Fourier transform of the finite-thickness
potentials divided by thickness $d$, i.e., $V(k)/d$ versus $kd$ for
the ZDS (solid line), SQ (dashed line), and FH (dashed-dotted line)
potentials.  Also shown is the Coulomb potential.}
\label{v-kspace}
\end{figure}

\begin{figure*}[t]
\begin{center}
\mbox{\includegraphics[width=5.5cm,angle=0]{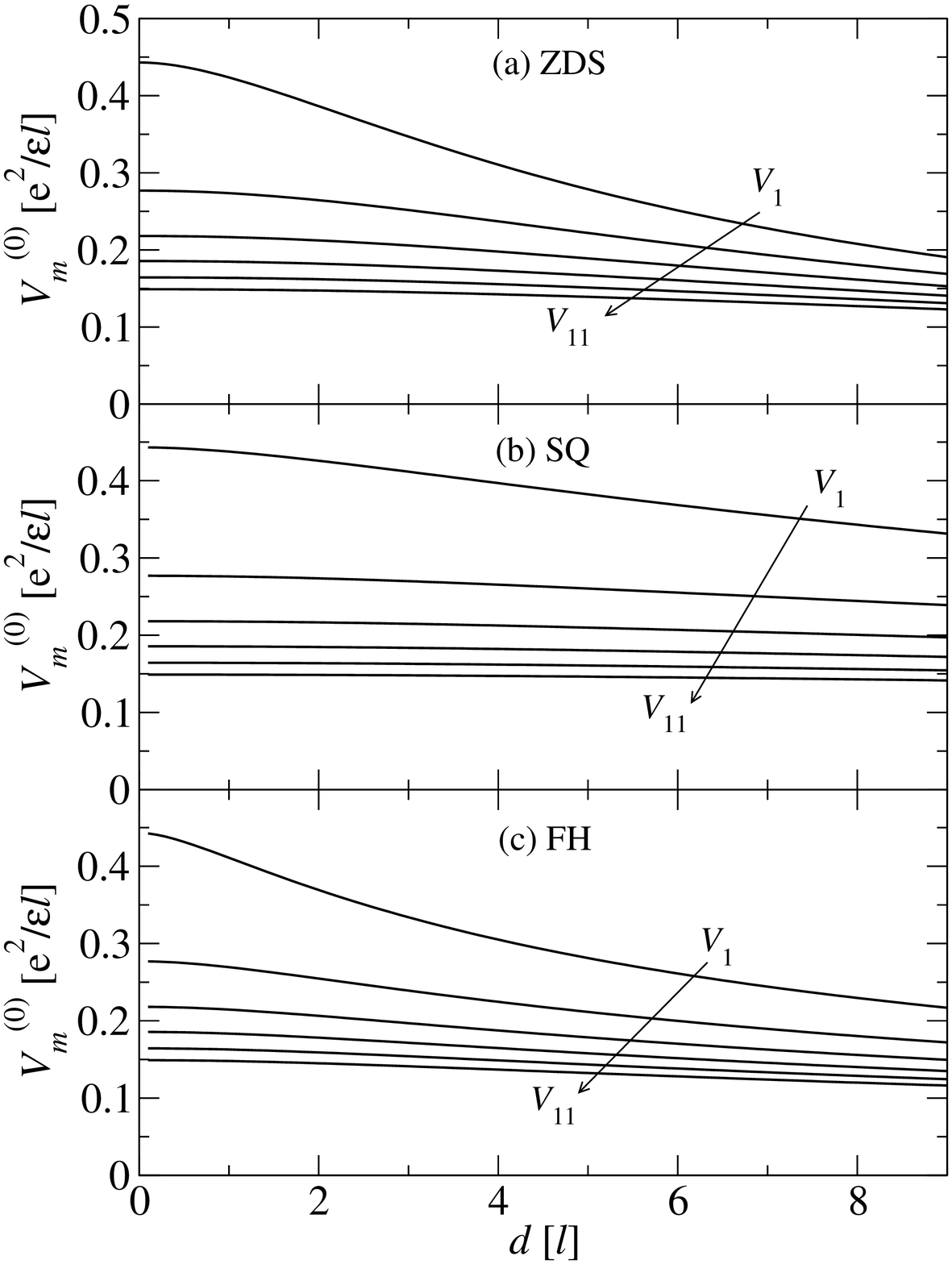}}
\mbox{\includegraphics[width=5.5cm,angle=0]{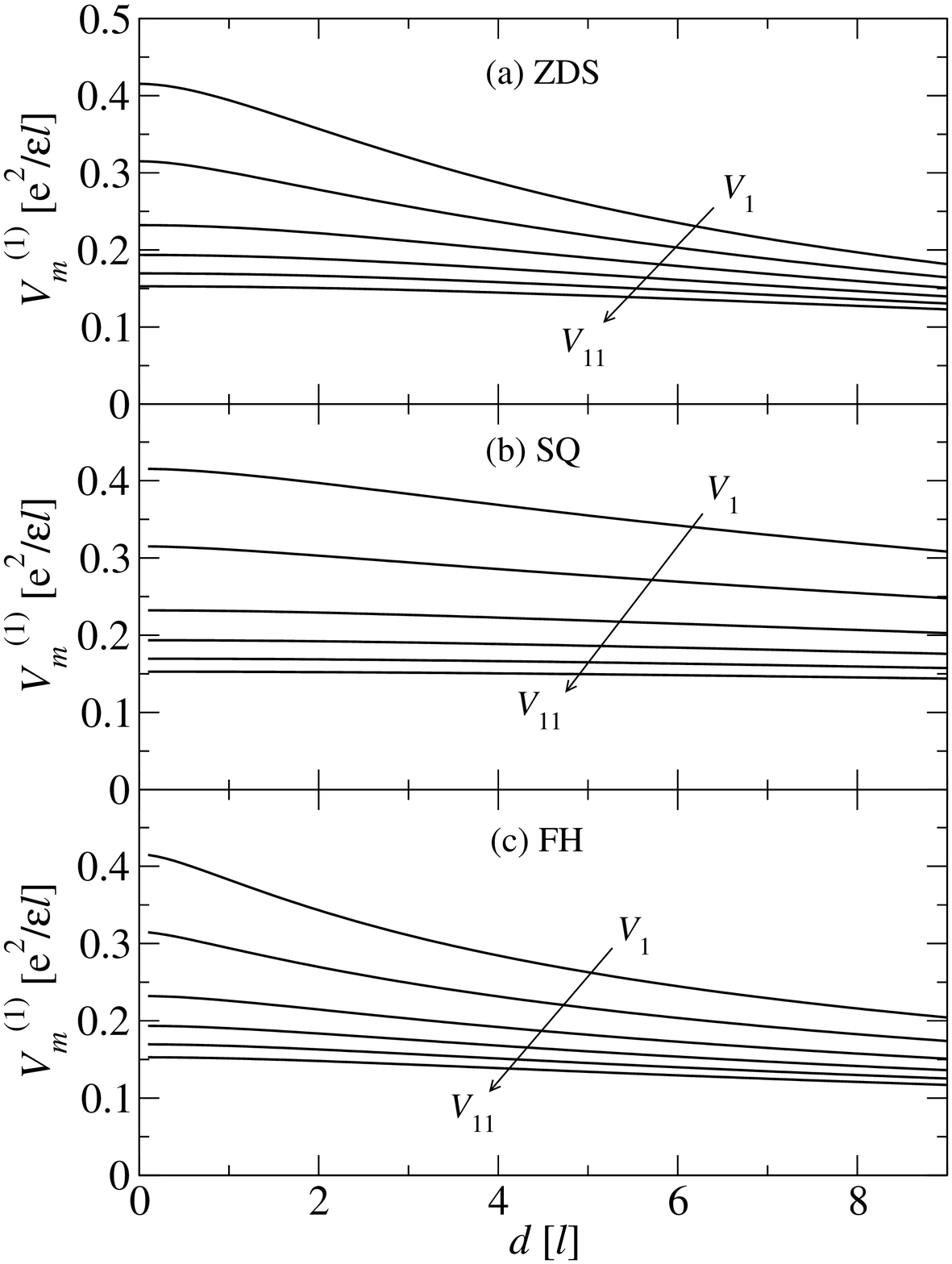}}
\mbox{\includegraphics[width=5.5cm,angle=0]{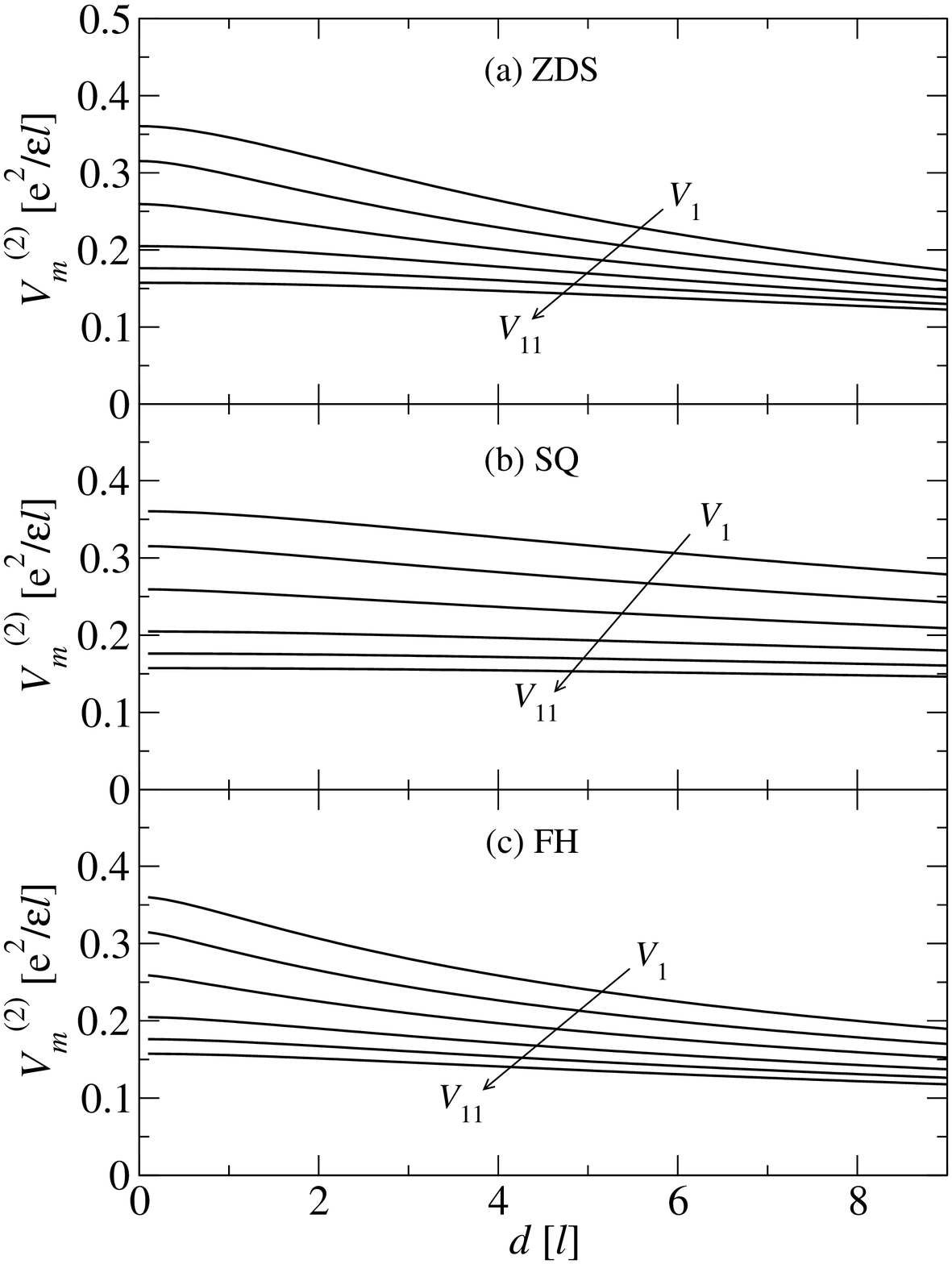}}\\
\end{center}
\hspace{1.cm}LLL\hspace{5.cm}SLL\hspace{5.cm}TLL
\caption{Pseudopotentials for the LLL (left), the SLL (middle), and
the TLL (right).  Each panel has three plots each displaying the
$V_m$'s for the three different finite-thickness modeling potentials,
which are labeled: (a) Zhang-Das Sarma (ZDS), (b) infinite square-well
(SQ), and (c) Fang-Howard heterostructure (FH).  The units of the
$V_m$'s are as in Fig.~\ref{vms-coulomb}.}
\label{vms}
\end{figure*}

(We are providing necessarily very brief discussions of the Laughlin
and Pfaffian Hamiltonians in order to facilitate easier discussions in
later sections of this work.)

In the pure 2D system, where the above discussion applies, the
electron interaction is just pure Coulomb, however, the finite extent
of the single-particle electron wavefunction in the direction
perpendicular to the plane in actual experimental quasi-2D systems
will modify the electron interaction as discussed below.

\subsection{Finite-thickness Modeling Potentials}
\label{sec-ft}

Depending on the details of the physical systems (i.e., quantum wells,
heterostructures, etc.), there are many possible models for the
inclusion of the finite-thickness effect with the 2D Coulomb
interaction.  We do not, however, anticipate much qualitative
difference among these different models since they all lead to the
softening of the Coulomb potential, with the softening depending
crucially on the largeness of the thickness parameter $d/l$.  In
particular, the pseudopotentials change in the presence of the
finite-thickness since, in addition to the magnetic length, a new
length scale $d$ characterizing the quasi-2D thickness becomes
relevant, and for $d\gg l$, the pseudopotentials are modified
substantially from their $d=0$ ideal 2D Coulomb values.

We consider the three most extensively used finite-thickness potential
models, for which we give the Fourier transforms, that seek to model
the effect of the quasi-2D nature of the experimental system, namely,
(1) the Zhang-Das Sarma (ZDS) potential~\cite{zds}, which was
introduced specifically to theoretically model finite thickness
effects on the FQHE,
\begin{eqnarray}
V_{ZDS}(k)=\frac{e^2l}{\epsilon}\frac{e^{-dk/2}}{k}\;,
\label{eq-zds}
\end{eqnarray}
(2) the infinite square-well (SQ) potential~\cite{sq-pot}, which is 
appropriate for 2D GaAs quantum well structures,
\begin{eqnarray}
V_{SQ}(k)=\frac{e^2l}{\epsilon}\frac{1}{k}\frac{\left\{3kd+
\frac{8\pi^2}{kd}-\frac{32\pi^4(1-e^{-kd})}{(kd)^2[(kd)^2+4\pi^2]}\right\}}{(kd)^2+4\pi^2}\;,
\label{eq-sq}
\end{eqnarray}
and (3) the Fang-Howard (FH) variational
potential~\cite{fh-pot1,fh-pot2} for a heterostructure
\begin{eqnarray}
V_{FH}(k)=\frac{e^2l}{\epsilon}\frac{9}{8k}\frac{(24+9kd+(kd)^2)}{(3+kd)^3}\;.
\label{eq-fh}
\end{eqnarray}
Potentials (2) and (3) are found by using single-particle electron
wavefunctions in the $z$-direction of $\eta(z)=\sqrt{2/d}\cos(\pi
z/d)$ and $\eta(z)=\sqrt{27/2d^3}z\exp(-3z/2d)$, respectively (these
functions are given merely to make our definitions of the thickness
parameter $d$, for each model, completely clear).  In the above, for
(1) and (2), $d$ is the width of the electron layer (in units of $l$)
in the $z$-direction, and for (3) it parameterizes the layer thickness
variationally.  Obviously, as $d\rightarrow0$, all of the above
finite-thickness potentials, describing quasi-2D systems, reduce to
the pure 2D Coulomb potential $V(k)=(e^2l/\epsilon)(1/k)$.
Fig.~\ref{v-kspace} shows the Fourier transforms divided by $d$ as
functions of $kd$ for each finite-thickness potential used clearly
indicating the ``softening'' of the Coulomb potential.  Note that we
keep the background lattice dielectric constant $\epsilon$ in our
definition of the 2D Coulomb interaction only for the sake of
completeness with $e^2/(\epsilon l)$ being our energy unit (and `$l$'
the length unit).  We note that our definition of $d$ as the relevant
thickness or width parameter for each model of quasi-2D confinement is
contained entirely in Eqs.~\ref{eq-zds}-\ref{eq-fh}--this is important
since later we introduce alternative width parameters $d^\prime$ and
$w$.

As previously mentioned, the purpose of this work is not to determine
the quantitative accuracy of some particular finite-thickness model
compared to experimental systems.  Rather, we are interested in the
possible non-trivial qualitative changes that can occur when
considering realistic potentials which are not pure 2D Coulomb.  We
note, however, that the SQ and FH models correspond to the two most
common quasi-2D experimental systems (quantum well and
heterostructure, respectively) whereas the ZDS model, while not
corresponding to any physical system, is extensively used in FQH
theoretical studies.

For the sake of completeness we provide the first six pseudopotentials
for all three finite-thickness potentials as functions of $d$ in the
LLL, the SLL, and the TLL shown in the left, middle, and right panels
of Fig.~\ref{vms}.  In all LLs shown, the finite-thickness has the
effect of reducing (or softening) all of the pseudopotentials in a
rather ``trivial'' way, that is, there is no crossing or non-monotonic
behavior: $V^{(n)}_1>V^{(n)}_3>V^{(n)}_5>\ldots$ remains for all $d$.
It is clear, however, that for the SQ and FH potentials the softening
as $d$ increases is less severe compared to the ZDS potential.  (We
mention as a cautionary note that although the same thickness
parameter $d$ has been used in Figs.~\ref{v-kspace} and~\ref{vms} for
our three quasi-2D models, the parameter has somewhat different
meaning in the three cases as can been seen from their strongly
different quantitative effects on the pseudopotential softening in the
three models as is obvious from
Figs.~\ref{overlaps-13}-\ref{overlaps-15}.)  Another qualitative
feature to note is that it is visually difficult to notice any
striking difference between the behavior of the pseudopotentials,
themselves, in different orbital LLs.

We have actually carried out calculations for a fourth model, the
parabolic quantum well (PQW) (or the Gaussian confinement model),
which we discuss in Sec.~\ref{sec-ip}.

\subsection{$f$- and $g$-functions}
\label{fg-funcs}

Only the relative differences in the pseudopotentials are important in
characterizing the physical nature of the FQH ground state.  Following
Ref.~\onlinecite{he} we form the dimensionless $f$-functions defined
through
\begin{eqnarray}
f^{(n)}_m = \frac{V^{(n)}_3-V^{(n)}_m}{V^{(n)}_1-V^{(n)}_3}\;
\end{eqnarray}
which quantitatively describe how close a given Hamiltonian 
is to $\hat H_L^{(3)}$ that produces the 1/3 Laughlin state as the
exact ground state.

From the definition, it is clear that $f^{(n)}_1=-1$ and
$f^{(n)}_3=0$.  As described in Ref.~\onlinecite{he} the $f$-function
is useful because for $\hat H_L^{(3)}$ all $f^{(n)}_m$=0 for $m\geq
3$, and hence, any interaction that produces $f$-functions with this
property will be exactly described by the Laughlin state for
$\nu=1/3$.  Further, if the $f^{(n)}_m$ are very small for $m>3$ then
the exact state will be well approximated by the Laughlin
state for $\nu=1/3$, so, by the simple computation of the
$f$-functions for some Hamiltonian (or interaction potential) one can
get an idea of how well the Laughlin state will describe the actual
ground state.  Thus, the $f$-functions far better manifest Laughlin
wavefunction-like correlations than the pseudopotentials (the
$V_m^{(n)}$ functions) themselves.

One can generalize this to functions which characterize how close a
given Hamiltonian is to $\hat H^{(q)}_L$ that produces the $\nu=1/q$
Laughlin state as the exact ground state.  To that end we define
$g$-functions to handle the $\nu=1/5$ Laughlin state as
\begin{eqnarray}
g^{(n)}_m = \frac{V^{(n)}_5-V^{(n)}_m}{V^{(n)}_1-V^{(n)}_5}\;.
\end{eqnarray}
The $g$-functions satisfy $g^{(n)}_1=-1$ and $g^{(n)}_5=0$.  The
``hard-core'' aspect of a potential is displayed by $g^{(n)}_m=0$ for
all $m\geq 5$.  One could go further with this procedure defining
$h$-functions for investigating the ``Laughlin-ness" of a Hamiltonian
at $\nu=1/7$, $i$-functions for $\nu=1/9$, etc.

\begin{figure*}[t]
\begin{center}
\mbox{\includegraphics[width=5.5cm,angle=0]{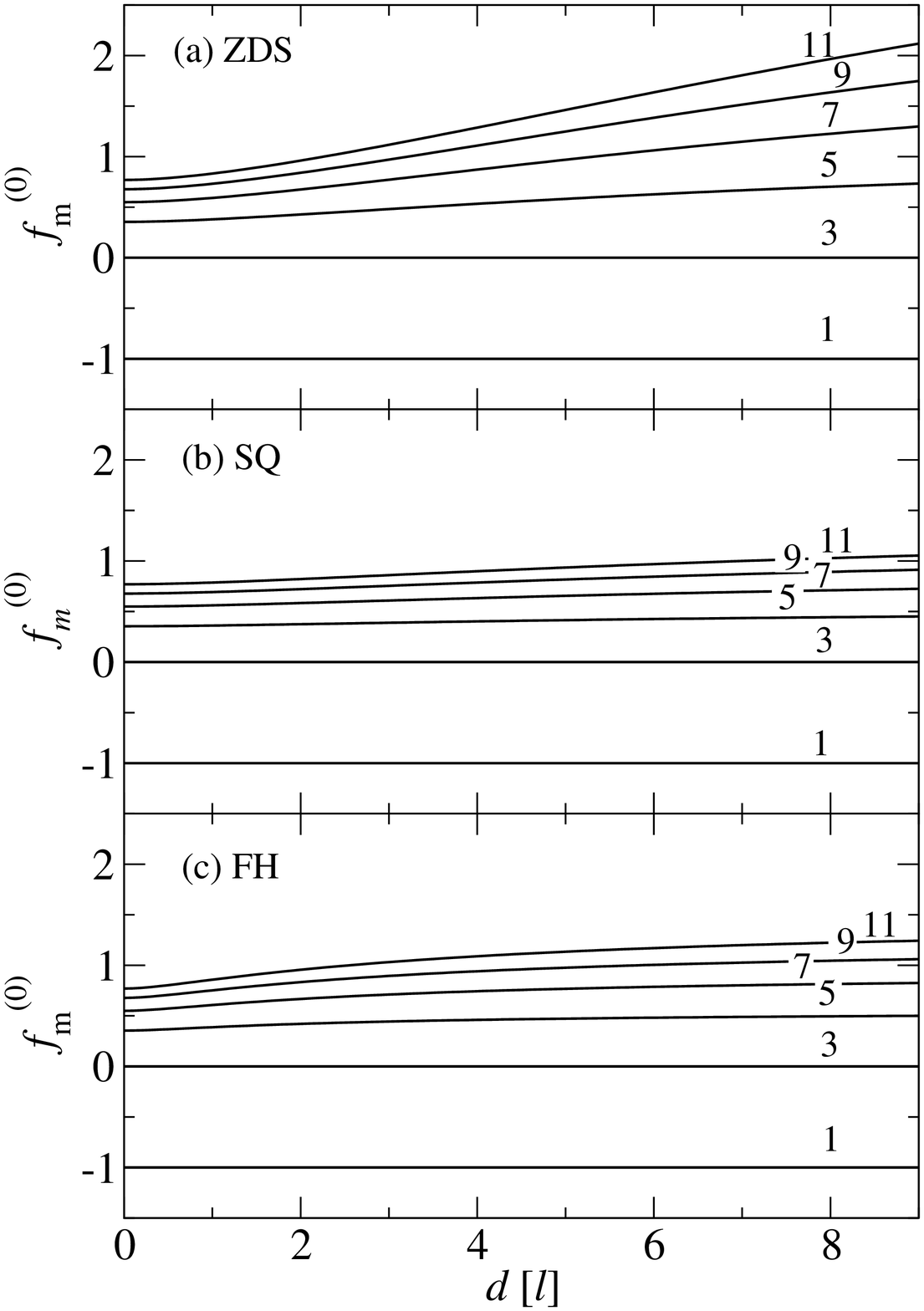}}
\mbox{\includegraphics[width=5.5cm,angle=0]{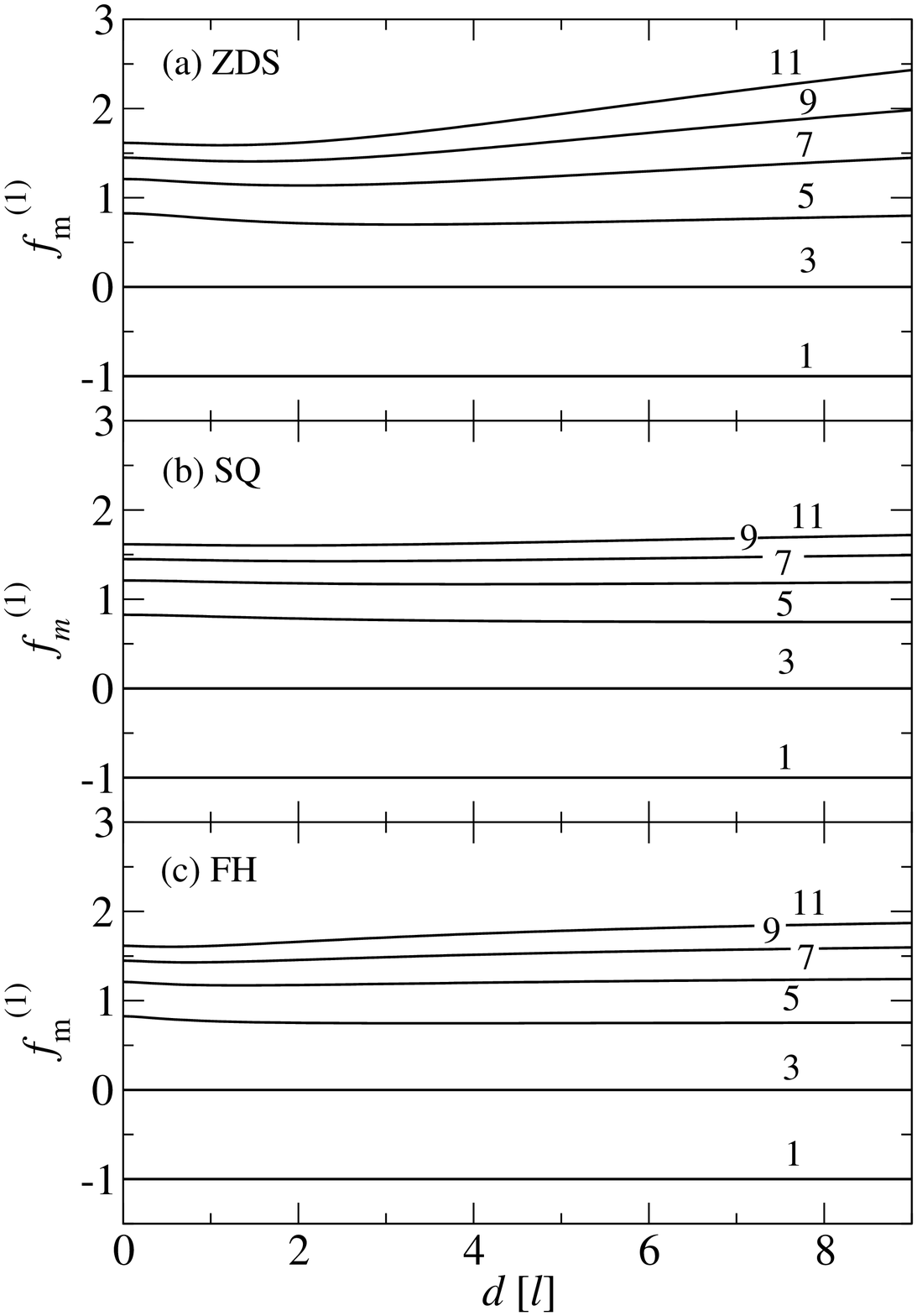}}
\mbox{\includegraphics[width=5.5cm,angle=0]{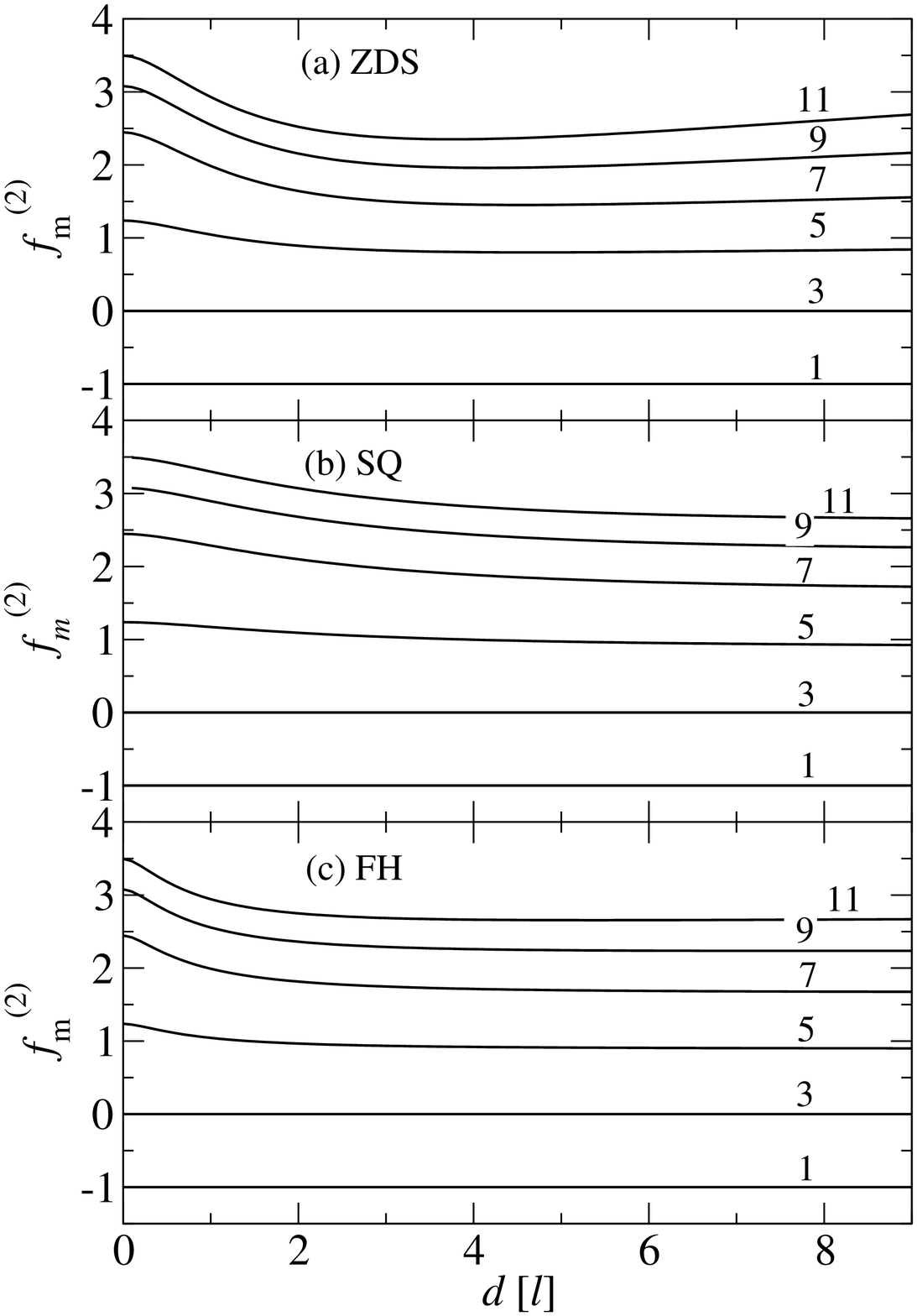}}\\
\end{center}
\hspace{1.cm}{LLL}\hspace{5.cm}{SLL}\hspace{5.cm}{TLL}
\caption{$f$-functions for the LLL (left), the SLL (middle), and the
TLL (right).  Each panel has three plots each displaying the
$f$-functions for the three different finite-thickness modeling
potentials which are labeled: (a) Zhang-Das Sarma (ZDS), (b) infinite
square-well (SQ), and (c) Fang-Howard heterostructure (FH).  The
$f$-functions, by construction, are dimensionless.}
\label{f-funcs}
\end{figure*}

Figure~\ref{f-funcs} shows the $f$-functions as a function of
thickness $d$ for the finite-thickness potentials we have considered,
namely, the ZDS (top labeled (a)), SQ (middle labeled (b)), and FH
(bottom labeled (c)) for the $n=0$ LLL (left panel), the $n=1$ SLL (middle panel),
and the $n=2$ TLL (right panel).  Note that the $f$-functions for the ZDS
potential in the LLL were shown previously in Ref.~\onlinecite{he} as
well as the $f$-functions for the FH potential in the SLL in
Ref.~\onlinecite{belkhir-jain}.  For the LLL (left panel of
Fig.~\ref{f-funcs}) we observe that for zero-thickness $d=0$ all
$f^{(0)}_m < 1$ and it is known (and shown below with other measures)
that the Laughlin state for 1/3 is a very good approximation to the
exact state.  However, as $d$ is increased all $f$'s that are free to
vary ($f^{(0)}_3$ through $f^{(0)}_{11}$) become larger monotonically.
It is interesting to note that for the SQ and FH potentials there is
very little increase in the $f$'s for increasing $d$ compared to the
ZDS potential, i.e., ZDS overestimates the finite-thickness effect
compared with the SQ and FH models.  This is also apparent 
in Figs.~\ref{v-kspace} and~\ref{vms}.

From the calculation of the $f^{(0)}_m$'s, as in Ref.~\onlinecite{he},
one can determine that for some finite-thickness modeling potentials
(e.g., ZDS) the Laughlin state is not a good description of the physics
for $d$ beyond some value whereas for other potentials (SQ and FH) the
Laughlin state most likely remains a good description of the physics
for all $d$, but the description becomes progressively poorer with
increasing $d$.  (This suggestion is further investigated below by
calculating the overlap between the Laughlin state and the exact
ground state as a function of $d$.)

We now investigate $f$-functions in the SLL: The middle panel of
Fig.~\ref{f-funcs} shows $f^{(1)}_m$ for all potentials considered.
These functions are very similar to those of the LLL, at least
qualitatively; they increase essentially monotonically as a function
of $d$, however, there is a difference.  Specifically, for $d=0$, only
$f^{(1)}_5$ is below unity while all $f^{(1)}_{7}$ through
$f^{(1)}_{11}$ are between $1$ and $2$.  This already indicates that
perhaps the Laughlin state will not be a very good description of the
physics here, and, in fact, this has been known for some
time~\cite{reynolds}.

However, there is another property the $f$-functions reveal as $d$ is
increased.  The $f$-functions for the SQ potential were previously
calculated in Ref.~\onlinecite{belkhir-jain} and it was remarked that
in the SLL they do not increase monotonically the way they do in the
LLL.  Instead there is a weak minimum for intermediate $d$ .
That being said, the minimum for the SQ potential is weak and
$f^{(1)}_{7}$ through $f^{(1)}_{11}$ are still greater than unity for
all $d$.  Hence, while the Laughlin state does become a \textit{better}
description for finite $d$ compared to $d=0$ it still never becomes essentially 
exact, as in the LLL.  In
our calculations we observe this sort of behavior for the ZDS and FH
potentials: an initial decrease in the $f$-functions before a
monotonic increase.

Finally, the right panel of Fig.~\ref{f-funcs} displays the
$f$-functions for the TLL where the behavior follows the trend seen
from the LLL to the SLL.  Namely, $f^{(2)}_{5}$ through
$f^{(2)}_{11}$ are all greater than unity for $d=0$ with
$f^{(2)}_{11}\sim3.5$.  Clearly, the Laughlin state will not be a good
description of the physics for $d=0$ in the TLL.  However, as $d$
increases the minimum is much more marked and, in fact, for SQ and
FH potentials at $d=0$ $f$-functions are maximum.

\begin{figure*}[t]
\begin{center}
\mbox{\includegraphics[width=5.5cm,angle=0]{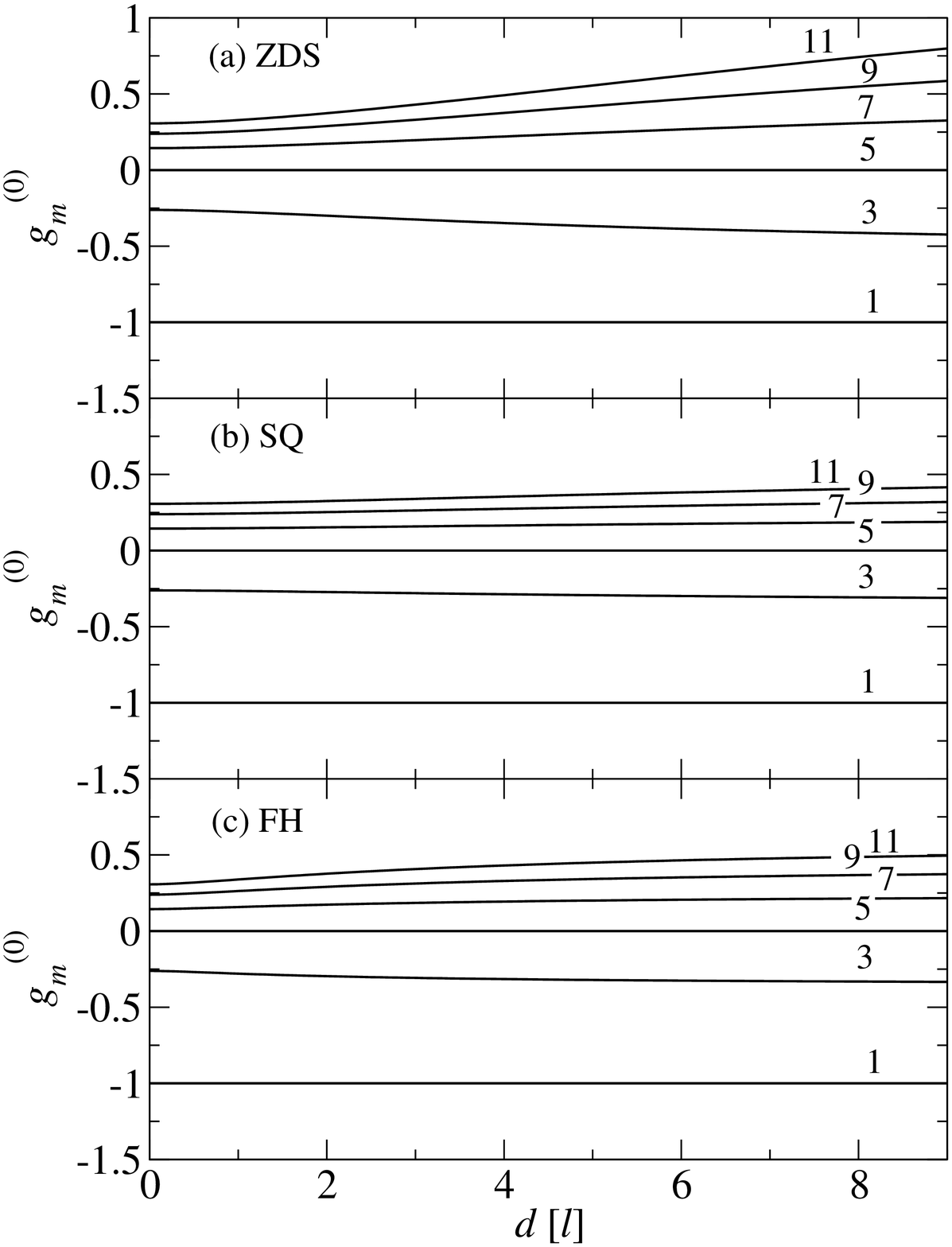}}
\mbox{\includegraphics[width=5.5cm,angle=0]{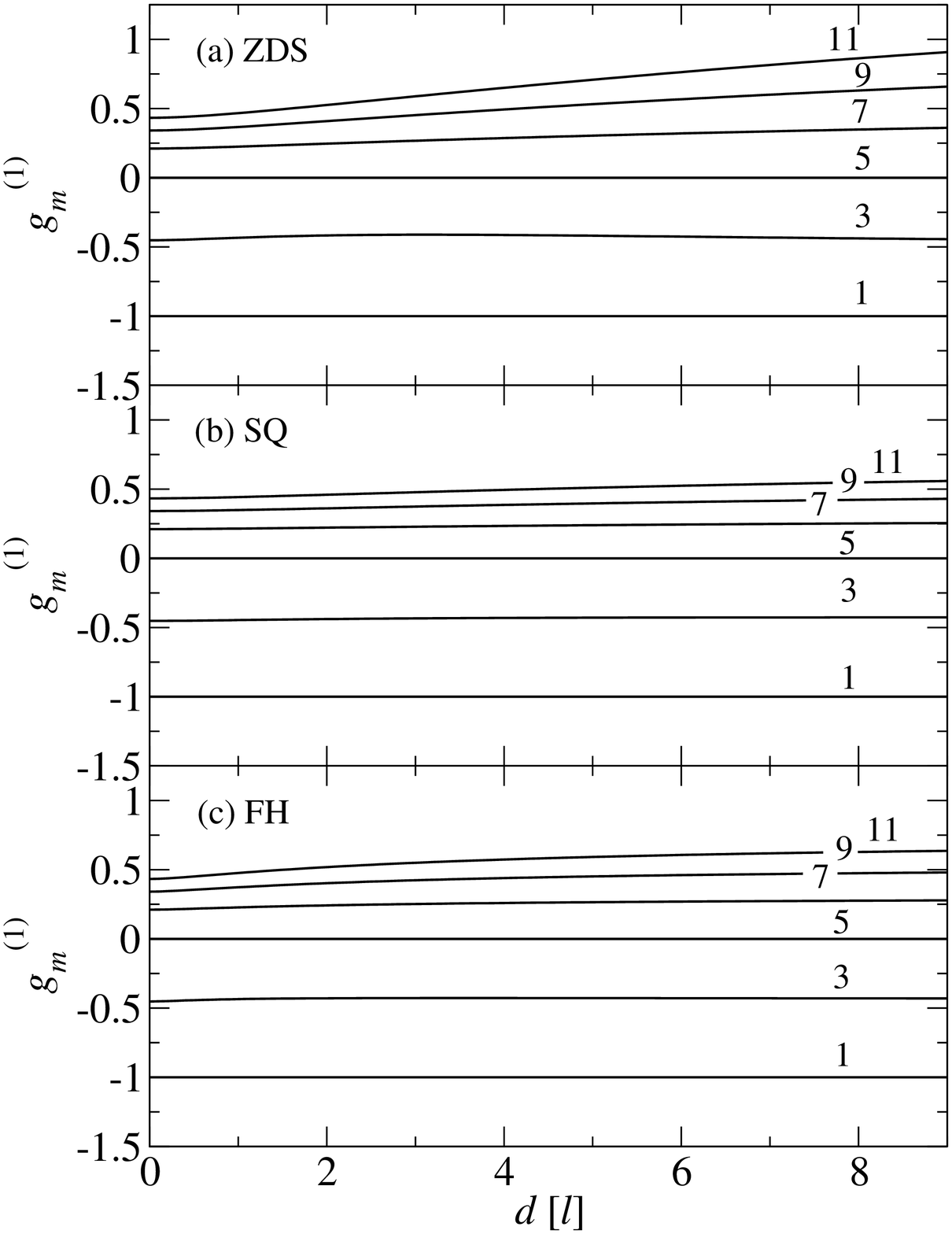}}
\mbox{\includegraphics[width=5.5cm,angle=0]{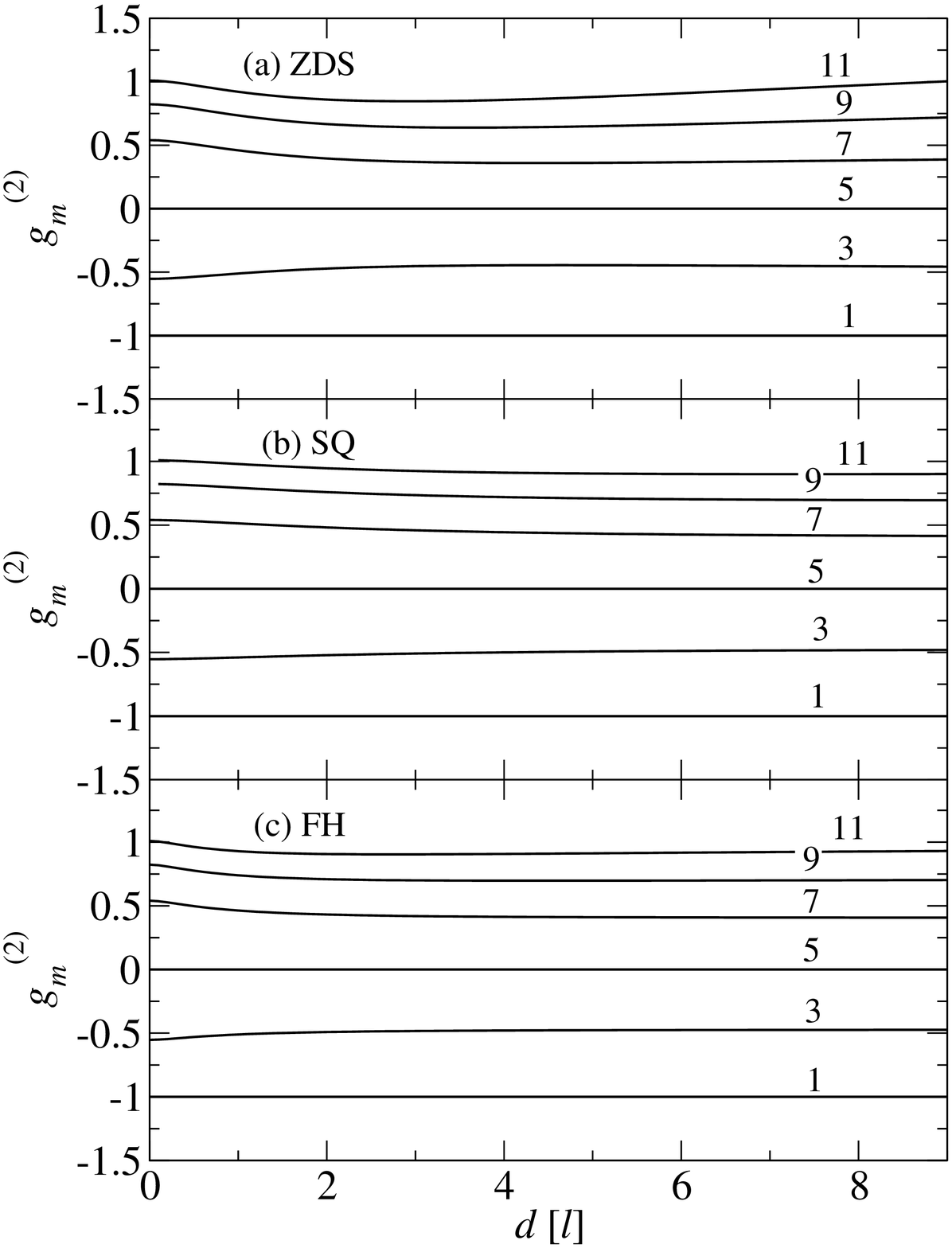}}\\
\end{center}
\hspace{1.cm}LLL\hspace{5.cm}SLL\hspace{5.cm}TLL
\caption{As in Fig.~\ref{f-funcs} except for $g$-functions.}
\label{g-funcs}
\end{figure*}

We now consider the $g$-functions in order to investigate the physics
at filling factor 1/5.  We would expect the Laughlin 1/5 state to
accurately represent the physics if $g^{(n)}_m \sim 0$ for $m\geq 5$.
The left and middle panels of Fig.~\ref{g-funcs} display the
$g$-functions for the LLL and SLL, respectively, for the three
model potentials.  It is clear that for $d=0$ in the LLL and SLL
the Laughlin state is likely a good representation of the exact
state.  This is because $g^{(0)}_m$ and $g^{(1)}_m$ are less than
$0.5$ for $m\geq 5$.  On the other hand, in the TLL (the right panel
of Fig.~\ref{g-funcs}) we observe that, for $d=0$, $1> g^{(2)}_m >
0.5$, indicating that perhaps the 1/5 Laughlin state is not a good
physical description of the exact $\nu=1/5$ state in the TLL.

As a function of the thickness parameter $d$ the
$g$-functions behave in much the same way as the $f$-functions, that
is, for the ZDS, SQ, and FH potentials they increase monotonically as
a function of $d$.  The main difference between the two is that the
local minimum in the $g$-function is obtained for finite $d$ only in the TLL as opposed to
the SLL for the $f$-function (cf. Fig.~\ref{f-funcs}).  A global minimum is obtained 
in $g$ for
the TLL (right panel) for the ZDS potential for finite $d$ while for
the SQ and FH potentials the minima are obtained for very large $d$.

It is clear from inspection that one can obtain a rough qualitative
idea of how well the Laughlin state at 1/5 will represent the physics
as a function of $d$, namely, in the LLL the 1/5 Laughlin state will
be a good description for up to quite large $d$ for the ZDS potential
before presumably losing out to some other state.  The SQ and FH
potentials, on the other hand, will mostly likely produce a state that
is very similar to the 1/5 Laughlin for all $d$.  A similar prediction
is made for the SLL.

For the TLL, however, the ZDS potential will produce a state that is
likely not Laughlin-like for small $d$, a state better described by
Laughlin for intermediate $d$, and becomes unlike Laughlin again for
large $d$.  The SQ and FH potentials are likely to produce a state
that is consistently unlike Laughlin for all $d$.

Our qualitative discussion based on the thickness-dependent behavior of 
$f$- and $g$-functions is consistent with experimental findings:  The 
Laughlin FQH state is abundant in the LLL, scarce in the SLL, and 
essentially non-existent in the TLL.  We emphasize that, by contrast, 
no such qualitative discussion is possible with respect to the 
relative abundance of the Moore-Read even-denominator state in 
various LLs since the MR state, unlike the Laughlin state, 
is \textit{not} an exact ground state of any known two-body Hamiltonian, 
and therefore pseudopotential-based functions such as 
$f$ and $g$ do not provide any direct insight into the MR state.

The $f$- and $g$-functions provide a guide to our intuition and,
perhaps, a qualitative understanding.  However, a way to
quantitatively understand the quality of the physical description of
the Laughlin state is provided by calculating overlaps with exact
wavefunctions which we discuss in the next section.

\section{Results}

\subsection{Overlaps}
\label{overlaps1}

A measure of how accurate a variational wavefunction (Laughlin or
Pfaffian) is compared to the exact wavefunction is encapsulated in the
calculation of the overlap between the two wavefunctions.  An overlap
of unity means the variational state is exact and a vanishing overlap
means the variational state is completely unlike the exact state,
perhaps due to different symmetries (different total angular momentum
for example).  Although the use of overlap calculations in
establishing the nature of the incompressible FQH states has been a
central conceptual and theoretical tool in FQHE studies, it should be
emphasized that the overlap calculation has its limitation since it can
only make statements about (necessarily small) finite systems and
specific FQH ansatz wavefunctions (e.g., Laughlin, Pfaffian).  In spite of these limitations 
the calculation of wavefunction overlap between exact numerical 
wavefunctions for small systems with ansatz variational wavefunctions 
has been a standard theoretical FQHE tool for almost 25 years.

\begin{table}[b]
\begin{center}
\caption{Overlap integrals between the exact ground state wavefunction
using spherical and planar pseudopotentials ($\langle
\Psi_{\mbox{sphere}}|\Psi_{\mbox{plane}}\rangle$), respectively.  Also
given are the overlap between the Laughlin or Pfaffian wavefunction
and the exact ground state wavefunction using spherical and planar
pseudopotentials ($\langle\Psi_{a}|\Psi_{\mbox{plane}}\rangle$ and
$\langle\Psi_{a}|\Psi_{\mbox{sphere}}\rangle$ where $a$ denotes either
$L$ for Laughlin or Pf for Pfaffian as appropriate).  This table
quantifies the similarities and differences between states using
planar or spherical pseudopotentials.}
\begin{tabular}{|c|c|c|c|c|}
\hline
  $N$ & $\nu$ & $\langle\Psi_{\mbox{sphere}}|\Psi_{\mbox{plane}}\rangle$ & $\langle \Psi_a|\Psi_{\mbox{plane}}\rangle$ & $\langle\Psi_a|\Psi_{\mbox{sphere}}\rangle$ \\ \hline
  6 & 1/3 & 0.998840 & 0.992129 & 0.996446 \\ \hline
    & 7/3 & 0.948005 & 0.736947 & 0.528481 \\ \cline{2-5}
    & 13/3 & 0.917779 & 0.021261 & 0.013854 \\ \hline
  5  & 1/5 & 0.999966 & 0.996919 & 0.997427 \\ \hline
    & 11/5 & 0.999974 & 0.997886 & 0.998198 \\ \cline{2-5}
    & 21/5 & 0.988710 & 0.000012 & 0  \\ \hline
  8  & 1/2 & 0.997841 & 0.895311 & 0.921297 \\ \hline
    & 5/2 & 0.968754 & 0.963623 & 0.867392 \\ \cline{2-5}
    & 9/2 & 0.978191 & 0.030311 & 0.002384 \\ \hline
  10 & 1/2 & 0 & 0.889655 & 0 \\ \hline
     & 5/2 & 0.972034 & 0.934183 & 0.837637 \\ \cline{2-5}
     & 9/2 & 0.986467 & 0 & 0 \\ \hline
\end{tabular}
\label{table1}
\end{center}
\end{table}

Before we calculate overlaps we make a technical point.  To calculate
properties of the states and diagonalize the Hamiltonians we have made
use of the spherical geometry~\cite{yang-wu,exact-laughlin,fano}--we
also use the torus geometry later in Sec.~\ref{sec-topo} for obtaining 
the ground state degeneracy at 5/2 since the ground 
state degeneracy does not show up in the spherical geometry.   
The spherical geometry is defined by confining $N$
electrons to the surface of a sphere with a radial magnetic field
produced by a magnetic monopole of strength $Q$ at the sphere center
(the total flux through the sphere is $2Q(hc/e)$).  $Q$ is an integer
or half-integer due to Dirac's quantization condition and is related to
the radius of the sphere through $R=\sqrt{Q}$.  The filling factor in
a particular LL with LL index $n$ is defined through its thermodynamic
limit $\nu=\lim_{N\rightarrow\infty}N/g$ where $g=2(Q+n)+1$ is the
total LL degeneracy.  The total angular momentum $L$ is a good quantum
number and a uniform state is the state with total angular momentum
$L=0$ and is considered incompressible if it has a non-zero energy
gap between the ground state and the low-lying excitation spectra.

We use the pseudopotentials calculated in the infinite
planar geometry for carrying out our spherical system finite 
size diagonalization.  The reasons for this are twofold: (1) it can be
argued that the planar pseudopotentials, since they are the
thermodynamic limit of the spherical pseudopotentials, better
represent the real physical 2D system, and (2) it is much more convenient
when considering finite-thickness modeling potentials to use planar
pseudopotentials.  At any rate, as discussed below, it makes very
little difference whether we use planar or spherical pseudopotentials,
and our conclusions (although perhaps not the precise values of the
overlap in each case) are completely independent of this
approximation.  We believe that all qualitative conclusions in 
this paper are independent of our planar pseudopotential approximation.

\begin{figure*}[t]
\begin{center}
\mbox{\includegraphics[width=5.5cm,angle=0]{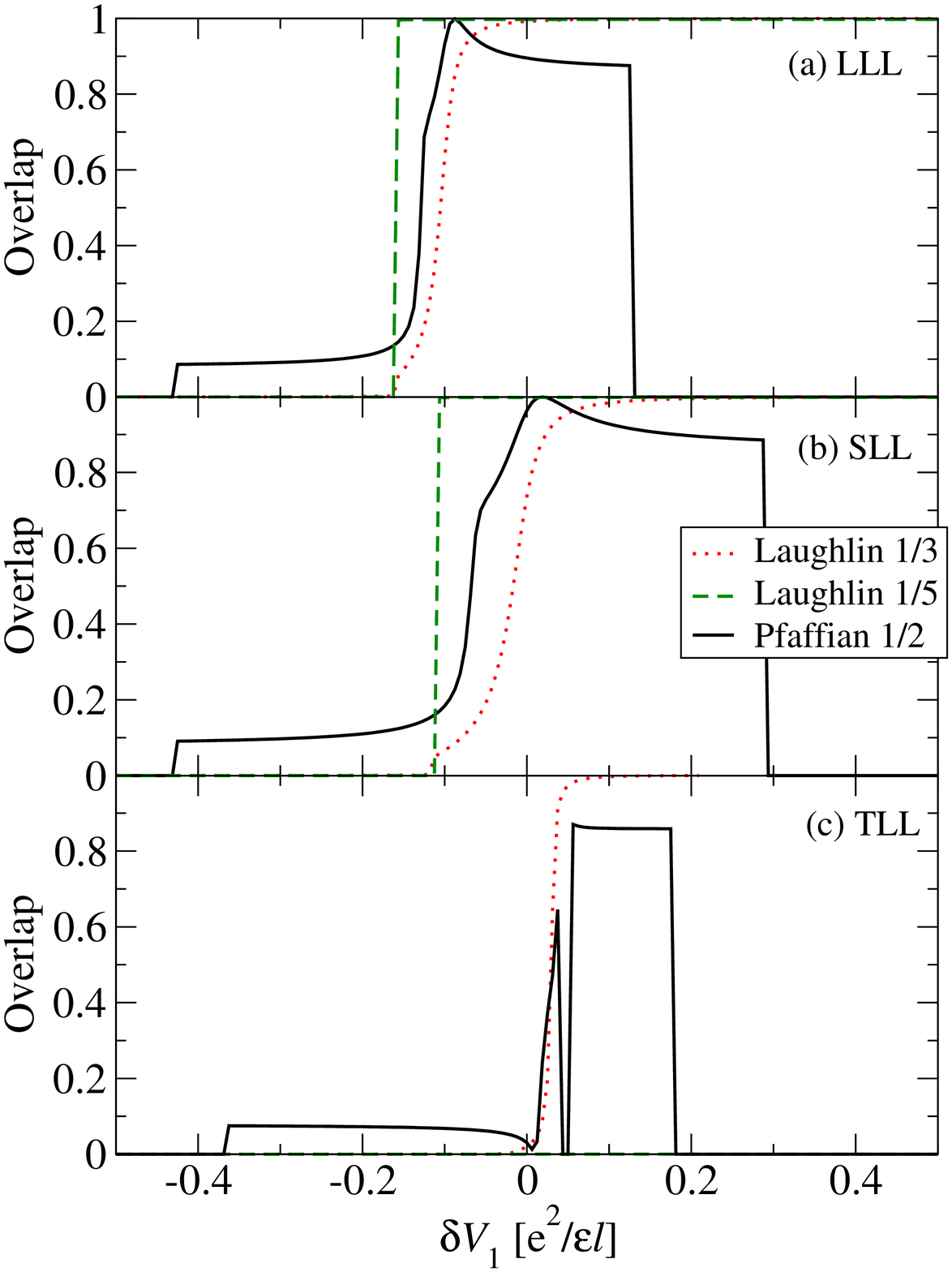}}
\mbox{\includegraphics[width=5.5cm,angle=0]{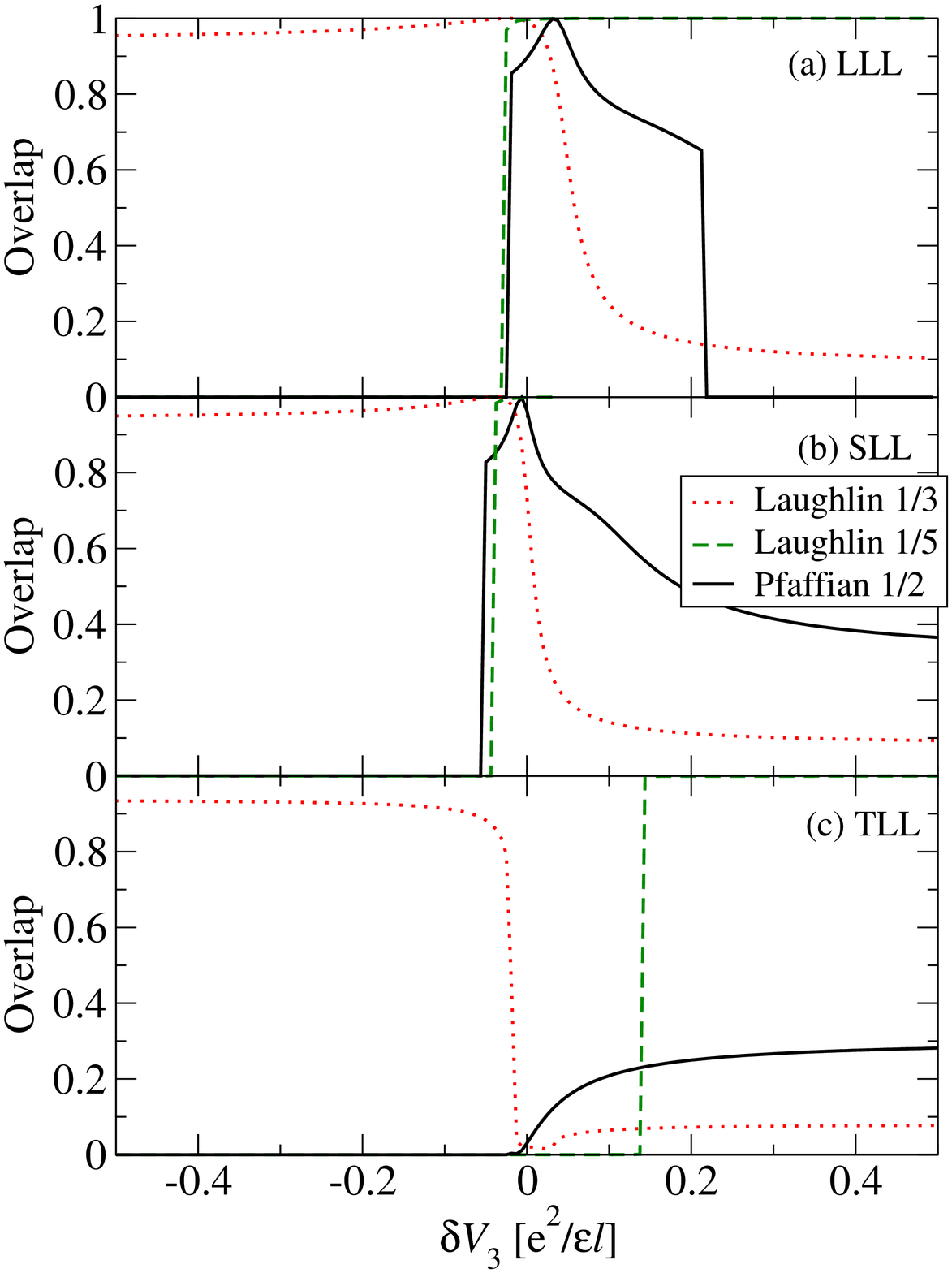}}
\mbox{\includegraphics[width=5.5cm,angle=0]{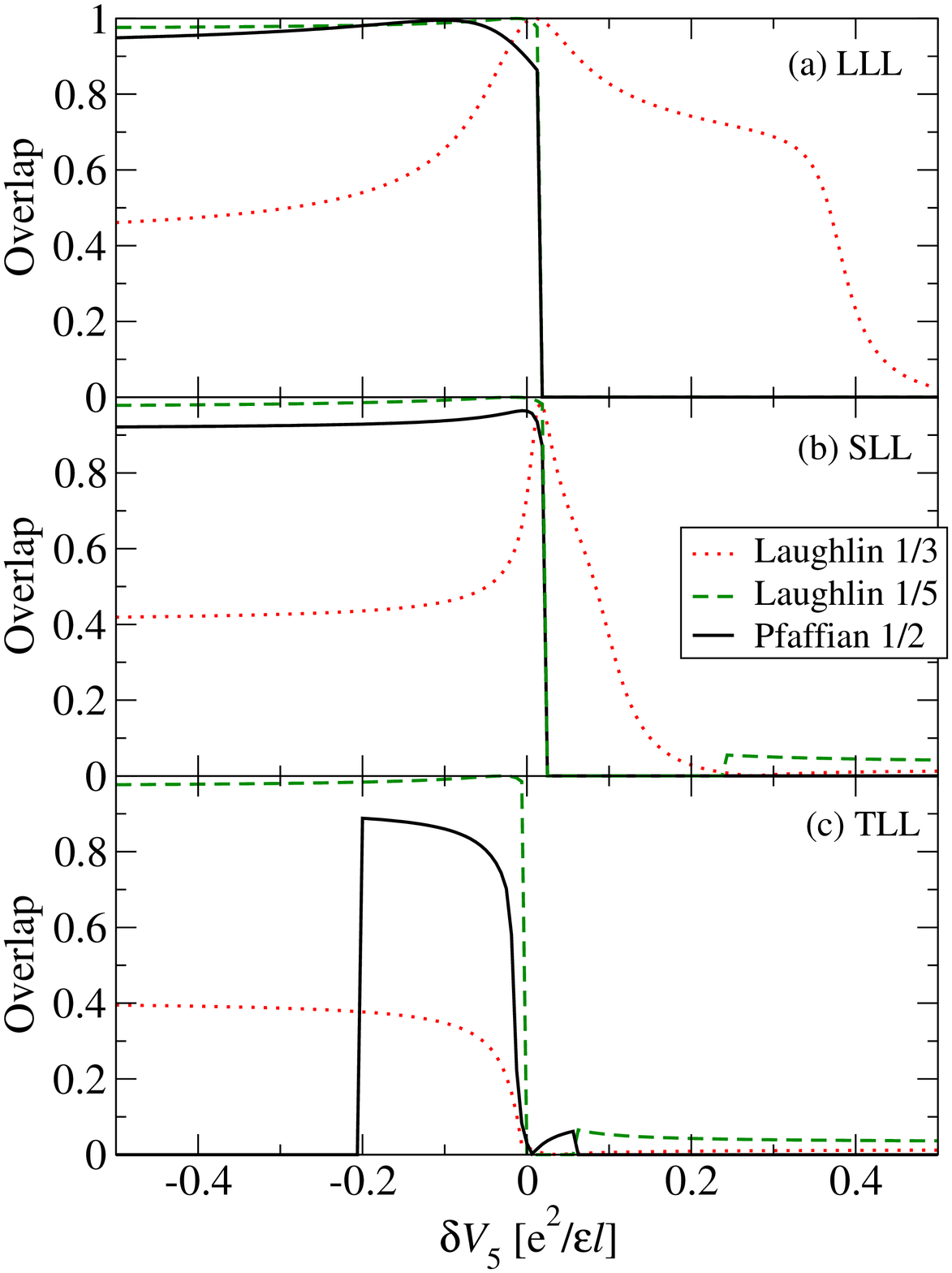}}\\
\end{center}
\hspace{1.5cm}varying $V_1$\hspace{4.cm}varying $V_3$\hspace{4.cm}varying $V_5$
\caption{(Color online) Overlaps of either the Laughlin or the
Pfaffian wavefunction with the exact ground state wavefunction of the
pure ($d=0$) Coulomb Hamiltonian where $V_1$ (left), $V_3$ (middle),
and $V_5$ (right) are varied by $\delta V_1$, $\delta V_3$, and
$\delta V_5$, respectively.  Of course, $\delta V_i=0$ corresponds to
the ``Coulomb point'' where the interaction is a pure 2D Coulomb
interaction.  In the above figures, the situations considered are
$\nu=1/3$ for $(N,l)=(6,7.5)$ (dotted line), $\nu=1/5$ for
$(N,l)=(5,10)$ (dashed line), and $\nu=1/2$ for $(N,l)=(8,6.5)$ (solid
line).}
\label{vm-varied-over}
\end{figure*}

For this work we are interested in the FQHE at filling factors 1/3,
1/5, and 1/2 in the LLL, SLL, and TLL.  As described in
Sec.~\ref{sec-hamil} all the calculations are done in the LLL and all
information about higher LLs is encapsulated by the pseudopotentials
$V_m^{(n)}$.  The relationship between the LL degeneracy $g=2l+1$ ($l$
is the single-particle angular momentum~\cite{foot-sphere}) and the
number of electrons $N$ for these states is as follows: $\nu=1/3$ has
$2l=3N-3$, $\nu=1/5$ has $2l=5N-5$, and $\nu=1/2$ has $2l=2N-3$. In
general, the relation between $l$ and $N$ for some LL filling factor
$\nu$ is $2l=\nu^{-1}N+\chi$ where $\chi$ is referred to as the
``shift''.  The shift is non-zero in the spherical geometry and a
consequence of the finite curvature of the spherical surface--this is
discussed in more detail in Sec.~\ref{sec-topo}.  The ($N,l$) relation
for $\nu=1/2$ was chosen to be the same as it is for the Pfaffian
wavefunction~\cite{pfaff} and the $(N,l)$ for $\nu=1/3$ and 1/5 was
chosen to be the same as for the Laughlin wavefunction.  We further
note that the FQH states considered in the work are not so-called
``alias'' states, that is, the $(N,l)$ relations used only correspond
to fillings 1/3, 1/5, or 1/2 (and 2/3 and 4/5 through particle-hole
symmetry).  This somewhat restricts our choice of particle numbers in
our finite size diagonalization and will be mentioned again in
Sec.~\ref{sec-gap} in relation to the calculation of excitation gaps.  We 
believe that the price of restricting $N$ in avoiding any aliasing problem 
is well worth paying in our work because our wavefunction overlap calculations 
become necessarily unique since no two distinct FQH states compete at 
the same ($N,l$) values.

Confusion about notation can sometimes arise when considering FQH states in higher
LLs.  Here we clarify our conventions.  In the LLL the 1/3, 1/5, and
1/2 states exist at an experimental filling factor of $\nu=1/3$,
$\nu=1/5$, and $\nu=1/2$, respectively.  In the SLL, the LLL (both
spin up and down) are filled and inert yielding filling factors for
1/3, 1/5, and 1/2 of $\nu=2+1/3=7/3$, $\nu=11/5$, and $\nu=5/2$,
respectively.  Finally, for the TLL, both the LLL and SLL (both spin
up and down) are filled and inert yielding, for 1/3, 1/5,
and 1/2, $\nu=4+1/3=13/3$, $\nu=21/5$, and $9/2$.  Hence, it should be
clear that the FQH state at filling 1/5 in the TLL corresponds to an
experimental $\nu=21/5$, for example, and so on.

Since we are using planar, instead of spherical, pseudopotentials we
quantify the difference between the two.  Table~\ref{table1} provides
a number of overlaps for the FQH states considered here for the pure
2D ($d=0$) Coulomb interaction.  What is clear from these results is
that the overlap between the exact ground states using either the
planar ($|\Psi_{\mbox{planar}}\rangle$) or spherical
($|\Psi_{\mbox{sphere}}\rangle$) pseudopotentials (Table~\ref{table1}
third column) are generally high
($\langle\Psi_{\mbox{sphere}}|\Psi_{\mbox{plane}}\rangle>0.9$).  The
major exception is the $N=10$ electron system.  Here the symmetry of
the ground state at filling factor 1/2 in the LLL is different, on the
sphere versus on the plane, yielding a vanishing overlap.  It is for
this reason~\cite{footnote-n10} that we do not consider the $N=10$
electron system in this work.  Furthermore, the overlap for
either the Laughlin or the Pfaffian state between the exact planar and
spherical states is qualitatively similar (Table~\ref{table1} columns
four and five).  One point of note is $N=5$ at $\nu=21/5$ (1/5 in the
TLL); here the overlap with the Laughlin state and the exact planar
state is nearly zero ($0.000012$) but retains the same symmetry.
However, the overlap with the Laughlin state and the exact spherical
state is exactly zero due to a different symmetry.  None of these
minor technical issues has any bearing on our main goal in this work,
which is to study the comparative \textit{qualitative} trends of the
quasi-2D finite layer thickness effect on the stability of the
Laughlin or the Pfaffian FQH state at $\nu=1/5$, 1/3, and 1/2 in the
lowest three orbital Landau levels.  In particular, we want to study 
how the finite thickness of the quasi-2D experimental systems 
affects the comparative stability of the Laughlin (for 
$\nu=1/3$ and 1/5) and the Pfaffian ($\nu=1/2$) 
state in different ($n=0$, 1, 2) orbital Landau levels.

We now return to the calculation of overlaps.  In the fourth column of
Table~\ref{table1} the calculated overlap between the exact state (using planar
pseudopotentials) at filling factor 1/3, 1/5, and 1/2 and the Laughlin
(1/3 and 1/5) and the Pfaffian (1/2) states, respectively, for the
pure 2D ($d=0$) system in the LLL, the SLL, and the TLL is shown 
for $N=5$, 6, 8, and 10.  In
the LLL the overlap at 1/3 and 1/5 is very high ($\sim0.99$).  For
1/2, however, the overlap is not nearly as high ($\sim0.9$).  An
overlap of $0.9$ or less is not considered 
particularly compelling in the FQHE
and is indicative of, perhaps, different physics.  In fact,
experimentally there is no FQHE observed in the LLL at 1/2 to date.
In the SLL the story changes.  For 1/3 the overlap is significantly
decreased ($\sim0.74$) while for 1/5 the overlap stays as high as it
is in the LLL ($\sim0.99$).  The overlap at 1/2 increases in the SLL
compared to the LLL to a respectable value ($\sim0.96$).  In the TLL
the overlap at 1/3, 1/5, and 1/2 is essentially zero although the
symmetry between the two remains the same.  The most straightforward 
conclusion following from the $d=0$ results shown in Table~\ref{table1} is that the 
Laughlin state is the stable FQH state at $\nu=1/3$ and 1/5 
in the LLL even in the strict 2D limit, but the other FQH states may not exist in the 
ideal 2D limit except for $\nu=1/5$ in the SLL.

A theoretical strategy often used in studying the FQHE is to vary the first few
pseudopotentials away from the ideal 2D Coulomb values, in an ad hoc way, to
investigate whether the overlap between the resulting exact states
and the Laughlin or Pfaffian states gets better or worse.  In
Fig.~\ref{vm-varied-over} we consider a Hamiltonian where we
have varied $V_1$, $V_3$, and $V_5$ independently symmetrically around
the Coulomb point in the LLL (left panel), SLL (middle panel), and TLL
(right panel) (the results at the Coulomb point are, of course, given
in Table~\ref{table1}).

Clearly, changes in $V_1$, $V_3$, $V_5$ that bring the effective
$\hat H$ closer to $\hat H_L^{(3)}(\hat H_L^{(5)})$ increase the overlap
between the Laughlin wavefunction at 1/3 (1/5) and the exact
wavefunction, while changes opposite to this decrease the overlap.  
These conditions are obtained, for example, when $V_1$ is
increased producing a high overlap for 1/3 and 1/5 in the LLL and SLL
(Fig.~\ref{vm-varied-over} left panel (a) and (b)).  (In the TLL,
however, 1/5 has a zero overlap due to a symmetry change.) Increasing
$V_3$ takes $\hat H$ into $\hat H_L^{(5)}$ which is evident by the
very high overlap between the 1/5 Laughlin state and the exact state
in LLL, SLL, and TLL ((a)-(c) in the middle panel of
Fig.~\ref{vm-varied-over}).  Nothing particularly non-trivial is
happening here because we know how $\hat{H}$ is connected to
$\hat{H}_L^{(q)}$, i.e., any changes in $\hat{H}$ towards (away from)
$\hat{H}_L^{(q)}$ makes the Laughlin state a better (worse)
description for the $1/q$ FQHE state.

The overlap variation with changing $V_1$, $V_3$, $V_5$ for the 
Pfaffian state ($\nu=1/2$) apparent in Fig.~\ref{vm-varied-over} cannot, 
however, be explained easily since there is no existing two-body Hamiltonian, 
e.g., $\hat{H}_L^{q}$ for the Laughlin state, for which the Pfaffian 
is an exact eigenstate.  
It is clear (and has been shown previously~\cite{rez-hald,scarola}) that in
the LLL and SLL changing $V_1$ and $V_3$ in particular ways can
produce a state with an overlap close to unity for the Pfaffian, i.e.,
Fig.~\ref{vm-varied-over} (a) and (b) in the left and middle panel.
It is also interesting to note that in the SLL the value of $\delta
V_1$ and $\delta V_3$ that gives this high overlap between 
the Pfaffian and the exact state at $\nu=1/2$ is very near the
Coulomb point.  Changing $V_5$ (right panel) can also produce a high
overlap for the Pfaffian.  Qualitatively it behaves similarly to the
1/5 Laughlin state as $V_5$ is decreased in the LLL and SLL.  The
Pfaffian never achieves a particularly high overlap in the TLL for any values of 
$V_1$, $V_3$, $V_5$ we looked at.

\begin{figure*}[t]
\begin{center}
\includegraphics[width=5.5cm,angle=0]{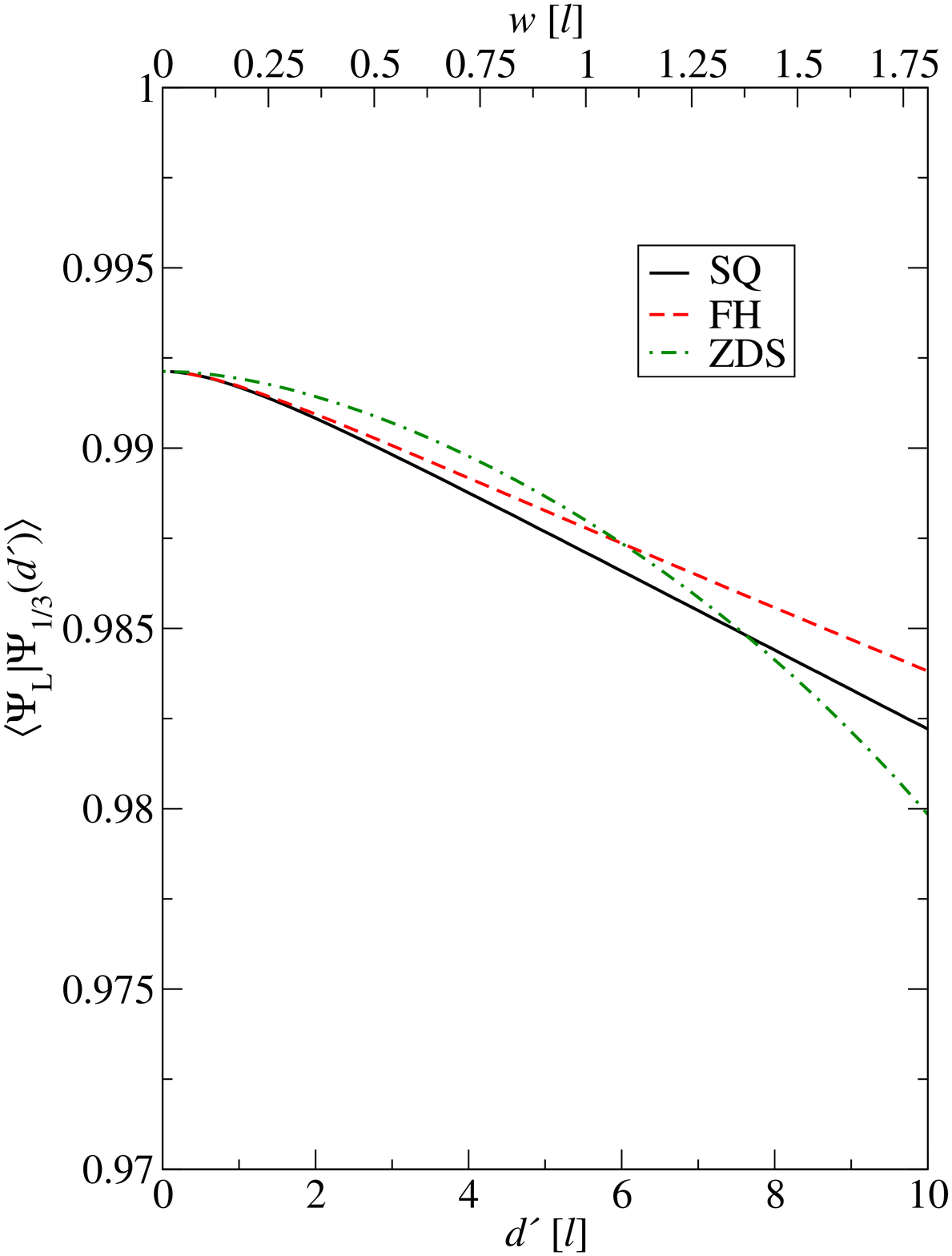}
\includegraphics[width=5.5cm,angle=0]{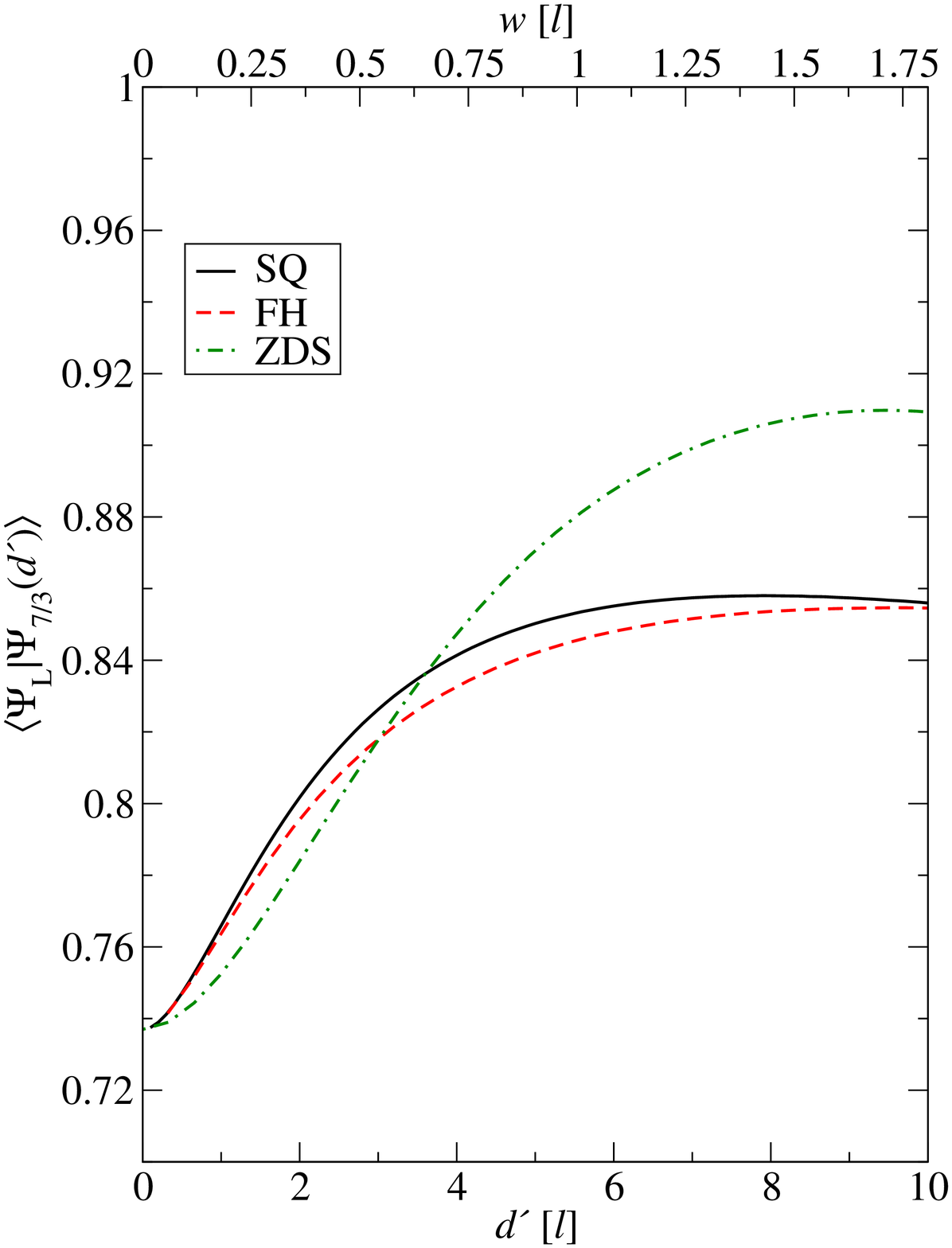}
\includegraphics[width=5.5cm,angle=0]{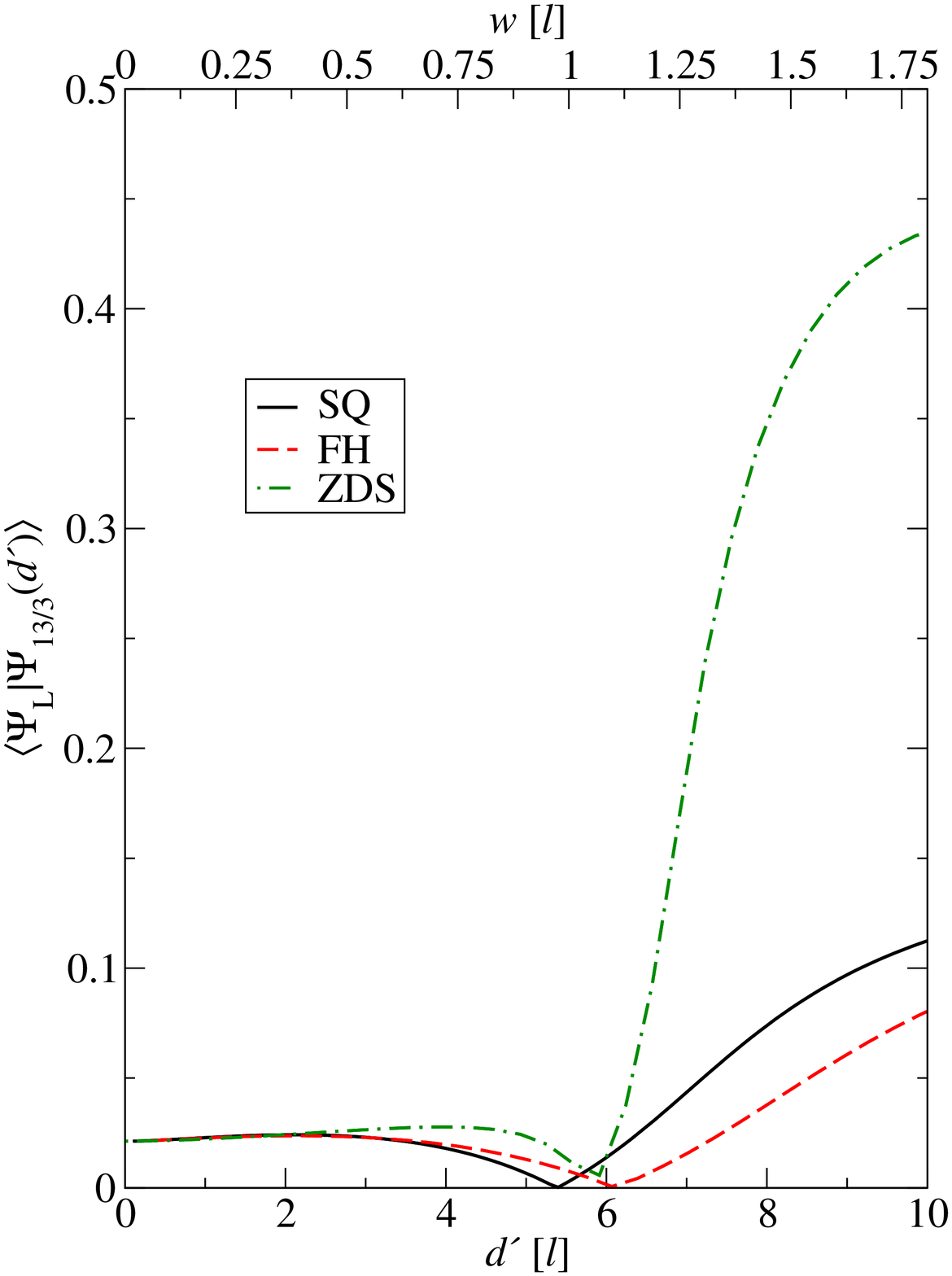}\\
\hspace{0.5cm}(a) LLL\hspace{5.cm}(b) SLL\hspace{5.cm}(c) TLL
\end{center}
\caption{(Color online) Overlap
($\langle\Psi_L|\Psi_{\nu}(d)\rangle$), as a function of thickness
$d^\prime$, between the Laughlin and the exact wavefunctions at the
fractional filling 1/3 in the $n$-th LL, i.e., the LLL ($\nu=1/3$)
(left), the SLL ($\nu=7/3$) (middle), and the TLL ($\nu=13/3$)
(right).  The overlap as a function of $w$ is also given on the
horizontal axis at the top of each plot.  The particular system has
$N=6$ electrons at $l=7.5$.  The different curves represent the
different finite-thickness modeling potentials: SQ (solid line), FH
(dashed line), and ZDS (dashed-dotted line).  Note the different scale
in each plot of the vertical axis.}
\label{overlaps-13}
\end{figure*}

However, the technique of changing $V_m$'s is arbitrarily artificial and ad
hoc, without shedding much light on how real experimental quasi-2D
systems, where $V_m$'s are determined by the layer thickness $d$, will
behave.  (In fact, {\em all} the $V_m$'s change when considering
finite thickness, a fact that is further investigated in
Sec.~\ref{contour-sec}--this means that tuning just one specific 
$V_m$, e.g., $V_1$ or $V_3$ or $V_5$, as done in 
Fig.~\ref{vm-varied-over} is purely a theoretical construct which is impossible 
to achieve in real 2D systems.)  To understand the variation in the states as
a function of $d$ and their incompressible or compressible nature we
carefully define the overlap calculated previously in
Ref.~\onlinecite{he} which is $\langle
\Psi_{L(Pf)}|\Psi_{\nu}(d)\rangle$.  This overlap quantifies exactly
how similar $|\Psi_{\nu}(d)\rangle$ is to the Laughlin wavefunction
$|\Psi_L\rangle$ ($\nu=1/3$ or 1/5) or the Pfaffian wavefunction
$|\Psi_{Pf}\rangle$ ($\nu=1/2$) as a function of the finite layer
thickness $d$.  We have some intuition (and previous results~\cite{he})
about how the overlap will behave with increasing $d$ after studying
the $f$- and $g$-functions (cf. Sec.~\ref{fg-funcs}).  In
Figures~\ref{overlaps-13},~\ref{overlaps-15}, and~\ref{overlaps-12} we
report overlaps in the LLL (left panel), SLL (middle panel), and the
TLL (right panel) of the Laughlin or Pfaffian wavefunction and the
exact wavefunction for the finite-thickness models of ZDS, SQ, and
FH for fillings 1/3, 1/5, and 1/2, respectively.  We emphasize that the 
variation in $d$ cannot be described in terms of a variation in the 
value of one (or a few) pseudopotentials ($V_m$).

In Figs.~\ref{overlaps-13},~\ref{overlaps-15},~\ref{overlaps-12},~\ref{overlap-exp},
and~\ref{o-pqw} we normalize $d$ defining a new width parameter
$d^\prime$ such that all the ``widths'' for each potential are defined
equivalently.  For the SQ and FH confinement (as well as the PQW
confinement) one can calculate $w=\sqrt{\langle z^2\rangle-\langle
z\rangle^2}$ which is the variance of the wavefunction (i.e., the
root-mean-square fluctuation in the electron position).  Using the
values of $w$ for each potential, $w_{SQ}$ for the SQ and $w_{FH}$ for
FH, we normalize the FH potential to the SQ potential.  That is, we
rescale $d$ in the FH potential to
$d^\prime=(w_{FH}/w_{SQ})d=(0.57735/0.180756)d$, while $d^\prime$ for
the SQ confinement is just the original $d$ in that model (since
$w_{SQ}/w_{SQ}=1$).  To rescale the ZDS potential we use a more ad hoc
but well defined method.  Using $\nu=1/2$ in the SLL we scale the
maximum in the overlap between the exact wavefunction at $d$ and the
Pfaffian wavefunction for the ZDS potential to be equal to the maximum
using the SQ potential, i.e., $d^\prime=(4.6/1.4)d$.  In this way the
behavior of the overlap as a function of $d^\prime$ for each
finite-thickness potential is quantitatively similar.  Further, in
Figs.~\ref{overlaps-13},~\ref{overlaps-15},~\ref{overlaps-12},~\ref{overlap-exp},
and~\ref{o-pqw} we also give the overlaps as functions of $w$ in units
of magnetic length.  With this parameterization one is able to
distinguish between two regimes of layer thickness, i.e., $w/l<1$ and
$w/l>1$.  Note that our rescaling of the width parameter from $d$ to
$d^\prime$ ($\propto w$) is a purely non-essential book-keeping
procedure which makes it explicit that, when the quasi-2D width
parameter is properly defined (i.e., $d^\prime$), then the different
quasi-2D models show similar quantitative trends in the calculated
overlaps as a function of layer width.  Theoretical descriptions in
terms of $d$ or $d^\prime$ are completely equivalent--the only
advantage of using the normalized thickness parameter $d^\prime$ is
that the calculated overlap is now quantitatively similar in all the
quasi-2D models we consider.

For the sake of completeness we provide below the formulae for 
$w=\sqrt{\langle z^2\rangle-\langle z\rangle^2}$ and $d^\prime$ 
for the four models (SQ, FH, ZDS, and PQW) respectively in 
terms of their wavefunction parameter, i.e., $d$, as 
given in Eqs.~\ref{eq-zds}-\ref{eq-fh} (and Eq.~\ref{eq-pqw})
\begin{eqnarray}
w_{SQ}&=&0.180756 d\nonumber\;,\\
w_{FH}&=&0.57735 d\nonumber\;,\\
w_{PQW}&=&0.5 d\nonumber\;,\\
d^\prime_{SQ}&\equiv&\frac{w_{SQ}}{w_{SQ}}d=d\nonumber\;,\\
d^\prime_{FH}&\equiv&\frac{w_{FH}}{w_{SQ}}d=3.194085d\nonumber\;,\\
d^\prime_{SQ}&\equiv&\frac{w_{PQW}}{w_{SQ}}d=2.76616d\nonumber\;,\\
d^\prime_{ZDS}&\equiv&\frac{4.6}{1.4}d = 3.285714 d\nonumber\;.
\end{eqnarray}
Note that for the ZDS model there is no single particle 
wavefunction in the $z$-direction, $\eta(z)$, that produces 
the effective potential of the form of the ZDS model, hence, 
we cannot define $w_{ZDS}$, and $d^\prime$ is calculated 
as described above.

\subsubsection{Filling factor 1/3 (Laughlin wavefunction)}

We first concentrate on filling factor 1/3 (Fig.~\ref{overlaps-13}).
In the LLL (left panel) the overlap between the Laughlin state and the
exact state decreases as $d^\prime$ increases.  In fact, this was
previously shown for the ZDS potential in Ref.~\onlinecite{he}.  The
SQ and FH potentials also show a decrease in overlap for increasing
$d^\prime$ but the change is mild compared to the ZDS potential in
that the overlap drops from near unity only to
$\sim0.975$-0.96~\cite{foot}.  Qualitatively, the finite-thickness
could eventually destroy the FQHE at 1/3 in the LLL, again, as first
reported in Ref.~\onlinecite{he}. But our results tend to support a
scenario where such a destruction is continuous as a function of
$d^\prime$, not abrupt at any particular value of $d^\prime$.  
More specifically, our finite system study, as shown in 
Figs.~\ref{overlaps-13}-\ref{overlaps-12}, does not indicate 
the occurrence of a thickness-driven sharp transition from an 
incompressible FQH state to a compressible one--any such transition 
seems continuous, although in practice the excitation gap may 
become very small for large thicknesses.

In the SLL (middle panel of Fig.~\ref{overlaps-13}) the overlap starts
at a modest value ($\sim0.74$) (cf. Table~\ref{table1}) for
$d^\prime=0$ and an increase in overlap is seen for increasing
$d^\prime$ until a maximum is obtained for large $d^\prime$
(increasing $d^\prime$ to unrealistic values produces an overlap that
approaches zero).  We note that the value of $d^\prime$ where the
highest overlap is obtained corresponds to $w/l\approx1$ which,
perhaps, more clearly shows the effect of the layer thickness, i.e.,
the overlap is seen to decrease as $w/l$ moves away (negatively or
positively) from unity--this qualitative feature is again seen when
filling factor 1/2 in the SLL is investigated,
cf. Fig.~\ref{overlaps-12}.  This result could be anticipated from the
observation that the $f$-functions for the SLL have a
local minimum for non-zero $d$ (or scaled $d^\prime$).  However, it
should be noted that the overlap, while increasing for non-zero
$d^\prime$, still only reaches a modest value of
approximately 0.84-0.92 (depending on the quasi-2D model).

In the TLL (right panel of Fig.~\ref{overlaps-13}) the overlap never
reaches a value greater than $0.5$ (for any model) indicating that
the Laughlin state is not a good description of the exact state in the
TLL.  Again, this result is evident in the Laughlin-unfriendly nature
of the $f$-functions in the TLL.

\subsubsection{Filling factor 1/5 (Laughlin wavefunction)}

\begin{figure*}[t]
\begin{center}
\includegraphics[width=5.5cm,angle=0]{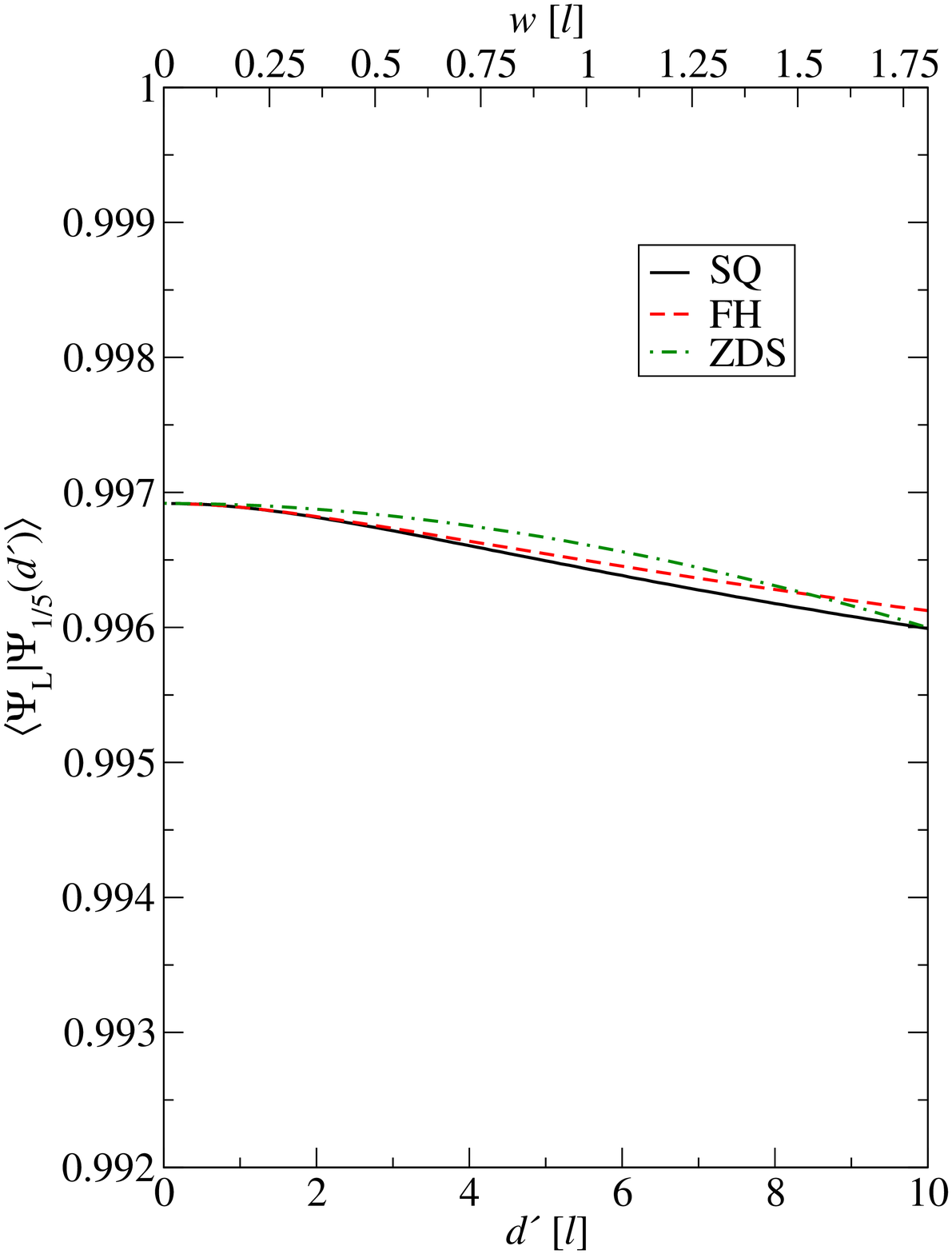}
\includegraphics[width=5.5cm,angle=0]{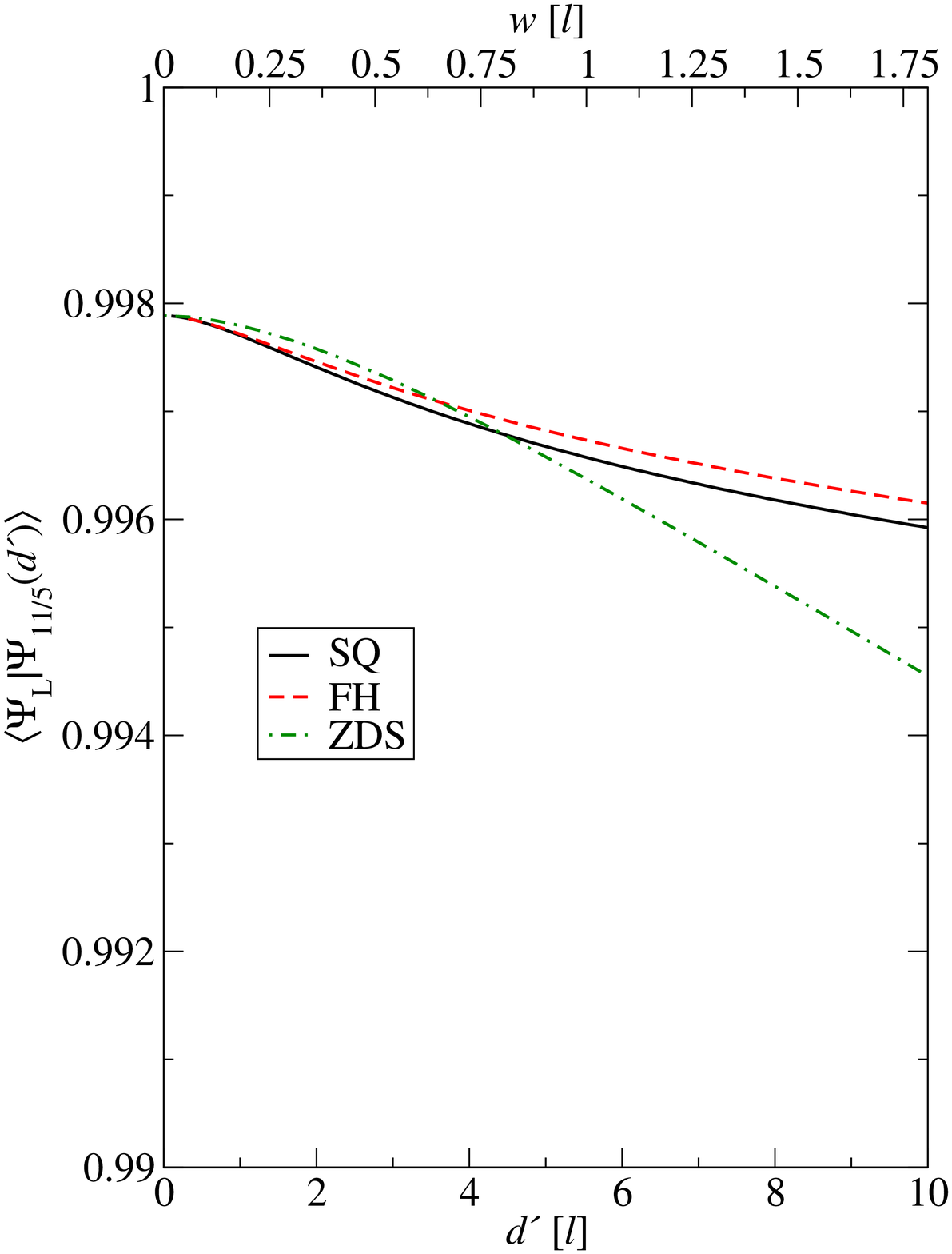}
\includegraphics[width=5.5cm,angle=0]{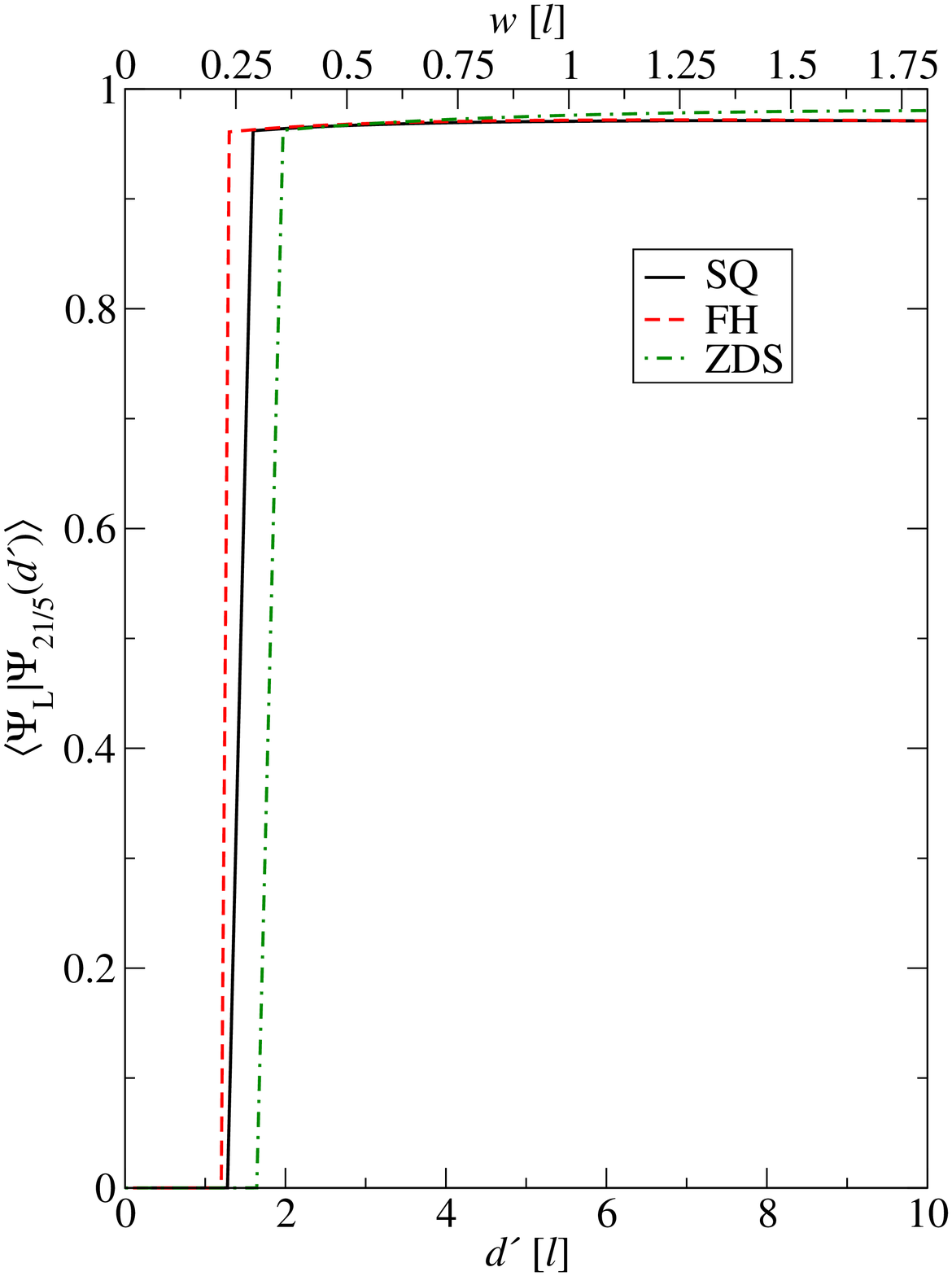}\\
\hspace{0.5cm}(a) LLL\hspace{5.cm}(b) SLL\hspace{5.0cm}(c) TLL
\end{center}
\caption{(Color online) Same as Fig.~\ref{overlaps-13} except for
fractional filling 1/5.  The particular system has $N=5$ electrons at
$l=10$.}
\label{overlaps-15}
\end{figure*}

Next we consider filling factor 1/5 in the LLL (left panel), 
SLL (middle panel), and TLL (right panel) shown in
Fig.~\ref{overlaps-15}.  From an investigation of the
$g$-functions we would expect the 1/5 Laughlin state to have a high
overlap with the exact state for a large range of $d^\prime$ in the
LLL and SLL and that is, in fact, what can been seen 
in Fig.~\ref{overlaps-15}.  In the LLL and SLL
the overlap is greater than $0.995$ for all models up to large
$d^\prime$ (the overlap in the LLL and SLL appear nearly identical in
their qualitative and quantitative behavior).  This result indicates
that the 1/5 FQH state is as strong in the SLL as it is in the LLL,
 which is consistent with earlier
results~\cite{reynolds,macdonald-SLL,toke,meskini}.  Note that nothing
interesting happens to the overlap in any of the first three LLs at
$w/l\approx1$.

The TLL provides a strange scenario for 1/5.  Here the exact ground
state has a different symmetry from the Laughlin state at
$d^\prime=0$ yielding a vanishing overlap.  However, as $d^\prime>0$
the overlap very abruptly becomes non-zero and large ($\sim0.95$).
Again, large $d^\prime$ reduces the overlap severely for the ZDS
potential (not shown) while only moderately for the SQ and FH
potentials (although in the reported range of $d^\prime$,
on this scale, the overlap appears essentially constant).

Note that the qualitative behavior of our 
calculated overlap for filling factor 1/3 and 1/5 in
the SLL is different in that for 1/3 increasing $d^\prime$
\textit{improves} the overlap while for 1/5 increasing $d^\prime$
\textit{worsens} the overlap, as it does in the LLL.  Based on our overlap 
calculation we would therefore predict a more (less) stable 
7/3 (11/5) FQHE with increasing quasi-2D layer thickness.

\subsubsection{Filling factor 1/2 (Pfaffian wavefunction)}
\label{ff12}

\begin{figure*}[t]
\begin{center}
\includegraphics[width=5.5cm,angle=0]{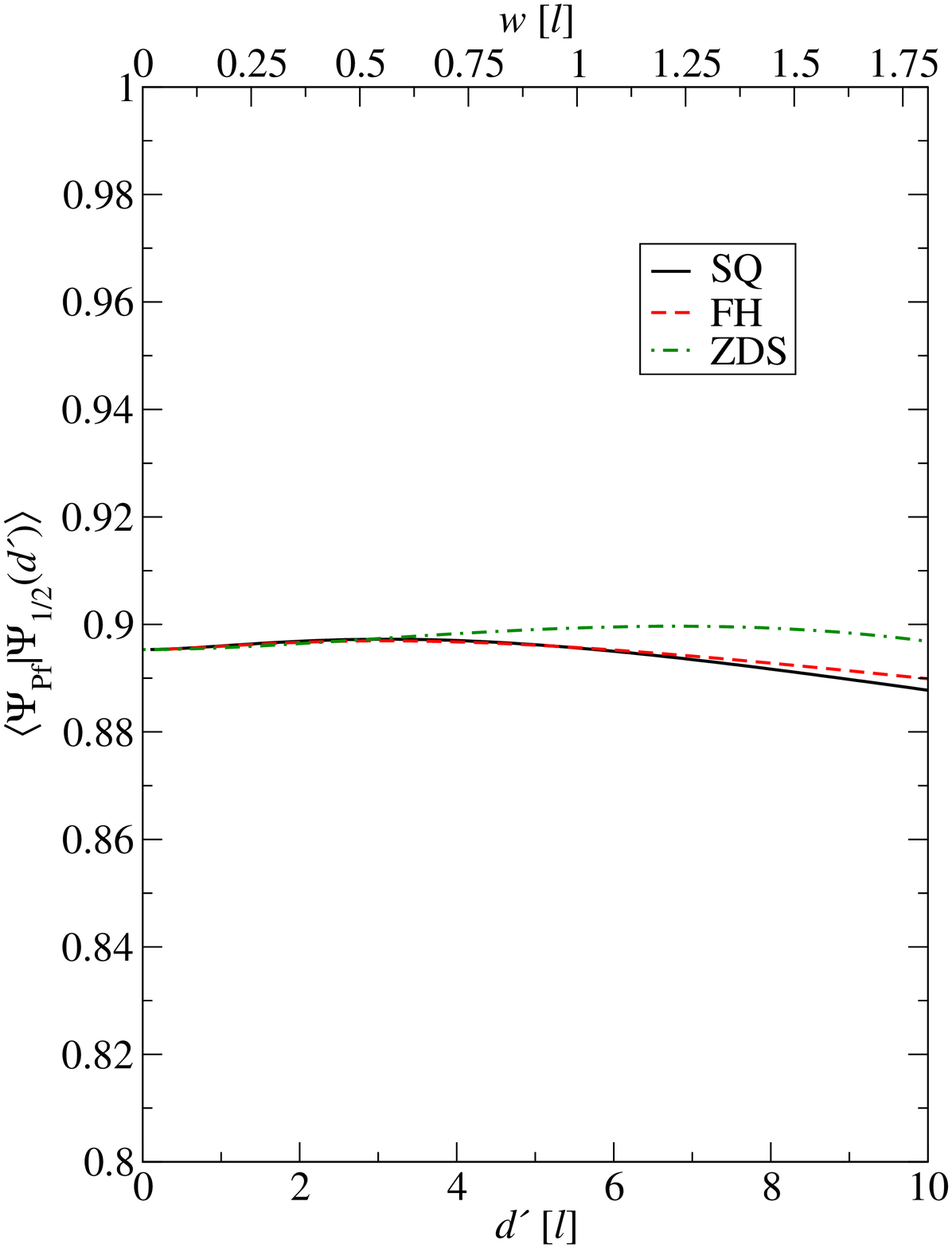}
\includegraphics[width=5.5cm,angle=0]{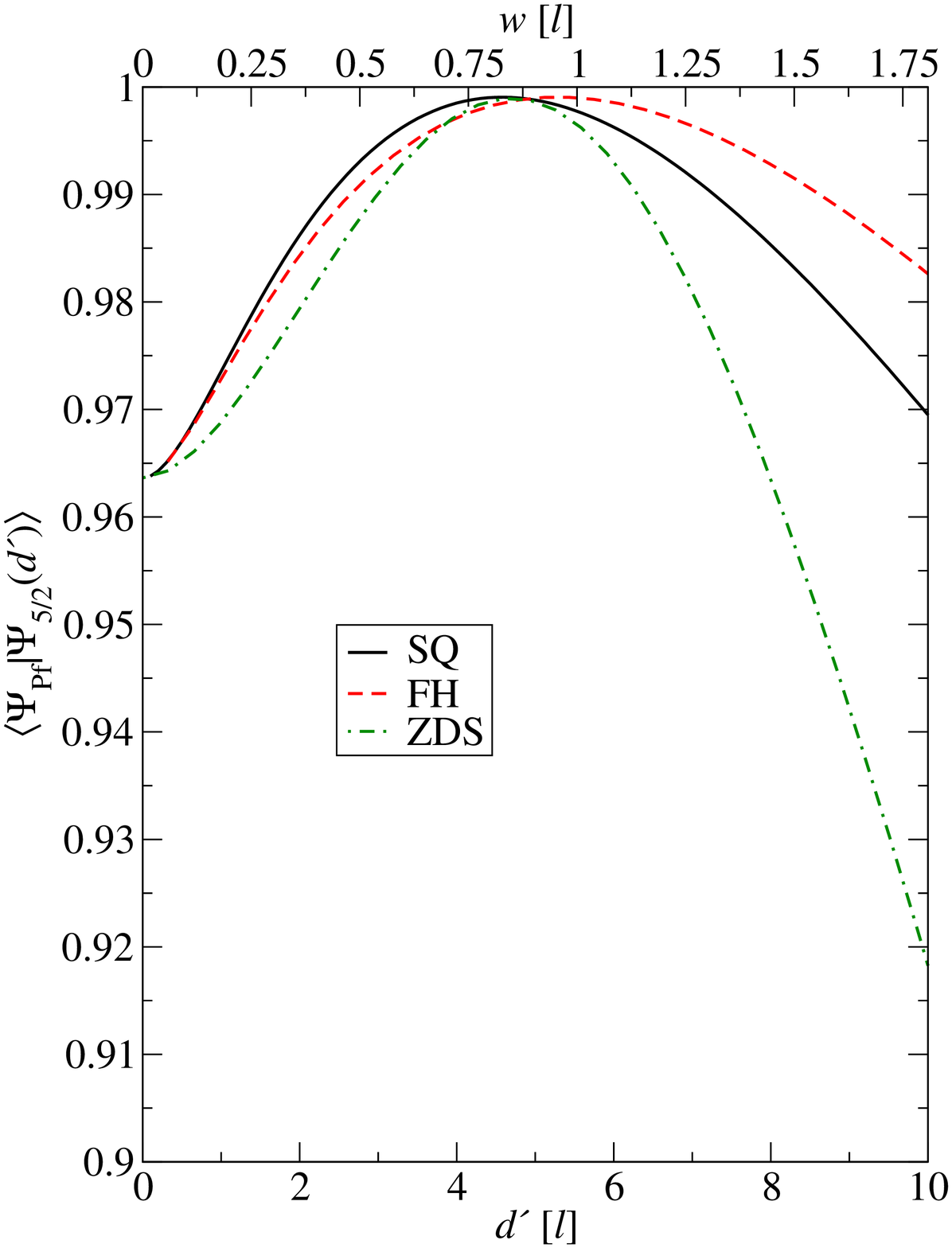}
\includegraphics[width=5.5cm,angle=0]{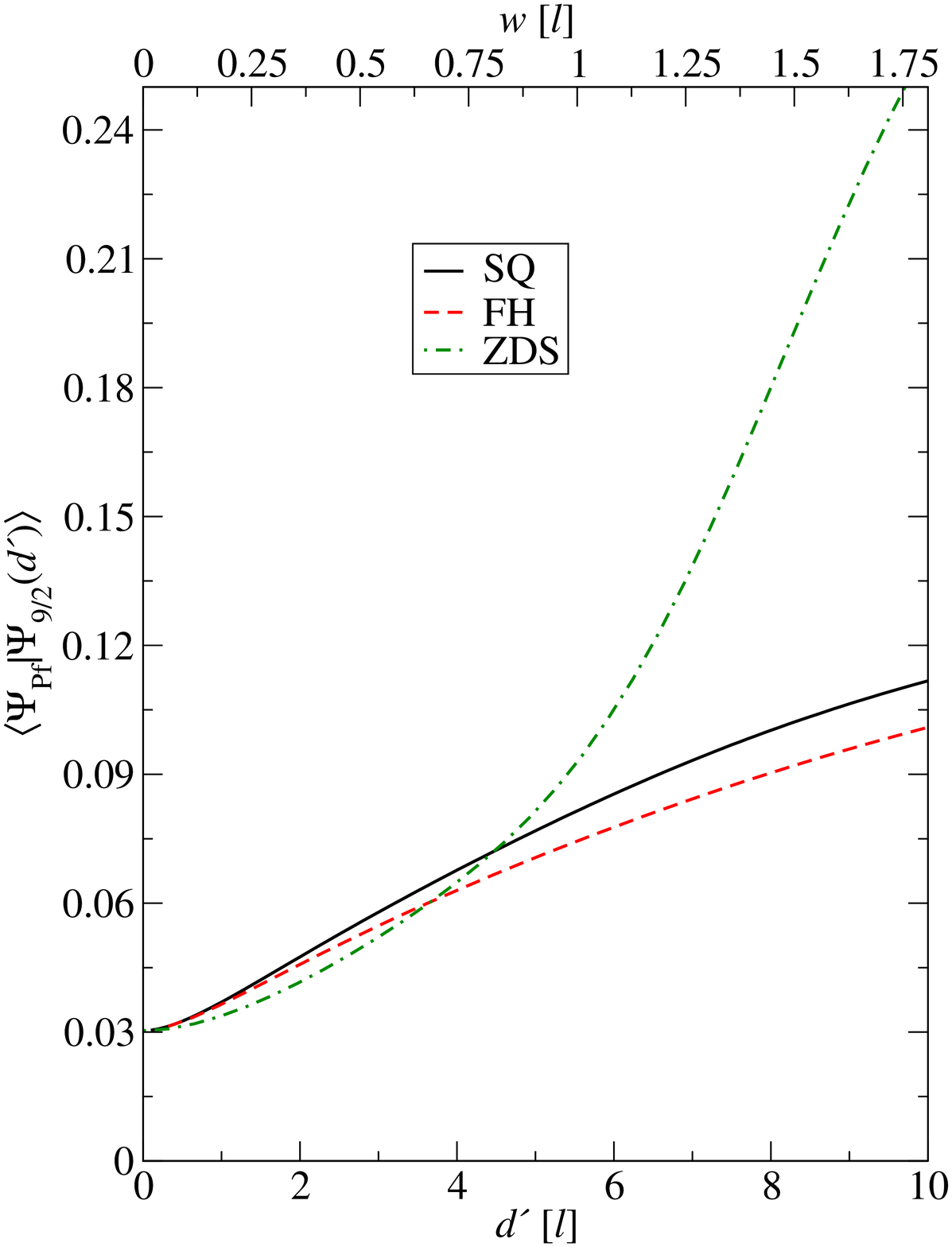}\\
\hspace{0.5cm}(a) LLL\hspace{5.cm}(b) SLL\hspace{5.0cm}(c) TLL
\end{center}
\caption{(Color online) Same as Figs.~\ref{overlaps-13}
and~\ref{overlaps-15} except for fractional filling 1/2.  Hence, the
relevant overlaps are between the exact ground state wavefunction at
some thickness $d^\prime$ ($|\Psi_{\nu}(d)\rangle$) and the Pfaffian
wavefunction ($|\Psi_{Pf}\rangle$).  The particular system has $N=8$
electrons at $l=6.5$.}
\label{overlaps-12}
\end{figure*}

Lastly, and most importantly, we turn to filling 1/2 in the LLL
(left panel), SLL (middle panel), and TLL (right panel) in
Fig.~\ref{overlaps-12}, considering now 
the overlap between the Moore-Read Pfaffian wavefunction 
and the exact numerical wavefunction at $\nu=1/2$.  In the LLL, the overlap starts at
approximately $0.9$ for $d^\prime=0$.  However, unlike the LLL
behavior for 1/3 and 1/5 where the overlap monotonically decreases as
$d^\prime$ increases, the overlap between the Pfaffian and the exact
wavefunction here increases to a weak maximum for some finite
$d^\prime$.  The maximum overlap for the ZDS, SQ, and FH potentials is
moderate and barely above the $d^\prime=0$ value of approximately
$0.9$, however.  Nevertheless, there is a distinct quasi-2D width
induced enhancement of the Pfaffian overlap here at $\nu=1/2$, not
apparent in the corresponding $\nu=1/3$, and 1/5 Laughlin states 
in the LLL where the overlap decreases monotonically with 
increasing layer thickness.

In the SLL the overlap of the Pfaffian wavefunction with
the exact wavefunction increases from 0.96 for $d^\prime=0$ to
essentially unity for finite $d^\prime/l\sim5$.  This result suggests
that the finite layer thickness of the
quasi-2D experimental system actually leads to an exact wavefunction
that is more like the Pfaffian at $\nu=5/2$ for finite $d^\prime$ than
for $d^\prime=0$.  Continually increasing $d^\prime$ beyond 
this optimal value produces
wavefunctions with decreasing overlaps (similar to 1/3 and 1/5).  This
finite width induced stabilization of the Pfaffian state at $\nu=5/2$
perhaps explains the fragility of the observed 5/2 FQHE in
experiments.

In both the LLL and SLL we see that the highest $\nu=1/2$ Pfaffian 
overlap occurs for
$w/l\approx1$.  This is similar to what was observed for filling
factor 1/3 in the SLL (Fig.~\ref{overlaps-13}) but the effect is more pronounced 
at $1/2$.

Filling factor 1/2 in the TLL behaves much the same as for 1/3.  The
 overlap starts below $0.05$ for $d^\prime=0$ and achieves a value
 which is model dependent, between 0.125-0.25 for some
 finite $d^\prime$ (for very large $d^\prime$ all overlaps eventually
 approach zero) .  From this calculation one would not expect to see
 the FQHE at $\nu=1/2$ in the TLL.

The preceding calculations establish that including finite
layer thickness of the realistic quasi-2D system produces non-trivial
behavior of the overlap between either the Laughlin or Pfaffian
wavefunction with the exact wavefunction.  The physical expectation
based on our overlap calculation 
would be that the FQHE in the SLL is stronger for the 1/5 and 1/2
state (provided there is some finite $d^\prime$) than it is for the
1/3 state.  Experimentally 1/3, 1/5, and 1/2 (and their particle-hole
conjugates) are all observed in the SLL~\cite{choi,pan}.

\begin{figure}[t]
\begin{center}
\mbox{\includegraphics[width=6cm,angle=-90]{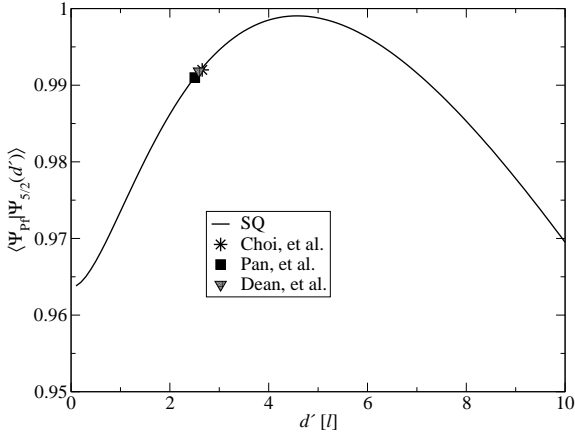}}
\end{center}
\caption{Same as Fig.~\ref{overlaps-12} (middle panel, SLL) except
only the SQ potential is shown.  The experimental $d^\prime$ of
Refs.~\onlinecite{choi} and~\onlinecite{pan} are shown as a asterisk
and solid square, respectively.  The particular system has 
$N=8$ electrons at $l=6.5$.  Also show is the result for a 
very recent experiment (Ref.~\onlinecite{fqhe-SLL-2008}) by Dean, et al.}
\label{overlap-exp}
\end{figure}

To make a connection between experiment and theory we consider the
recent experimental observations of $\nu=5/2$ reported by Choi, et
al.~\cite{choi} and Pan, et al.~\cite{pan} which were obtained in
quantum well structures of width $30$ nms.  In Ref.~\onlinecite{choi}
and~\onlinecite{pan} $\nu=5/2$ was observed at a $B$ field of
approximately $B\approx4.6$ T and $5.2$ T, respectively.  Using the
standard formula for magnetic length in GaAs/AlGaAs quantum wells
($l\approx25\mbox{nm}/\sqrt{B[\mbox{T}]}$) we plot in
Fig.~\ref{overlap-exp} the value of $d^\prime$ these particular
experiments correspond to on our overlap plot at 1/2 in the SLL
($\nu=5/2$) for the SQ potential.  From this comparison we see that
the experimental systems are not {\textit{optimized}} to observe the
strongest possible Pfaffian state at $\nu=5/2$.  Somewhat increasing
the value of the quantum well width so that $d^\prime/l$ (or
equivalently $d/l$) is increased to the optimal value 
should give more a stable 5/2 FQHE!  This is further elaborated (and 
reinforced) by studying the ground state topological degeneracy 
on a torus, a defining hallmark of non-Abelian states in the 
next section.

\subsection{Threefold topological degeneracy signature of the 
Pfaffian state on the torus}
\label{sec-topo}

Recall (cf. Sec.~\ref{overlaps1}) that in the spherical geometry, the
signature of an incompressible FQH state is the existence of a
rotationally symmetric uniform state with total angular momentum $L=0$
with a finite excitation gap to higher energy states.  A given FQH state
generally has a ``shift'' in the equation relating the number of
electrons $N$ to the total flux $2Q$ through the finite sphere.  The MR Pfaffian
wavefunction is written on the sphere as
\begin{equation}
\label{PfaffianSphere}
 \Psi_{Pf}={\rm Pf}\left( \frac{1}{u_i v_j-u_j v_i}\right) \prod_{i<j}(u_i v_j-u_j v_i)^2,
\end{equation} 
where the spinor coordinates are $u_j =\cos(\theta_j/2) {\rm e}^{-i\phi_j/2}
$ and $v_j =\sin(\theta_j/2) {\rm e}^{i\phi_j/2} $ with $(\theta,\phi)$
being the coordinates on the surface of a sphere.  The Pfaffian symbol
above corresponds to
\begin{equation}
{\rm Pf}\left(A_{ij} \right) =\sum_{\sigma}\epsilon_{\sigma}
A_{\sigma(1)\sigma(2)}\dots A_{\sigma(N-1)\sigma(N)},
\end{equation} 
where $\sigma$ are permutations of the N particle indices.  It is
found that the wavefunction in Eq.~\ref{PfaffianSphere} requires a
flux $2Q=2N-3$. While this corresponds to filling factor 1/2 in the
thermodynamic limit ($N\rightarrow\infty$) the nontrivial constant
shift of $-3$ in this relation is characteristic of the Pfaffian state
and arises from the curvature of the spherical surface. For example,
a possible competing state at $\nu=1/2$ is the composite
fermion Fermi sea~\cite{cffs} with $2Q=2N-2$. While the ability to
discern between competing states via the ``shift'' may be considered
an advantage of the spherical geometry, it also has the drawback that
one cannot directly address the competition between different phases
without moving to a different Hilbert space with a different flux.  (In addition, 
complications may also arise from distinct thermodynamic FQH 
states occurring at the same shift for given ($Q,N$) values--the so-called 
`aliasing' problem.)

This problem is resolved using the torus geometry where the shift is
zero and all states are uniquely defined by the filling factor alone.
Thus, exact diagonalization on the torus provides additional
information on the nature of the ground state.  We have performed
calculations in periodic rectangular domains with unequal sides $a$
and $b$. In the presence of a magnetic field, the standard translation
operators no longer commute but they do satisfy the
so-called magnetic translation algebra. This non-commutation of the
standard translation operators prevents a simple straightforward
construction of conserved quantum numbers.  Haldane~\cite{Haldane85} has,
however, shown how to construct many-body states
that have conserved pseudomomenta corresponding to the magnetic
translations along the two periodicity directions. These pseudomomenta
are bi-dimensional $(K_x,K_y)$ and they reside in a two-dimensional
Brillouin zone containing exactly $N_0^2$ points where $N_0$ is the
greatest common divisor of $N$ and $N_\phi$, where $N_\phi$ is the
number of flux quanta through the system (here we denote the total
flux as $N_\phi$ compared to $2Q$, as in the spherical geometry, to
distinguish the geometries more readily and because $N_\phi$ is more
commonly used in the torus geometry literature).  The pseudomomenta
are of the form $K_x=2\pi\hbar s/a$ and $K_y=2\pi\hbar t/b$ with
$s,t=0,\dots,N_0$.

On the torus, there is a degeneracy due to the center of mass motion
given by $q$ at a filling factor $p/q$. In the construction of
Haldane, the Hamiltonian is block-diagonal with exactly $q$ identical
blocks--this holds independently of the Hamiltonian and, hence, it has
nothing to do with the physics of the system.  We have systematically
discarded this trivial degeneracy in all that follows.  The construction
of conserved quantities has the practical advantage that it reduces
the size of the Hilbert space in which we search--through exact
diagonalization--for the few low-lying eigenstates.  

At least some of the candidate states for a half filled LL have
characteristic signatures in the quantum numbers of these low-lying
eigenstates.  In the case of the composite fermion Fermi sea, the
effective theory is that of quasi-free fermions with an interaction
induced mass.  In a finite system, with discrete energy levels, one
expects to find closed shell effects which, in addition, should be
quite sensitive to the aspect ratio of the unit cell.  This is what is
observed~\cite{rez-hald} in the LLL at $\nu =1/2$. There are also
other competing (compressible) phases with broken translational
symmetry~\cite{Rezayi99,Haldane00,Yang01} which are close in
energy--the so-called stripe and bubble phases. The stripe phases in
the LLL are equivalent to charge-density waves with unidimensional
electronic density modulation. The spectral signature of such a phase
is a set of low-lying states with pseudomomenta all related by a
single wavevector which has the periodicity of the stripe. In the SLL,
Rezayi and Haldane~\cite{rez-hald} have numerically shown that such a
stripe phase is the ground state for electrons interacting with the
pure Coulomb potential with zero width.

On the other hand, the Pfaffian state has a very different spectral
signature which can be obtained by translating the wavefunction
(Eq.~\ref{PfaffianSphere}) in the torus geometry.  One of the key 
ingredients of the Pfaffian is the Laughlin-Jastrow correlation factor
$\prod_{i<j}(u_iv_j-u_jv_i) $ which, if the coordinates on the torus
are $z=x+iy$, can be written as $\prod_{i<j}(z_i - z_j)$.  The correlation
factor with the proper periodicity is given by $\theta_1(z_i-z_j|\tau)
$ with $\tau = ib/a$ ($b/a$ is the aspect ratio of the rectangle) and
$\theta_1$ the first Jacobi theta function. This allows a
construction~\cite{HR85} of the standard Laughlin state for filling
factor $1/m$ (with an additional factor to treat the center of mass
motion).

This same construction does not work for the Pfaffian state. Indeed,
the presence of a denominator in the formula (Eq.~\ref{PfaffianSphere})
invalidates the periodicity properties if one simply introduces
$\theta_1(z_i-z_j|\tau)$ factors. The correct
procedure~\cite{Greiter92,Chung07} involves the {\it four} Jacobi theta
functions $\theta_{1}$, $\theta_2$, $\theta_3$, and $\theta_{4}$
through the following substitution,
\begin{equation}
\label{PfTorus}
 \frac{1}{(z_i-z_j)}\rightarrow  
\frac{\theta_a(z_i-z_j|\tau)}{\theta_1(z_i-z_j|\tau)},\quad
{\rm a=2,3,4}\;.
\end{equation} 
This construction gives \textit{three} ground states instead of only
one in the spherical geometry.  This remarkable degeneracy is of
topological origin.

If one considers a physical Hamiltonian whose ground state is
adiabatically connected to the Pfaffian state one expects to find the
threefold degenerate multiplet whose quantum numbers on the torus may
be deduced from Eq.~\ref{PfTorus}. In a finite system, there will be a
splitting of the multiplet of states, and it is only in the
thermodynamic limit that these states will become truly
degenerate. For electrons at filling 1/2 the three Pfaffian ground
states have pseudomomenta $\mathbf K=(0,N_0/2),(N_0/2,0),(N_0/2,N_0/2)$
(in units of $2\pi\hbar/a$ and $2\pi\hbar/b$) and this set of quantum
numbers clearly differentiates the Pfaffian from the other ground
states mentioned above.  Hence, to probe for this quasi-degeneracy one
has to use a rectangular unit cell since, in that case only, the
degeneracy is non-trivial--square or hexagonal cells have additional
discrete (geometric) symmetries leading to the equivalence of some or all
Brillouin zone points potentially masking the non-Abelian topological 
degeneracy.

To fully understand the properties of the Pfaffian state, one further
has to take into account the fact that it breaks particle-hole (p-h)
symmetry~\cite{Levin07,Lee07}, either due to explicit p-h symmetry 
breaking terms present in real systems (e.g., Landau level mixing 
or coupling) or due to a spontaneous p-h symmetry breaking in 
theories neglecting LL mixing~\cite{mrp-kp-sds}.  If we consider the wavefunction
written on the sphere, its flux $2Q = 2N-3$ leads automatically to the
p-h conjugate state (the so-called anti-Pfaffian) at $2Q = 2N+1$,
which, since it exists at a different flux for the same $N$, precludes
any mixing between the states. On the torus these two states will mix,
for finite size systems, which should lead to a lower energy symmetric
Pfaffian-Anti-Pfaffian combination and a higher energy antisymmetric
combination. This reasoning applies to each of the three states with
differing {$\mathbf K$}'s expected on the torus, so we expect to find
three doublets if we are, in fact, dealing with a system well
described by the Pfaffian state, corresponding to a non-trivial topological 
degeneracy of 6 in the thermodynamic limit.

\begin{figure}[t]
\begin{center}
\mbox{\includegraphics[width=5.0cm,angle=-90]{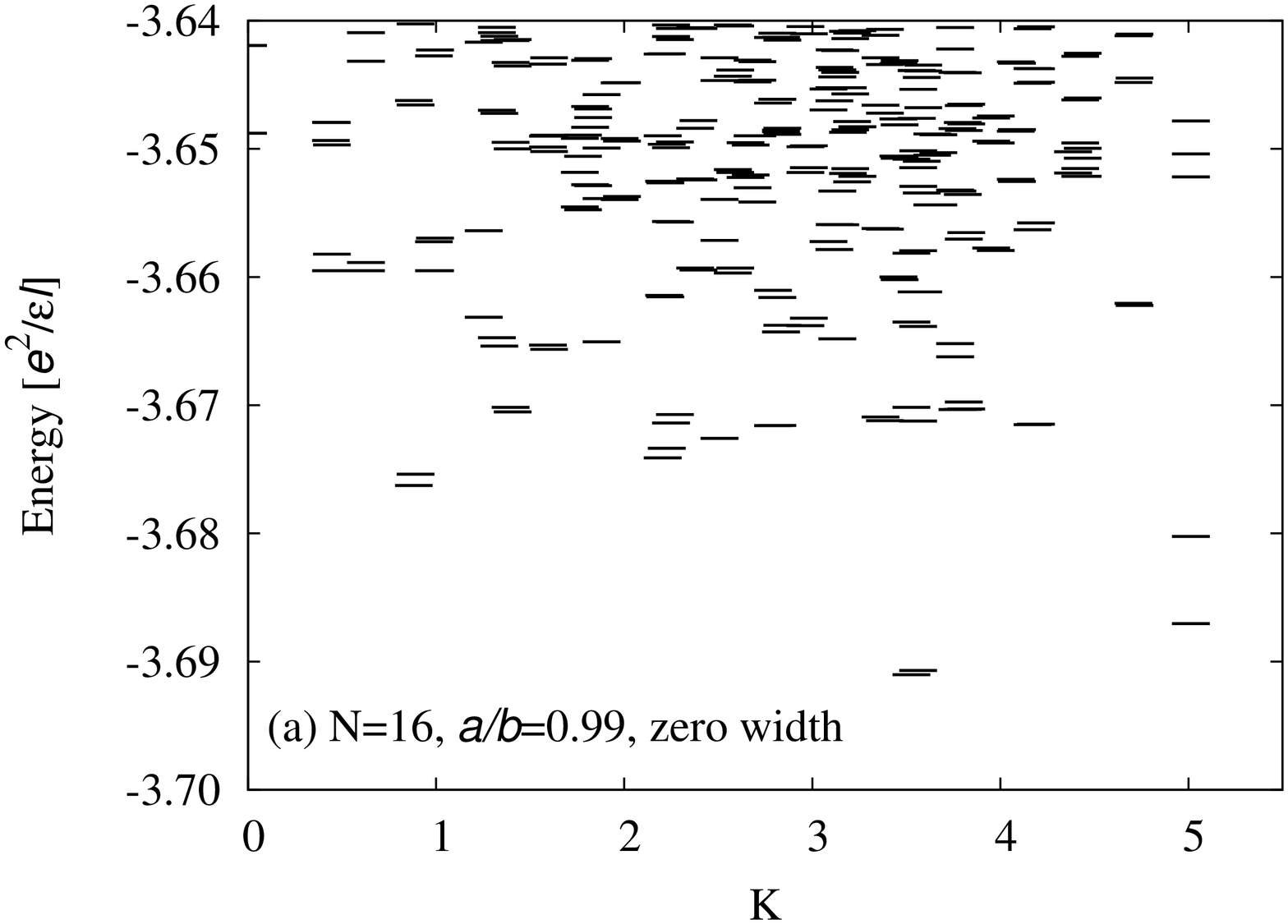}}
\mbox{\includegraphics[width=5.0cm,angle=-90]{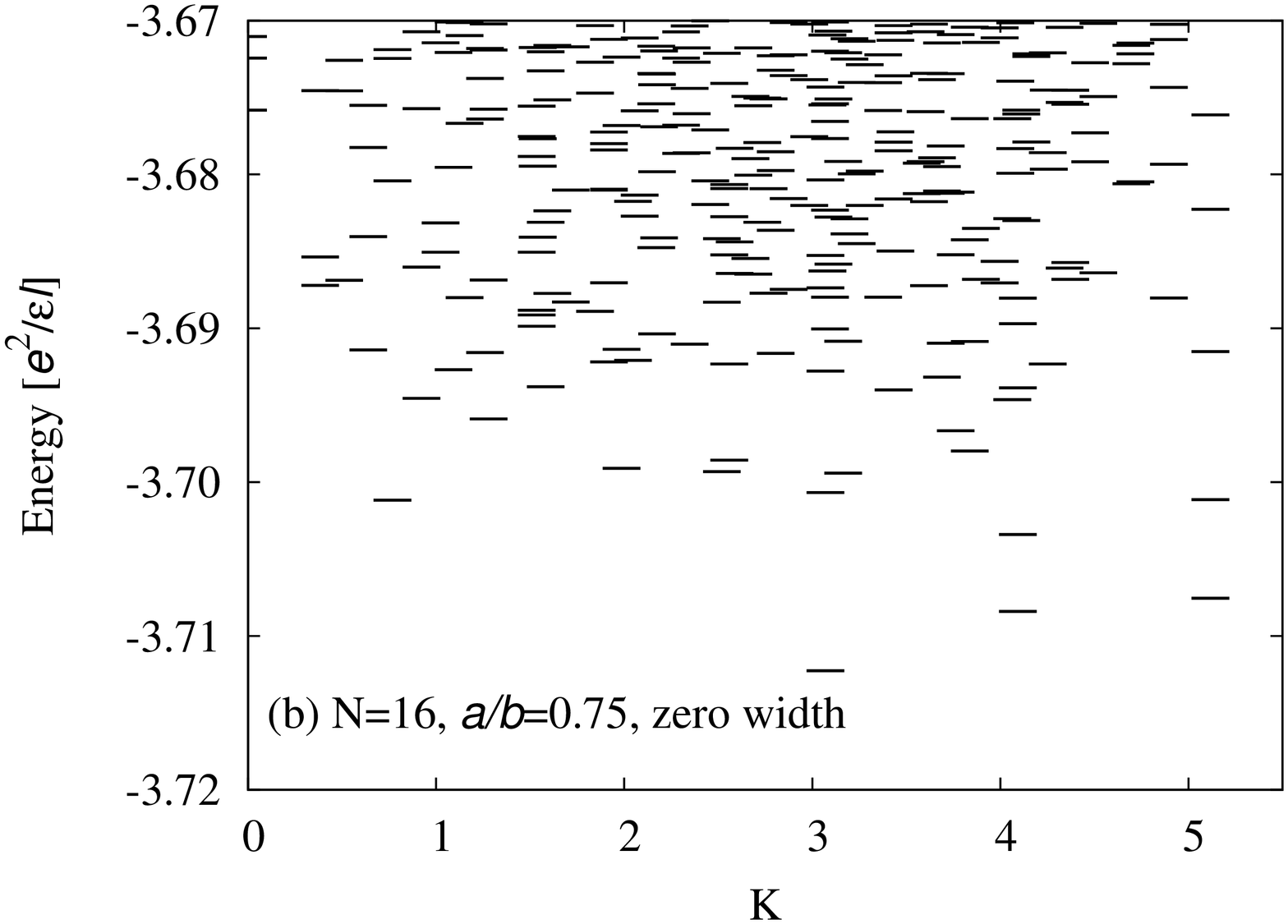}}
\end{center}
\caption{Exact energy in units of $e^2/\epsilon l$ (only the low
energy sector is being shown) versus pseudomomentum
K$=\sqrt{(a/b)K_x^2+(b/a)K_y^2}$ for $N=16$ electrons, using the torus
geometry, interacting via the SLL Coulomb Hamiltonian (zero width).
The pseudomomenta $K_x$ and $K_y$ are given in units of $2\pi\hbar/a$
and $2\pi\hbar/b$, respectively.  Panel (a) corresponds to an aspect
ratio of the rectangular unit cell equal to 0.99 while (b) has an
aspect ratio of 0.75.}
\label{pf0}
\end{figure}

To investigate the spectral signature of the Pfaffian we have performed
exact diagonalizations on the torus from $N=10$ to 16 electrons using the SLL
Coulomb interaction with finite thickness modeled by the three 
quasi-2D confinement models used before--FH, SQ and ZDS.

\begin{figure}[t]
\begin{center}
\mbox{\includegraphics[width=5.0cm,angle=-90]{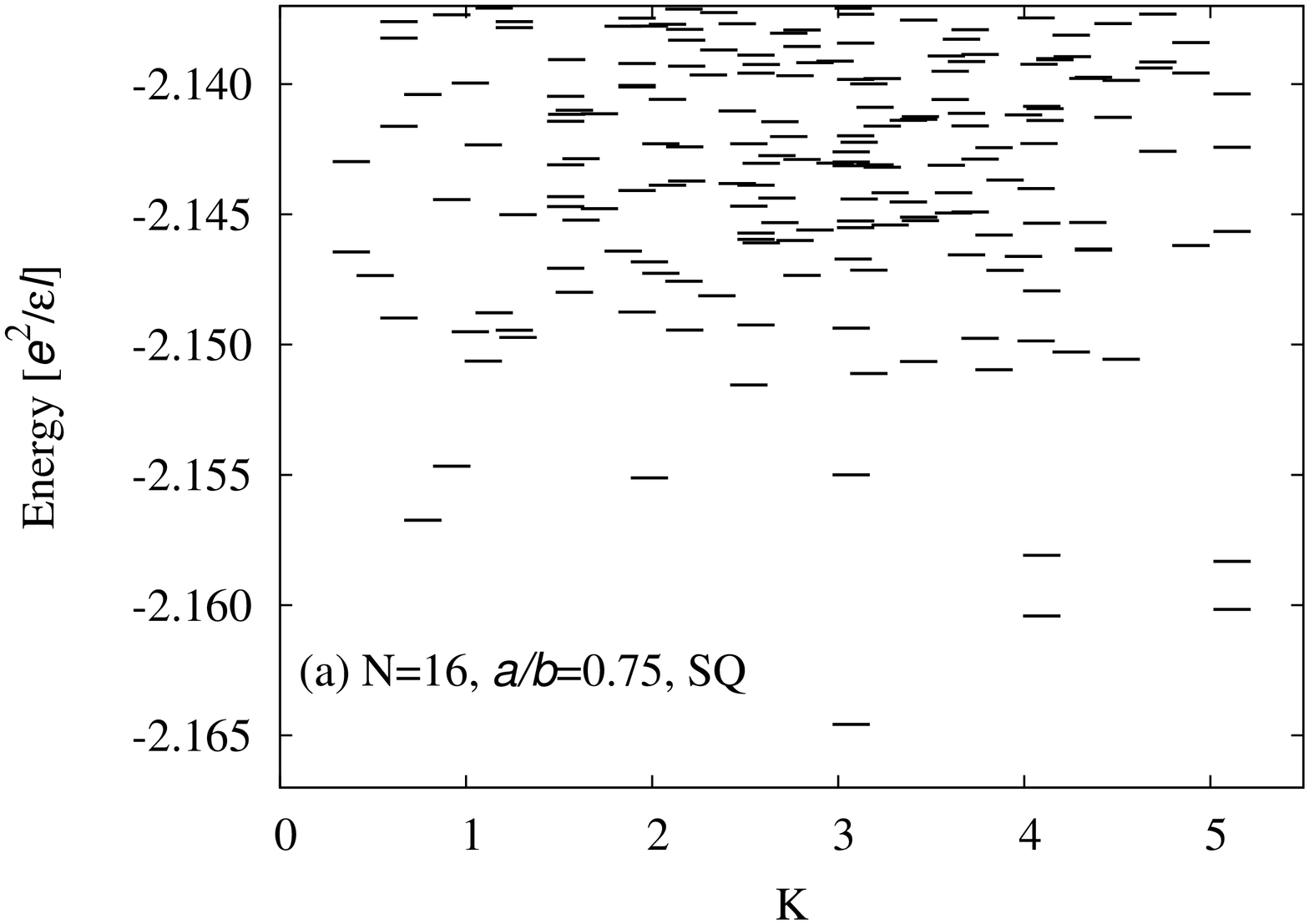}}
\mbox{\includegraphics[width=5.0cm,angle=-90]{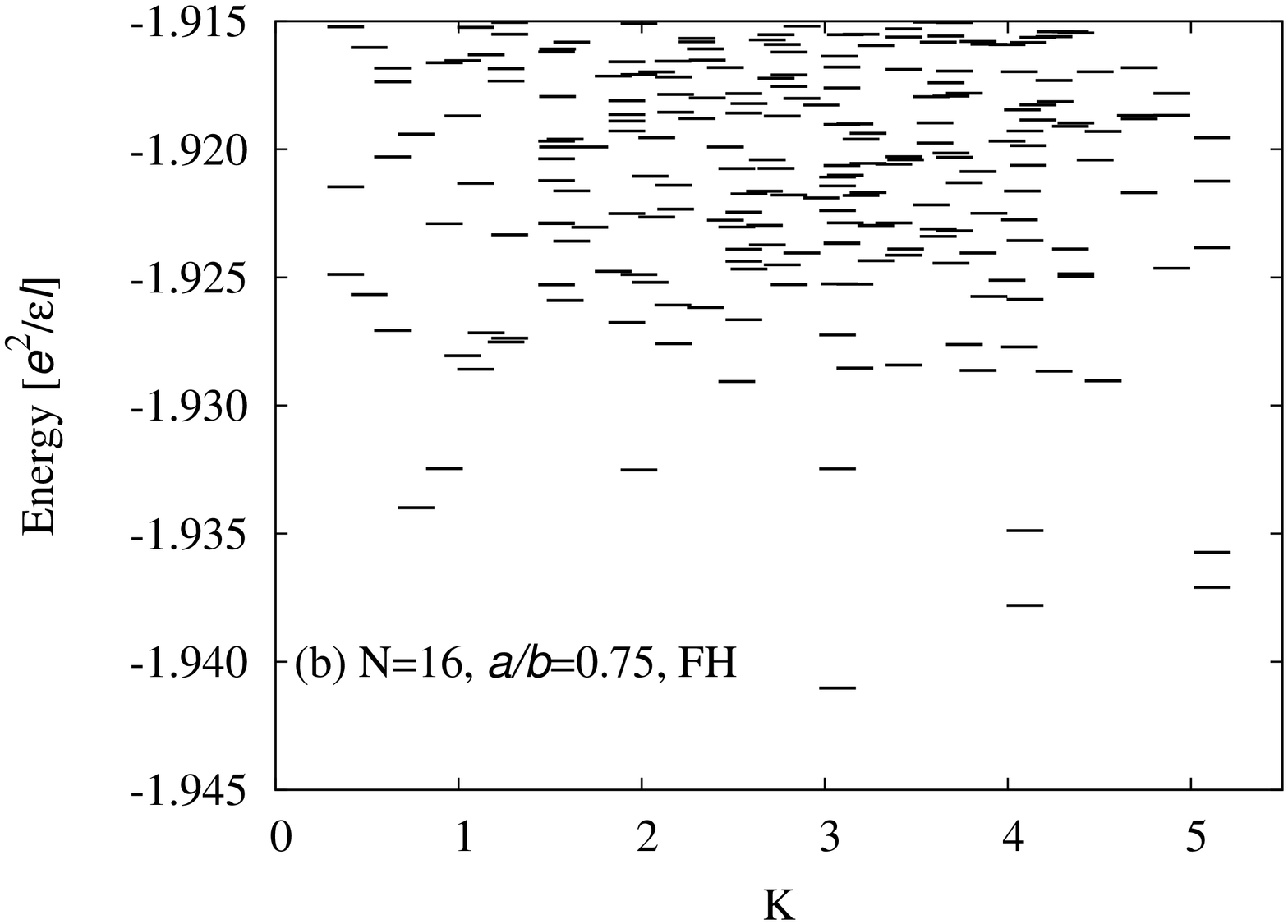}}
\mbox{\includegraphics[width=5.0cm,angle=-90]{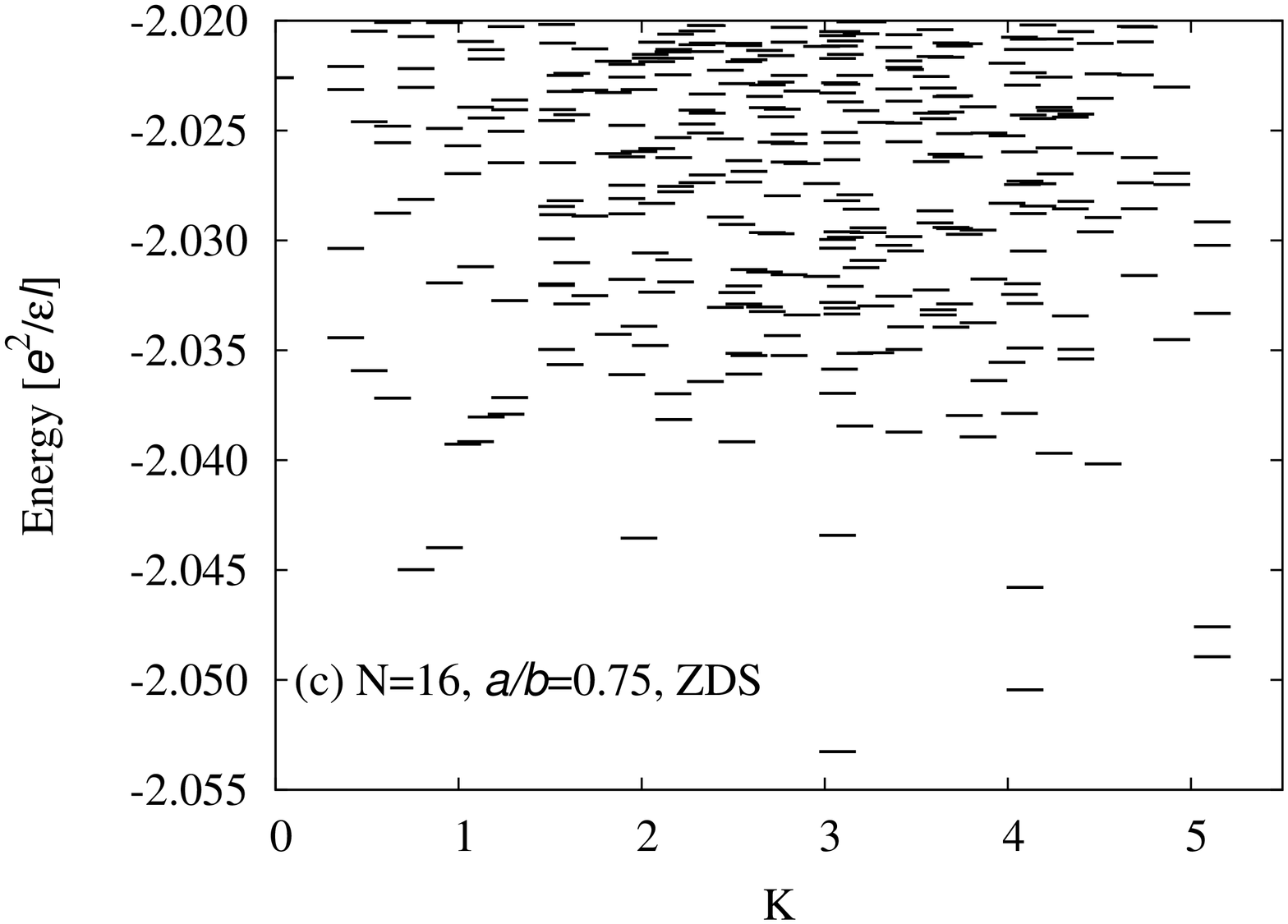}}
\end{center}
\caption{Same as Fig.~\ref{pf0} except here we have included finite
thickness effects using the SQ (a), FH (b), and ZDS (c) potentials,
respectively.  The aspect ratio is 0.75.}
\label{pffh}
\end{figure}

\begin{figure}[t]
\begin{center}
\mbox{\includegraphics[width=5.cm,angle=-90]{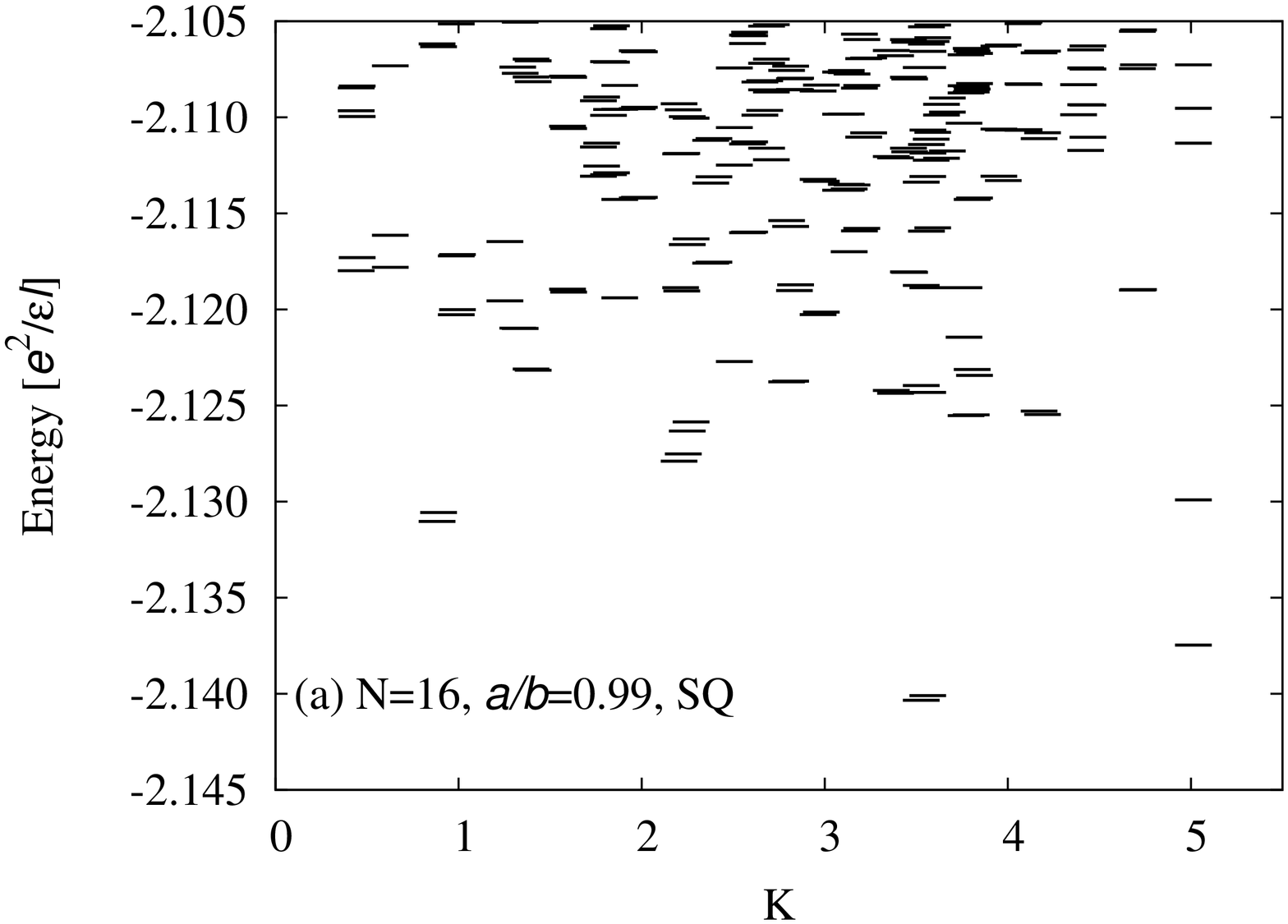}}
\mbox{\includegraphics[width=5.cm,angle=-90]{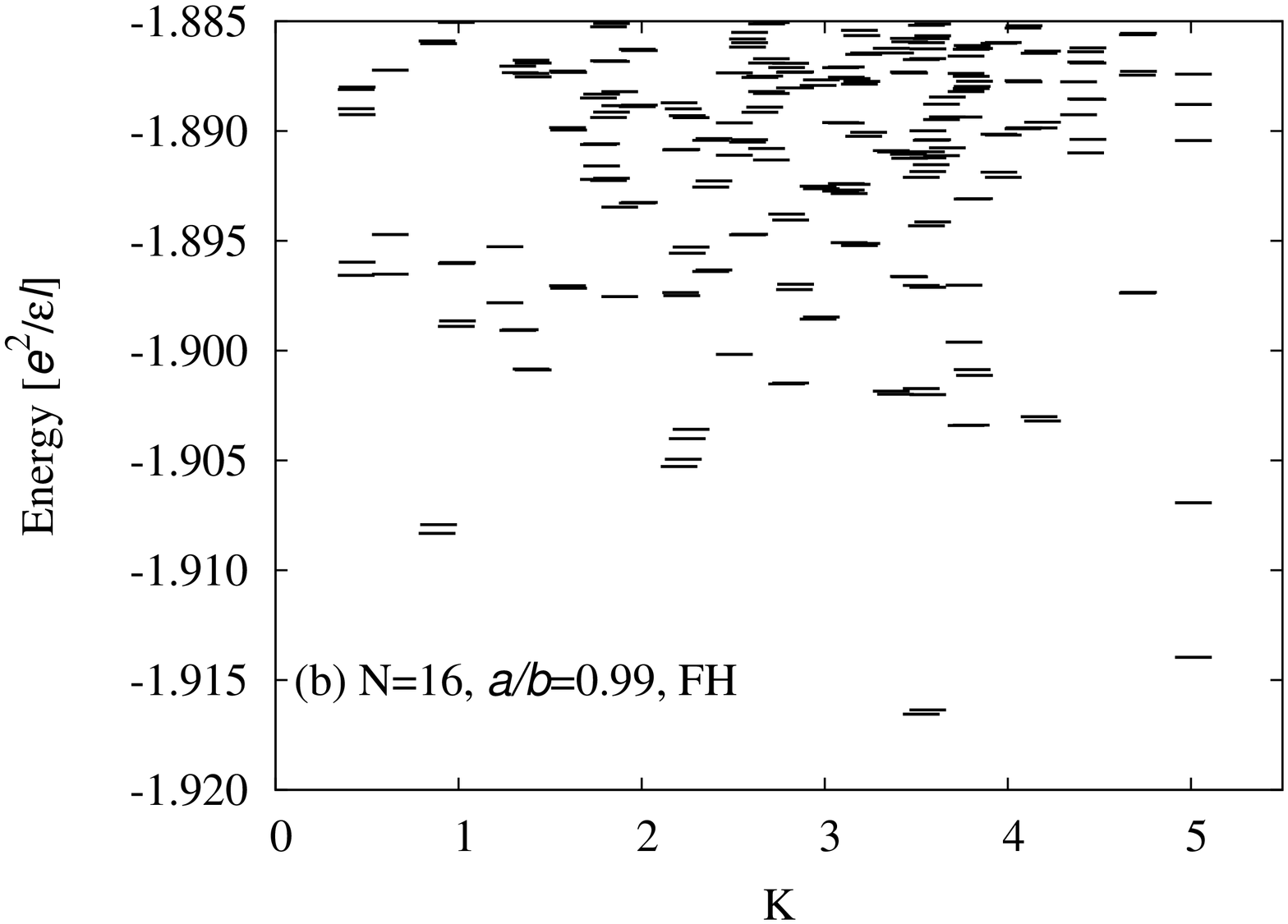}}
\mbox{\includegraphics[width=5.cm,angle=-90]{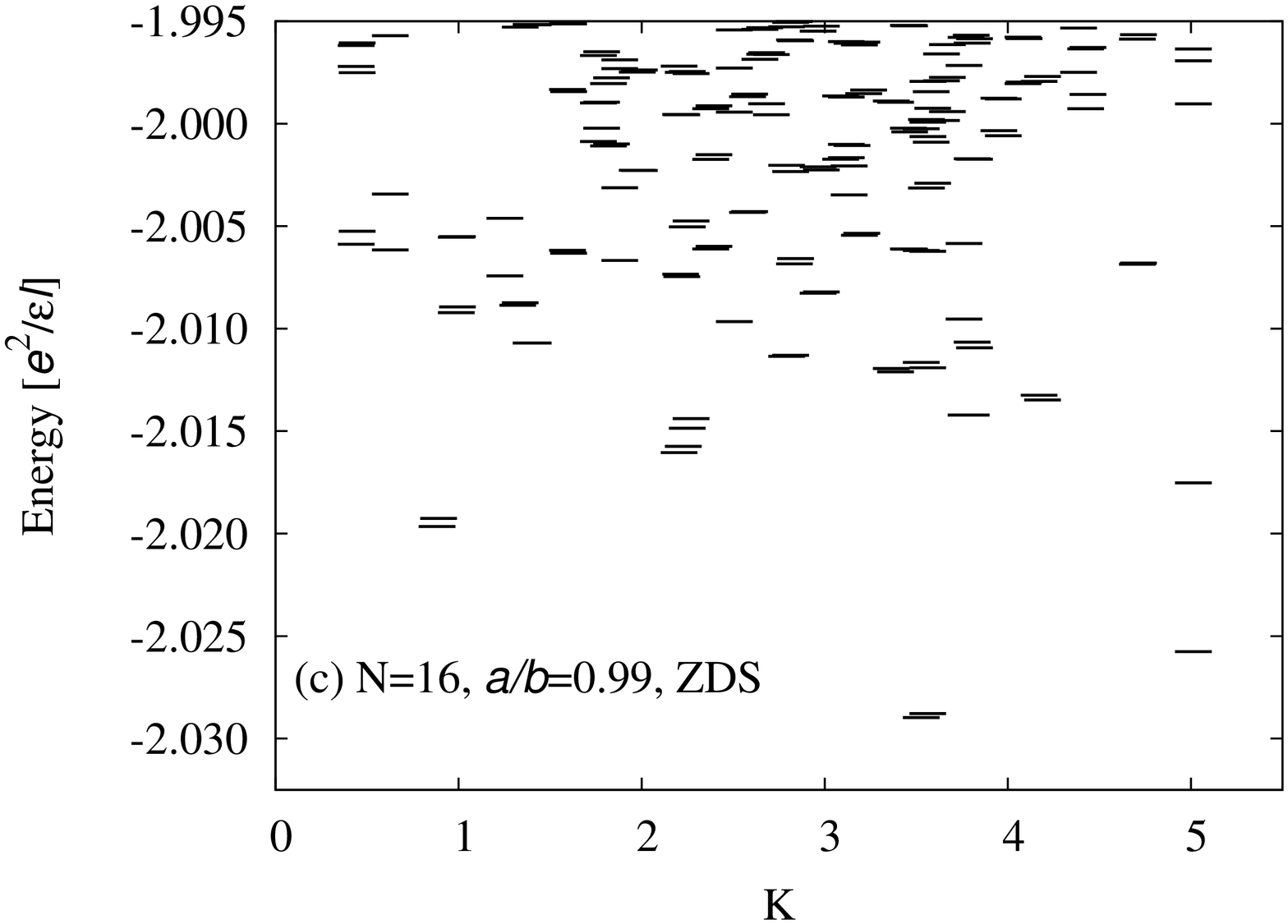}}
\end{center}
\caption{Same as Fig.~\ref{pffh} except the aspect ratio 
has been tuned to 0.99.}
\label{pffh2}
\end{figure}

For small systems, there is no obvious threefold degeneracy at zero
width while the threefold degeneracy is clearly seen when finite
thickness is included~\cite{Peterson08}.  For the largest system we
have been able to study i.e. $N=16$, the picture becomes clearer.  For
$N=16$ electrons interacting via the pure Coulomb potential in the SLL
we find that there is a threefold quasidegenerate set of ground states
with quantum numbers of the Pfaffian for aspect ratio 0.99: see
Fig.~\ref{pf0}(a).  However, we do not observe candidate,
higher-lying, states forming their p-h doublets that are expected from
the higher energy combination of Pfaffian-Anti-Pfaffian states.  If we
tune the aspect ratio from 0.99 to 0.75, then the three
quasidegenerate ground states are still obtained but there is a less
clear separation from the higher-lying states, see Fig.~\ref{pf0}(b).

If we now consider systems with a finite width taken to be the value
of maximum overlap ($d/l\sim4$-$5$), as determined from the
calculations on the sphere, then we find the Pfaffian signature
(threefold degeneracy and p-h partner states) is qualitatively
enhanced.  At an aspect ratio equal to 0.75 the three finite thickness
models considered all lead to the threefold quasidegeneracy for the
three states with the correct quantum numbers: see
Figs.~\ref{pffh}(a)-(c).  Now it is plausible to identify the p-h
partners for each of the states with the same quantum numbers
predicted by the Moore-Read wavefunction slightly higher in energy
than the ones forming the threefold quasidegenerate ground state
manifold.  If the aspect ratio, with finite thickness included, is
changed from 0.75 to 0.99 we find that things change somewhat, see
Figs.~\ref{pffh2}(a)-(c); while there is still a threefold
quasidegeneracy of the ground state, the doublet structure is no
longer clear. The reason for this sensitivity to the aspect ratio is
not known and deserves further study. We emphasize, however, that even
if the enhancement of the Pfaffian signature is restricted to some
range of aspect ratio, its presence is clearly enhanced by finite
thickness.

The results presented in this subsection, taken with those in
Sec.~\ref{overlaps1}, provide a satisfying picture where finite
thickness effects produce a ground state that is described by the MR
Pfaffian wavefunction.  The fact that two different geometries produce
the same qualitative conclusion using two different, complementary,
signatures is quite convincing.  On the sphere (Sec.~\ref{overlaps1}
and Figs.~\ref{overlaps-13}-\ref{overlaps-12}) we find that the
overlap between the Pfaffian and the exact finite-system numerical
wavefunction for the $\nu=5/2$ state is enhanced substantially as the
quasi-2D thickness parameter is increased.  On the torus
(Sec.~\ref{sec-topo} and Figs.~\ref{pf0}-\ref{pffh2}) we find that the
expected non-Abelian topological degeneracy, a characteristic
signature for both Pfaffian/anti-Pfaffian states, shows up precisely
where the wavefunction overlap is large.

\subsection{In-Plane magnetic field effects: Overlaps}
\label{sec-ip}

The consequence of the application of an in-plane magnetic field to
the $\nu=5/2$ FQH state is a very important question.  Experiments
have shown\cite{tilt1,tilt2} that the FQHE at $\nu=5/2$ is suppressed 
with increasing in-plane component of the magnetic
field.  Originally, this was thought to point
towards a spin-unpolarized FQH state at 5/2, since traditionally the
in-plane field is assumed to couple only to the spin degrees of
freedom, and increasing the in-plane field is supposed to 
enhance the spin-polarization of the system.  
However, subsequent theoretical work~\cite{morf} seemed to
settle the debate regarding the spin-polarization of the 5/2 state,
and the 5/2 state is now considered to be spin-polarized.  A 
question, therefore, arises about the suppression of the 5/2 
FQHE induced by the finite in-plane field since spin-polarization 
presumably cannot play a role in a spin-polarized FQH state.  It is 
speculated that the suppression arises from the in-plane 
field induced orbital effects leading to modifications of the 
pseudopotentials $V_m$.

The in-plane field serves to squeeze the single-particle electron
wavefunction in the direction perpendicular to
the plane.  Thus, the application of an in-plane field serves to
effectively decrease the width of the quantum well.  From the previous
section (i.e., section~\ref{overlaps1}) it is clear that such a
reduction in the effective width could lead to the destruction of the
FQHE at $\nu=5/2$ and 7/3 whereas it could seemingly strengthen the
FQHE at $\nu=11/5$.

\begin{figure}[t]
\begin{center}
\mbox{\includegraphics[width=6.5cm,angle=0]{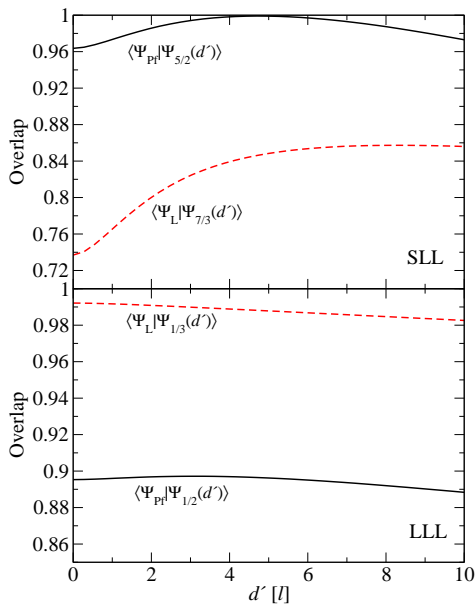}}
\end{center}
\caption{(Color online) Overlap between the exact ground state
wavefunction using a parabolic confinement at filling factor 1/3
(dashed line) and 1/2 (solid line) and the Laughlin and Pfaffian
wavefunction, respectively, as a function of $d^\prime$, in the SLL
(top panel) and the LLL (bottom panel). }
\label{o-pqw}
\end{figure}

\begin{figure*}[t]
\begin{center}
\mbox{\includegraphics[width=4.5cm,angle=-90]{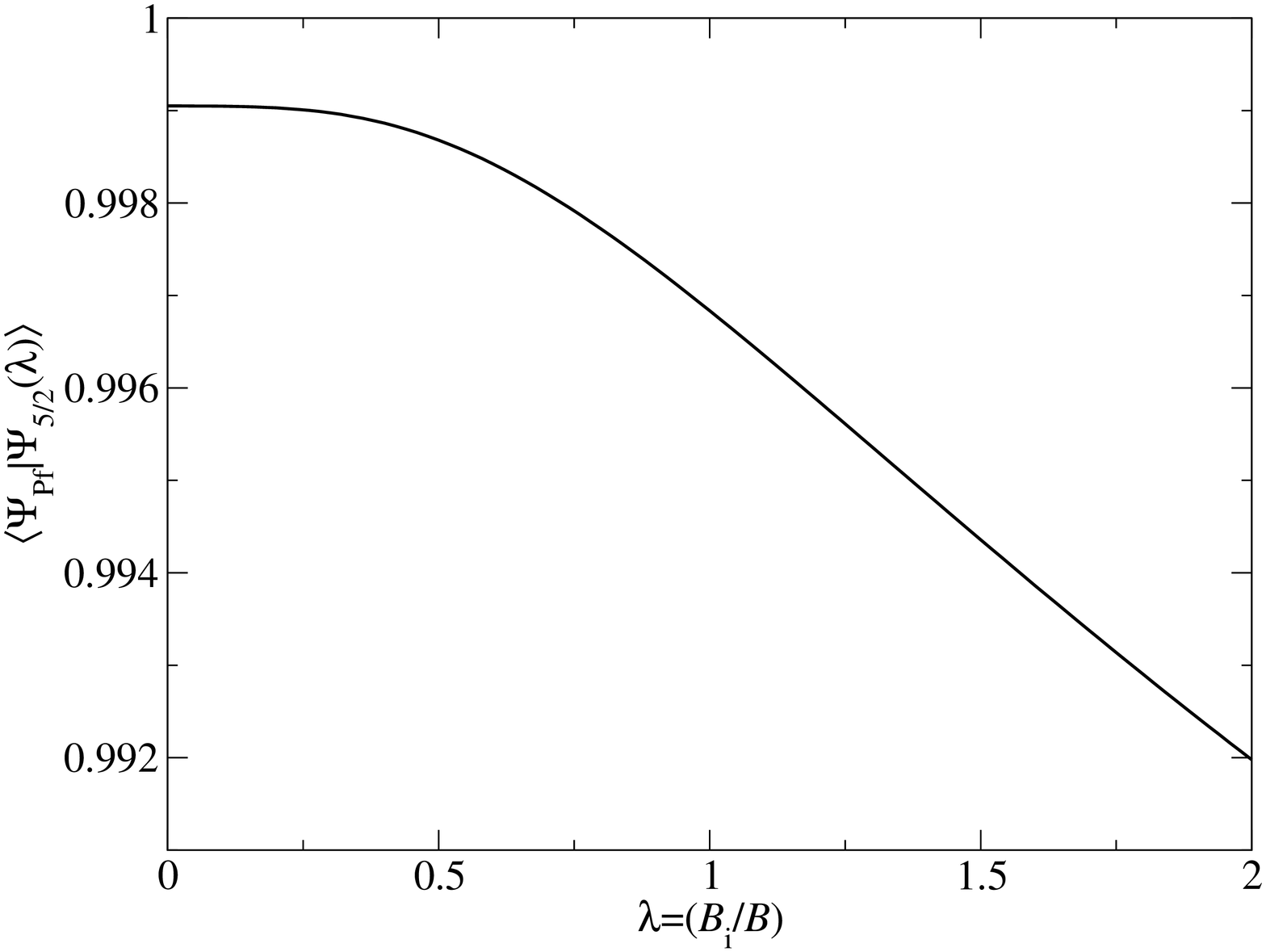}}
\mbox{\includegraphics[width=4.5cm,angle=-90]{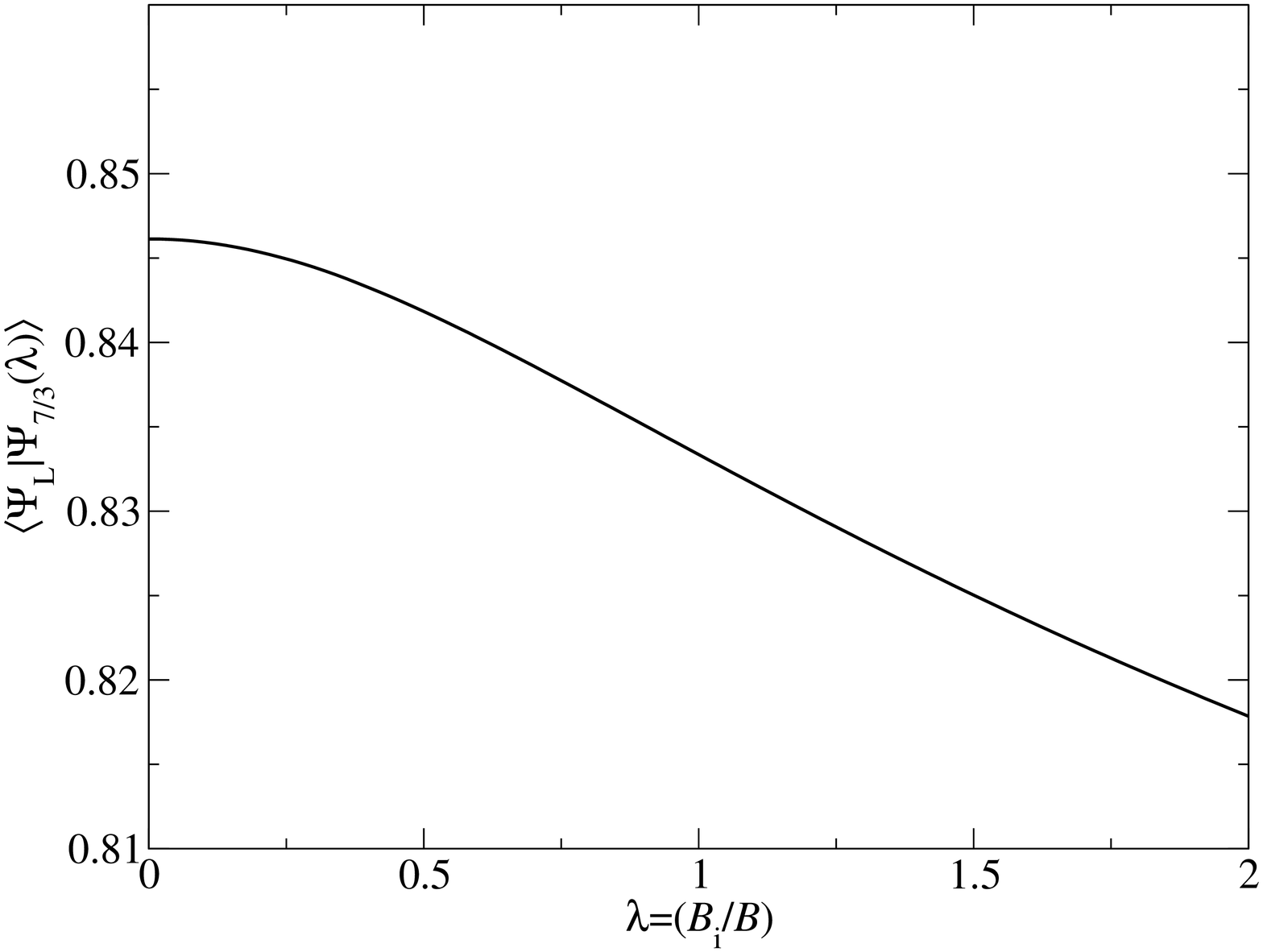}} 
\mbox{\includegraphics[width=4.5cm,angle=-90]{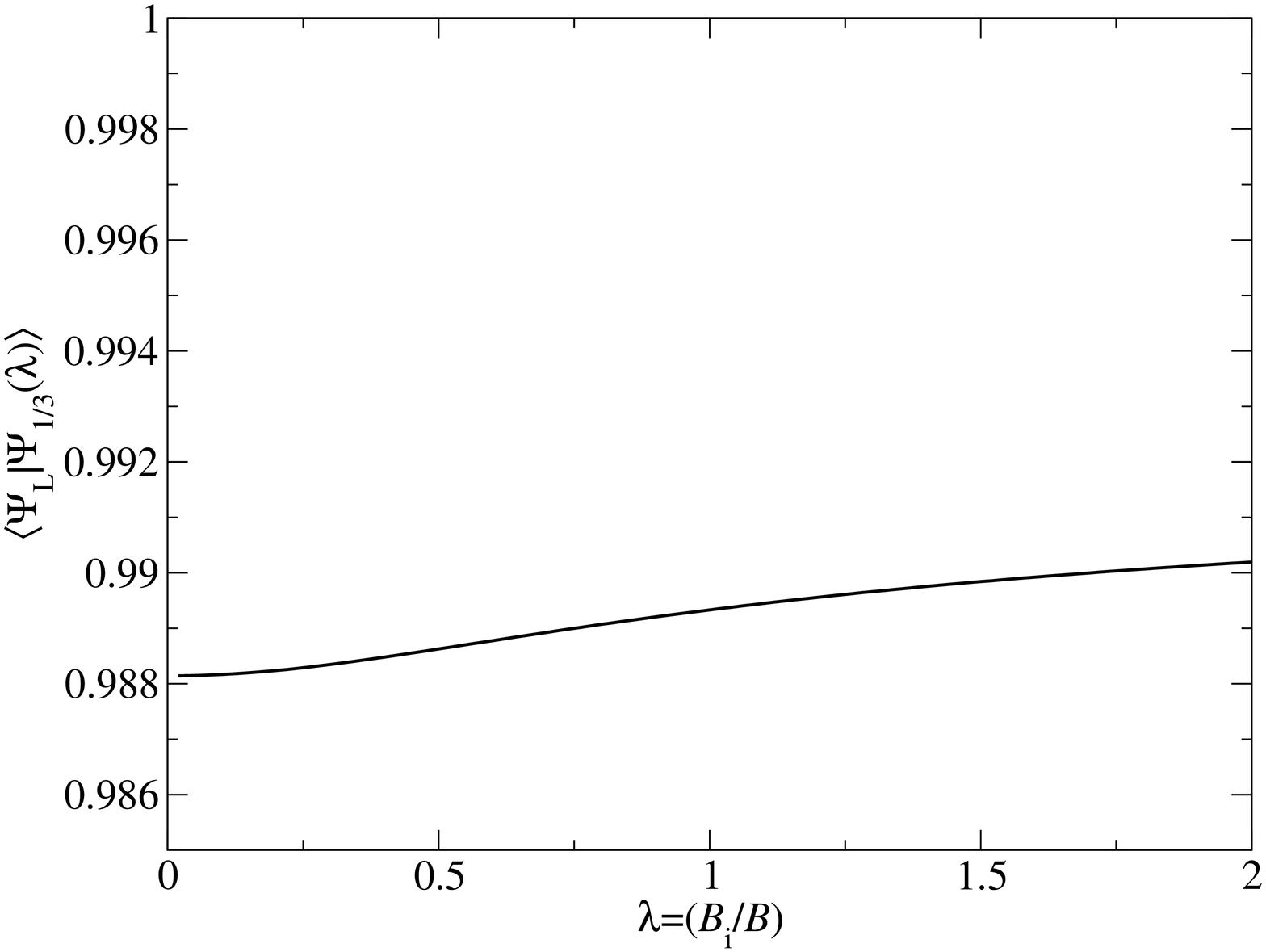}}
\hspace{0.25cm}{SLL ($\nu=5/2$)}\hspace{4cm}{SLL ($\nu=7/3$)}\hspace{4cm}{LLL ($\nu=1/3$)}
\end{center}
\caption{Overlap between Pfaffian and the exact ground state
 wavefunctions at 5/2 (left), the Laughlin and the exact wavefunctions
 at 7/3 (middle), and the Laughlin and the exact wavefunctions at 1/3
 (right) as a function of $\lambda$.}
\label{overlap-Bi}
\end{figure*}

We now explore this in detail.  Here we find that a parabolic
confinement model is convenient, i.e., the $z$-direction
confinement potential is $V(z)=(2\hbar^2/md_0^4)z^2$ where $d_0$ is
the thickness of the quantum well. (In this section we assume that the
$d_0$ of the actual system is such that the overlap between the exact
ground state and the Pfaffian wavefunction is maximum, that is, $d_0$
is a constant and not a variable, cf.  section~\ref{ff12}.)  The
single-particle ground state electron wavefunction in the $z$-direction is thus a
Gaussian
\begin{eqnarray}
\label{eta-pqw}
\eta(z)=\left(\frac{2}{\pi d_0^2}\right)^{\frac{1}{4}}e^{-z^2/d_0^2}\;.
\end{eqnarray}
We then apply an in-plane magnetic field of strength $B_i$ with the
vector potential $\vec A_{i}=(B_i z,0,0)$ and now make a simplifying
assumption, namely, we ignore the cross terms (i.e., we consider the
limit $B_i\ll B$ where $B$ is the perpendicular magnetic
field strength) thus arriving at a slight modification, from the
original quantum confinement, of the Schrodinger equation for
$\eta(z)$ as
\begin{eqnarray}
\left[\frac{\hat p_z^2}{2m}+V(z)+\frac{2\hbar^2}{m}\frac{1}{4l^4_i}\right]\eta(z)=E\eta(z)\;
\end{eqnarray}
where $l^2_i=\hbar c/eB_i$.  This equation has a solution of the 
same form as Eq.~\ref{eta-pqw}, i.e., 
\begin{eqnarray}
\eta(z)=\left(\frac{2}{\pi d^2}\right)^{\frac{1}{4}}e^{-z^2/d^2}
\end{eqnarray}
with $1/d^4=1/d_0^4+1/4l_i^4$.  This implies that the quasi-2D layer 
thickness $d$ in the presence of $B_i\neq 0$ is less than the original 
thickness $d_0>d$, i.e., the 2D layer is squeezed by $B_i$.

The potential used when calculating the planar pseudopotentials is 
\begin{eqnarray}
\tilde{V}(q)=\frac{e^2}{\epsilon}\frac{\mbox{erfc}(qd/2)e^{(qd)^2/4}}{q}\;,
\label{eq-pqw}
\end{eqnarray}
which includes the effects of a parabolic quantum well confinement and
an in-plane magnetic field through the definition of $d$ above.  Note
that the effective $B_i$-dependent apparent width $d$ decreases as $B_i$
increases, squeezing the layer width parameter in the
presence of the in-plane field.

The parabolic confinement is qualitatively similar to 
the confinement models used previously (SQ, FH,
ZDS) and, in fact, quantitatively almost identical to the
Fang-Howard confinement (see Ref.~\onlinecite{morf}).  In
Fig.~\ref{o-pqw} we show the calculated overlap between the exact
ground state wavefunction for 1/3 and 1/2 in the LLL and SLL with the
Laughlin and Pfaffian wavefunction, respectively, as a function of
$d^\prime$, to convince the reader that the difference between using a
parabolic confinement and the Fang-Howard confinement is very small
provided $d^\prime$ is defined as it was in section~\ref{overlaps1}
(here $d^\prime=(0.5/0.180756)d$ scaling it again with respect to the
square-well confinement).

Obviously, $V(k)/d$ vs. $kd$, $V_m$ vs. $d$, and $f$- and
$g$-functions vs.  $d$
(cf. Figs.~\ref{vms-coulomb},~\ref{vms},~\ref{f-funcs},
and~\ref{g-funcs}) are very similar to the results shown for the
Fang-Howard confinement and are, therefore, not shown.

To investigate the effect of a finite in-plane magnetic field we find
it convenient to plot overlap as a function of the dimensionless
variable $\lambda = (B_i/B)$ between the range $0<\lambda<2$.  This
variable is related to the effective thickness $d$ as $\lambda =
2\sqrt{1/d^4-1/d_0^4}$, where $d_0$ is chosen to be the value of
thickness (without any in-plane magnetic field) at which the overlap
is the highest, i.e., $d_0/l\sim5$.

Finally, in Fig.~\ref{overlap-Bi} we plot the overlap between the
exact ground state wavefunction for a parabolic confinement at a
finite in-plane magnetic field $B_i$ and the Pfaffian wavefunction for
1/2 in the SLL ($\nu=5/2$), the Laughlin wavefunction at 1/3 in the
SLL ($\nu=7/3$), and the Laughlin wavefunction at 1/3 in the LLL
($\nu=1/3$), respectively.  The application of an in-plane magnetic
field (i.e., $B_i,\lambda \neq 0$) causes the exact state at $\nu=5/2$ or 7/3 to become more unlike
the Pfaffian or Laughlin state, respectively, since the overlap 
goes down monotonically with increasing $B_i/B$, and hence, the in-plane
field could eventually destroy the FQH at $\nu=5/2$ and 7/3 by
effectively enhancing the confinement, making the system
more two-dimensional.  On the other hand, the in-plane field causes the
Laughlin state to become a better candidate for the exact state at 1/3
since it makes the system more 2D (this is also true for $\nu=1/5$ in
the LLL and SLL, which is not shown).  Note that all of our in-plane
field results follow directly from our findings in
Sec.~\ref{overlaps1} where we showed that increasing the layer width
stabilizes the Pfaffian 5/2 and the Laughlin 7/3 states whereas it
destabilizes the Laughlin 1/3 state.  In this section, we explicitly
show that an applied parallel field, by squeezing the quasi-2D layer
width, could suppress the 5/2 and 7/3 states but strengthen the 1/3,
1/5, and 11/5 states.

As important word of caution is, however, in order with respect to our
in-plane field results presented in this section.  We have only
considered one particular aspect of the applied in-plane field,
namely, the quasi-2D confinement effect through Eqs.~\ref{eta-pqw}-\ref{eq-pqw}.
There are several additional effects induced by the in-plane field,
which, although not considered in our work, may very well be important
in practice.  Two obvious spin effects of the in-plane field,
neglected in our work because we are only considering completely
spin-polarized states, are the Zeeman coupling induced spin
polarization of the ground (and possibly excited) states.  More subtle
magneto-orbital effects neglected in our work are the in-plane field
induced orbital anisotropy in the 2D plane and the subband-Landau
level coupling induced enhanced scattering due to the finite in-plane
fields, and the possible in-plane field induced stabilization of
competing compressible states (e.g., stripe or bubble phases) which
may have lower ground state energies than the incompressible FQHE
states under our exclusive consideration in this work.  A full
consideration of all possible effects of the in-plane magnetic field
is well beyond the scope of this paper, where we have concentrated
entirely on the wavefunction overlap effect for FQH states arising
from the wavefunction squeezing (i.e., $d<d_0$) by the applied
in-plane field.  In real experimental situations, some of these
neglected effects may very well be significant or perhaps even
dominant.

\section{Connection to prior work}
\label{contour-sec}
The main question that we address in this section is the following.
Do the first few Haldane pseudopotentials determine the FQHE 
physics in the SLL?  This has been earlier studied in the literature, 
and therefore this section connects our work to the existing 
work of Morf~\cite{morf-overlap} and Rezayi and Haldane~\cite{rez-hald}, 
which have been influential in theoretical studies of the 5/2 
state during the last 10 years.

As mentioned in the Introduction (Sec.~\ref{intro}), early important
work was done by Morf~\cite{morf-overlap} and Rezayi and
Haldane~\cite{rez-hald} regarding the Pfaffian description of the 5/2
FQHE.  In particular, the overlap between the exact ground state of the
SLL Coulomb potential, where the first pseudopotential $V^{(1)}_1$ was
varied around the Coulomb point, and the Pfaffian was calculated as a function of this
variation of the pseudopotential, i.e., $\delta V^{(1)}_1$.
Rezayi-Haldane utilized the torus geometry varying both $V^{(1)}_1$
and $V^{(1)}_3$, and as such, their results are not directly
comparable to the results presented here.  Morf, however, varied
$V^{(1)}_1$ using the spherical geometry and his results are
directly applicable to ours.  In fact, Fig.3(b) in
Ref.~\onlinecite{morf-overlap} is equivalent to our
Fig.~\ref{vm-varied-over}(b) (middle plot)--of course, we have also
included overlaps with respective Laughlin states as well as
considered two (LLL and TLL) other LLs.  A general result of
Refs.~\onlinecite{morf-overlap} and~\onlinecite{rez-hald} was that a
positive $\delta V^{(1)}_1$ enhanced the overlap.  However, as
discussed in Sec.~\ref{overlaps1}, for the Pfaffian, the change in
overlap via the variation in the short range pseudopotentials is not
easily motivated since the Pfaffian is not an exact eigenstate 
of a $V_1$-only two-body Hamiltonian, and, in fact, an increase in $V^{(1)}_1$ (leaving
all other $V^{(1)}_m$'s constant) is physically untenable, i.e., 
there is no experimental or physical way one can effect an 
increase only in $V_1^{(1)}$ in real systems.

\begin{figure}[t]
\begin{center}
\mbox{\includegraphics[width=6.cm,angle=-90]{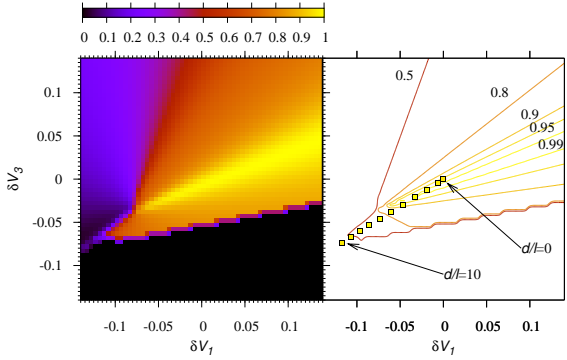}}
\end{center}
\caption{(Color online) Contour plot of the overlap between the
Pfaffian wavefunction and the 5/2 ground state wavefunction of a SLL
Coulomb Hamiltonian, where $V^{(1)}_1$ and $V^{(1)}_3$ have been
varied from their pure Coulombic values, as a function of the
variations $\delta V^{(1)}_1$ and $\delta V^{(1)}_3$.  The system
considered is the $N=8$ electron system shown in
Fig.~\ref{overlaps-12}.  The left panel is the color contour plot
while the right panel contains only contour lines at values of the
overlap equal to 0.5, 0.8, 0.9, 0.95, and 0.99.  Also shown in the
left panel is the path traced out in the $(V^{(1)}_1,V^{(1)}_3)$ plane
of the finite thickness of the SQ potential from $d/l=0$ to 10.}
\label{contour-v1v3}
\end{figure}

\begin{figure}[t]
\begin{center}
\mbox{\includegraphics[width=6.cm,angle=-90]{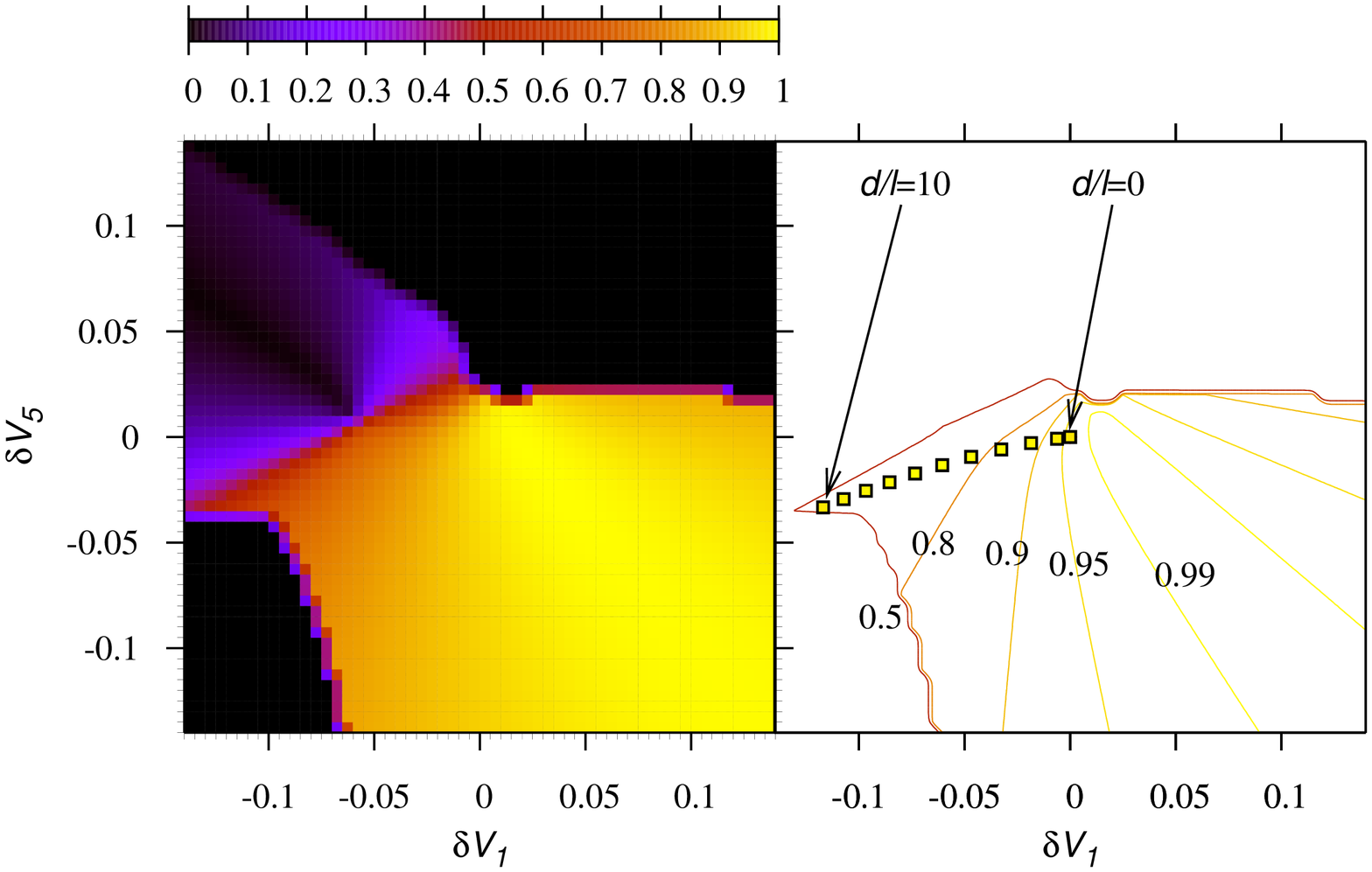}}
\end{center}
\caption{(Color online) Same as Fig.~\ref{contour-v1v3} expect here we
consider variations in $V^{(1)}_1$ and $V^{(1)}_5$. }
\label{contour-v1v5}
\end{figure}

\begin{figure}[t]
\begin{center}
\mbox{\includegraphics[width=6.cm,angle=-90]{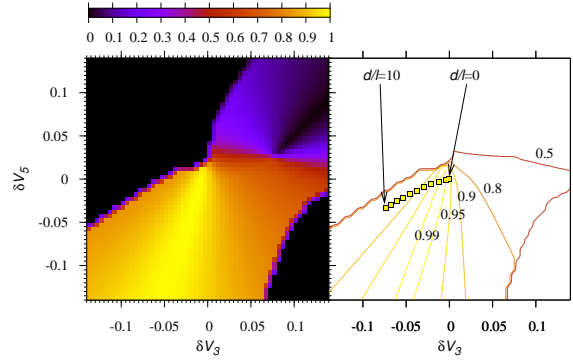}}
\end{center}
\caption{(Color online) Same as Fig.~\ref{contour-v1v3} expect here we
consider variations in $V^{(1)}_3$ and $V^{(1)}_5$.}
\label{contour-v3v5}
\end{figure}

Here we show that the first few pseudopotentials {\em do not} by
themselves determine the physics.  In
Figs.~\ref{contour-v1v3},~\ref{contour-v1v5}, and~\ref{contour-v3v5}
we show the calculated overlap between the Pfaffian wavefunction and
the exact ground state wavefunction of the SLL Coulomb Hamiltonian
where $(V^{(1)}_1,V^{(1)}_3)$, $(V^{(1)}_1,V^{(1)}_5)$, and
$(V^{(1)}_3,V^{(1)}_5)$ are allowed to vary away from their original
SLL values by $\delta V^{(1)}_m$ for $N=8$ electrons and $2(Q+n)=13$,
i.e., for the 5/2 state.  Each plot has left and right panels.  The
left panel is a color contour plot of the overlap as a function of the
change in pseudopotentials.  The right panel displays the same contour
plot but with only contour lines at values of the overlap equal to
0.5, 0.8, 0.9, 0.95, and 0.99.  Also on this panel is a series of
square points that show the overlap for the SQ potential at different
values of $d/l$ (we use $d$ here since $d^\prime=d$ for the SQ by
definition) from zero to $d/l=10$ in unit steps.  Finite thickness
changes all the $V^{(1)}_m$'s but we are only showing the path traced
out in the $V^{(1)}_1$-$V^{(1)}_3$ or $V^{(1)}_1$-$V^{(1)}_5$ or
$V^{(1)}_3$-$V^{(1)}_5$ pseudopotential space of the contour plot.

Two main observations can be made in the parameters space of
$(V^{(1)}_1,V^{(1)}_3)$, $(V^{(1)}_1,V^{(1)}_5)$, and
$(V^{(1)}_3,V^{(1)}_5)$.  First, there are always regimes where the
overlap is above 0.99.  That this occurs in the
$(V^{(1)}_1,V^{(1)}_3)$ space is not particularly surprising
considering the previous work of Rezayi-Haldane~\cite{rez-hald}
(although this is shown here for the first time in the spherical
geometry).  However, the other two contour plots in the
$(V^{(1)}_1,V^{(1)}_5)$ and $(V^{(1)}_3,V^{(1)}_5)$ spaces go against
the conventional wisdom.  In particular, Fig.~\ref{contour-v3v5} shows
that a large overlap with the Pfaffian can be obtained by leaving
$V^{(1)}_1$ constant and varying $V^{(1)}_3$ and $V^{(1)}_5$ only.
Hence, it is clear that the first two pseudopotentials do not dominate
the physics of the 5/2 FQHE.

The second observation that can be gleaned is that, since the values
of {\em all} the pseudopotentials change upon including finite
thickness effects, one cannot parameterize finite thickness
corrections in terms of only two pseudopotentials, be they the first
and third, first and fifth, third and fifth, or any other combinations
of two.  This effect is clearly shown in the right panel of each
contour plot where the path of the overlap, as a function of finite
width, in the $(V^{(1)}_1,V^{(1)}_3)$, $(V^{(1)}_1,V^{(1)}_5)$, and
$(V^{(1)}_3,V^{(1)}_5)$ plane, respectively, does not coincide with a
high overlap region shown in the contour plot when {\em only} changing
two pseudopotentials.  Note that the squares of the finite thickness
overlap results are shaded according to the color contour plot on the
left panel (as well as being more clearly visible in
Fig.~\ref{overlaps-12}).

We have thus demonstrated that the investigation of the overlap as a
function of changes (or ratios) between two pseudopotentials does not
elucidate the physics of the 5/2 state and produces ambiguous results.
In particular, in the inset of Fig.5 of Rezayi-Haldane~\cite{rez-hald}
the path of the finite thickness is plotted in the
$(V^{(1)}_1,V^{(1)}_3)$ plane and shown to cross a boundary between a
striped state and a paired FQHE state.  However, from the above
considerations, one cannot conclude that the effects of finite
thickness drive the system across boundaries in this way.
Specifically, finite thickness corrections involve variations in
\textit{all} the pseudopotentials from the ideal Coulomb point, and
tuning one or even a few pseudopotentials does not, under any
circumstances, mimic the finite thickness effect.  Tuning one or two
pseudopotentials around the Coulomb point in order to study the
stability of the 5/2 FQHE state with respect to the Pfaffian is,
therefore, somewhat misleading in our opinion, particularly since,
unlike the LLL Laughlin states, there is no theoretical two-body
Hamiltonian for which the Pfaffian is an exact ground state.

\section{Finite thickness effects on excitation gaps}
\label{sec-gap}

The FQHE transport activation gap can be readily measured
experimentally and is an extremely relevant quantity that
characterizes the incompressibility of a FQH state.  In this section
we discuss the effects of finite thickness on the excitation gaps for
the FQHE at $\nu=7/3$ and 5/2.  Usually the experimental activation
gap and the theoretical excitation gap are considered to be the same
although this may not be necessarily true in the presence of disorder.
It is, however, well-accepted that larger FQHE gaps imply stronger FQH
states associated with larger FQHE excitation energies.

For finite size systems there are a few ways to calculate the gap and
we only consider the gap in the spherical geometry.  The gap is
considered to be the energy of a well separated
quasiparticle-quasihole pair where the initial ground state is
incompressible.  Hence, one only considers the excitation gap if the
ground state of the $N$ particle system at flux $Q$ is a uniform state
with total angular momentum $L=0$.  If this is the case then the
excitation gap can be calculated as the energy of a quasiparticle
$E^{(Q-1/2)}_0$ and a quasihole $E^{(Q+1/2)}_0$ where $E^{(Q)}_0$ is
the ground state energy of a system of $N$ electrons at flux
$Q$--hence, the quasihole state has $Q+1/2$ while the quasiparticle
has $Q-1/2$.  With these energies the gap is given as $\Delta =
E^{(Q+1/2)}_0+E^{(Q-1/2)}_0-2E^{(Q)}_0$.  Note that this definition
involves only calculations of ground state energies albeit at
different flux values.

\begin{figure}[t]
\begin{center}
\mbox{\includegraphics[width=6.5cm,angle=-90]{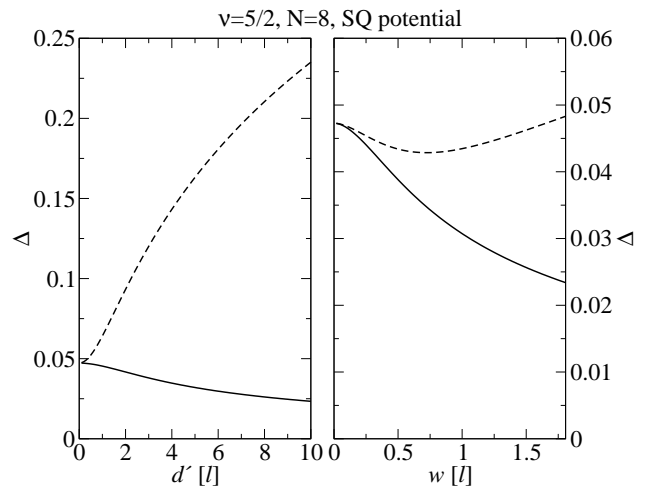}}
\end{center}
\caption{Energy gap $\Delta$ for $\nu=5/2$ using the SQ potential as a
function of $d^\prime/l$ (left) and $w/l$ (right).  The solid line is
the gap in units of $e^2/\epsilon l$ while the dashed line is the gap
in units of $e^2/\epsilon\sqrt{l^2+{d^\prime}^2}$ (left) and
$e^2/\epsilon\sqrt{l^2+w^2}$ (right).  The system considered has $N=8$
electrons.}
\label{gap-52}
\end{figure}

Another way to find the gap is to calculate the energy of the $N$
electron system at flux $Q$ as a function of angular momentum $L$.
The ground state (again only if the state is uniform and thus
incompressible) will have $L=0$, and a branch of low energy excited
states at different $L\neq0$.  The gap is then given as the energy in
the long wavelength limit which corresponds to the lowest energy
excitation with $L=N$ for the state at $\nu=7/3$ and $L=N/2$ for the,
presumably, paired state at $\nu=5/2$.  In other words,
\begin{eqnarray}
\Delta = E^{(Q)}(L_{ex})-E^{(Q)}(L=0)\;,
\end{eqnarray}
where $L_{ex}=N$ ($N/2$) for filling factor 7/3 (5/2).  This
definition involves obtaining the low-energy spectra of the system at
a given $Q$.  This is the method we use in this work to investigate
the gap.  The reason is that, as discussed by Morf~\cite{morf} and in
Ref.~\onlinecite{Peterson08}, the first method, described above, leads
to some ambiguity.  This is because for $\nu=5/2$ and $N=8$ electrons
the quasiparticle state is aliased to a FQHE state with filling factor
$2/3$ in the SLL.  Thus, what is being assumed as the quasiparticle
energy from the 5/2 FQH state may in fact be a 2/3 filled
incompressible FQH state.  Since we are also not calculating the gap
in the thermodynamic limit, it is not particularly important which
method we use since we are only interested in broad qualitative
features.  See Refs.~\onlinecite{morf} and~\onlinecite{feiguin} for
more thorough numerical investigations of the energy gaps in the FQHE.
We emphasize that our results are only qualitative and should not be
compared quantitatively with experimental activation gaps.

\begin{figure}[t]
\begin{center}
\mbox{\includegraphics[width=6.5cm,angle=-90]{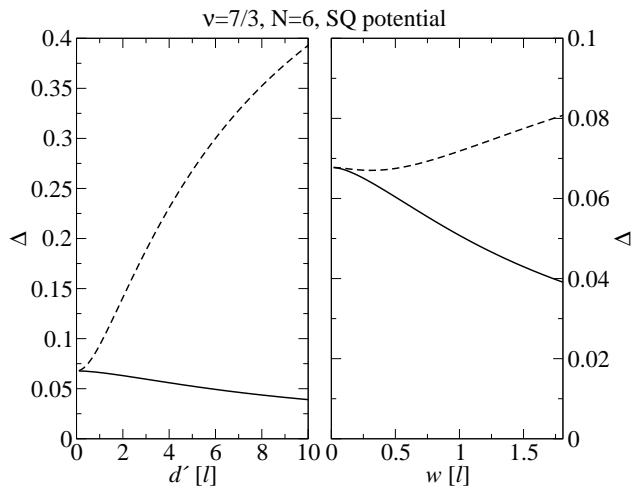}}
\end{center}
\caption{Same as Fig.~\ref{gap-52} except for $\nu=7/3$ and 
a system of $N=6$ electrons.}
\label{gap-73}
\end{figure}

In Figs.~\ref{gap-52} and~\ref{gap-73} we plot the calculated gap for $\nu=5/2$
and 7/3 for the SQ model for systems with $N=8$ and 6 electrons,
respectively.  The left panel shows the gap versus the width
$d^\prime$ and the right panel shows the gap versus $w$, both in units of $l$.
The solid line is the gap $\Delta$ in units of $e^2/\epsilon l$, and we
see that for both filling factors the gap (in units of $e^2/\epsilon l$) decreases with increasing
width (in units of $l$).  This is not surprising since the scale of the energy itself is being
reduced as the Coulomb energy is suppressed below $e^2/\epsilon l$ in the 
presence of finite thickness, e.g., see Fig.~\ref{vms} where the pseudopotentials
themselves are shown to decrease with increasing width.  In fact, this
introduces a subtle point when considering the theoretical energy gap including finite
thickness effects.  For every value of $d^\prime$ we are,
essentially, considering a different Hamiltonian, so it is not quite
appropriate to think of the gap as a function of the well width.  This is
most clear when one considers that the energy scale itself is changing
with $d^\prime$.  To readily incorporate the effect of a varying energy 
scale with varying $d^\prime$ we should rescale the gap energy.  The
dashed lines in the left (right) panels of Figs.~\ref{gap-52}
and~\ref{gap-73} show the gap in units of
$e^2/\epsilon\sqrt{l^2+{d^\prime}^2}$ ($e^2/\epsilon\sqrt{l^2+w^2}$).
In these units the gap is seen to increase with increasing width, i.e., 
in reduced energy units, the excitation gap is enhanced with 
increasing width.

We note that two alternative, and not necessarily equivalent,
signatures exist for the theoretical characterization of the stability
or the strength or the robustness of FQH incompressible states.  These
are (1) the overlap between the ground state finite-size numerical
wavefunction and a candidate incompressible state, and, (2) the FQH
excitation gap calculated directly numerically.  The relationship
between these two signatures of incompressibility is not obvious at
all, and both definitions have their problems.  The definition of the
excitation gap is ambiguous (particularly due to the aliasing problem)
since the two ways of defining it, as discussed above (using the
ground state energy at different $Q$ and the excitation spectra at the
same $Q$), are inequivalent.  Also, a finite size system always has
discrete energy levels, and thus always has a gap.  In addition, the
excitation energy $\Delta$, being an energy, is by definition
\textit{not} dimensionless, and therefore is somewhat problematic as a
signature for the ground state compressibility.  (For reasons
discussed above, we believe that the widespread practice of expressing
$\Delta$ in units of $e^2/\epsilon l$, the so-called Coulomb energy,
is arbitrary, and $(\Delta/(e^2/\epsilon l))$ may not necessarily be a
faithful representation of the stability of ground state
incompressibility, particularly in situations, e.g., 5/2 state, where
finite quasi-2D width is essential for the existence of the FQHE.)  On
the other hand, the signature of the overlap with a candidate
analytical incompressible state (i.e., a variational ansatz such as
the Laughlin or the MR wavefunction) has several intrinsic problems:
(1) such an analysis is necessarily limited by the constraint of the
variational ansatz, i.e., if a different (unknown) incompressible
state describes the ground state better, the overlap calculation would
miss that; (2) finite-size effects inherent in small system
calculations, where a large overlap may turn out to be a misleading
finite-size artifact; (3) the competition with nearby compressible
states is not captured.

It is our contention that when one investigates the FQHE theoretically
at some filling factor $\nu$ it is most illuminating to first
determine the nature of the state responsible for the FQHE.  This is
usually done via an overlap calculation with some trial wavefunction
(Laughlin, composite fermion, MR Pfaffian, etc.) such as was done in
Sec.~\ref{overlaps1}.  Once a satisfying identification is made for
the FQH ground state using the wavefunction overlap signature then
experimentally relevant quantities, such as the excitation gap, can be
calculated and compared with experiment.  In fact, this is essentially
the historical record of events following the discovery of the
$\nu=1/3$ FQHE and subsequent explanation by Laughlin.  This is also
the strategy we have followed in the current work.

\section{Conclusion}
\label{conc}

In this work, we investigate the relative stability of the FQHE at the
most important primary filling factors $\nu=$1/2, 1/3, and 1/5 in the
three lowest orbital Landau levels $n=0$ (LLL), 1 (SLL), and 2 (TLL)
by calculating, as a function of the quasi-2D layer width parameter,
the wavefunction overlap between the directly diagonalized exact
many-body wavefunction (for small spherical systems) with the
corresponding candidate theoretical ansatz for incompressible states,
namely, the Laughlin wavefunction for $\nu=1/3$ and 1/5 and the MR
Pfaffian wavefunction for $\nu=1/2$.  Rather surprisingly, we find the
layer thickness parameter, often neglected in theoretical studies of
the FQHE mostly carried out in the idealized zero-thickness strict 2D
limit, to be a key parameter in stabilizing the incompressible states
in the SLL.  In particular, we find that the SLL states at
$\nu=2+1/3=7/3$ and $2+1/2=5/2$ have larger exact overlap with the
corresponding theoretical incompressible states for finite values of
the layer thickness parameter $d^\prime$; typically, the overlap is
maximum (of the order of unity, in fact) for $d^\prime/l\sim$4-5 in
the SLL whereas in the LLL the overlap decreases monotonically with
increasing $d^\prime/l$, i.e., the incompressibility is the strongest
at $d^\prime=0$ (note that the width parameter $d^\prime\propto d$ is the ``normalized''
width parameter, and $d^\prime=d$ only for the square-well confinement
potential).  Furthermore, we also find that finite width is essential
to the observation of the threefold degenerate ground state
signature of the MR Pfaffian state in the torus geometry.  In particular, 
the expected threefold non-Abelian topological degeneracy for $\nu=5/2$ 
only shows up on the torus for $d^\prime\sim4$-$5l$, where the 
corresponding overlap with the MR wavefunction is optimal.  In the
TLL, the overlap is always (i.e., for any $d^\prime/l$) very small,
indicating the essentially generic absence of incompressible FQH
states in $n>1$ orbital LLs.

Our finding of the ``absence'' (i.e., small overlap) of the Laughlin
$\nu=1/3$ state in the $d^\prime=0$ limit in the SLL is consistent
with earlier theoretical work carried out in the ideal $d^\prime=0$
limit~\cite{macdonald-SLL,toke,reynolds,meskini}.  The experimental
FQHE observations~\cite{fqhe-SLL,fqhe-SLL-1,fqhe-SLL-2} at $\nu=7/3$
and $8/3$ occur, of course, at finite values of $d^\prime/l$, and the
experimental values of $d^\prime/l$ in the real samples is
$d^\prime/l\sim$2-5, which is in agreement with our theoretical
finding.  Our work indicates that tuning the quasi-2D width parameter
significantly (either far above or below the optimal value where the
overlap is maximum) should destroy incompressibility at $\nu=7/3$ (or
$8/3$).

The most important motivation for studying the SLL FQHE is to
understand the nature of the enigmatic 5/2 FQH state, the only
even-denominator FQH state ever observed in single-layer 2D systems.
What does our work imply for the stability of the anomalous FQHE at
half-filling?  We briefly discuss below the qualitative answer to this
question.

Our work clearly shows the importance of the finite quasi-2D layer
width in producing the incompressible FQH state at $\nu=5/2$, assuming
the state to be described by the MR Pfaffian wavefunction which is the
only available spin-polarized candidate wavefunction for the 5/2
state.  Consistent with earlier theoretical work in the
literature~\cite{morf-overlap,rez-hald,scarola}, we find the overlap
between the exact (small-system) many-body wavefunction at $\nu=5/2$
to have a modest overlap ($\sim$0.8-0.9) with the Pfaffian
wavefunction, which increases monotonically to a large overlap of
almost unity as $d^\prime/l$ increases.  For larger $d^\prime/l$, the
overlap decreases again.  Thus the behavior of the 5/2 state as a
function of finite width is similar to the $2+1/3=7/3$ Laughlin
state--both are most stable at a finite layer width in contrast to the
LLL Laughlin states at 1/3 and 1/5, which are most stable (i.e.,
maximum overlap $\sim1$) in the strict 2D ($d^\prime/l=0$) limit.

The behavior of the LLL 1/2 state is interesting in this context.  The
overlap with the Pfaffian here shows a very slight increase as a
function of $d^\prime/l$ before decreasing again similarly to the SLL
5/2 state except that the Pfaffian overlap never approaches unity for
the LLL 1/2 state, indicating that an incompressible $\nu=1/2$ FQH
state, at least one that is well-described by the MR Pfaffian, is
unlikely to occur in the LLL.  We emphasize that there is no
compelling fundamental reason for the LLL 1/2 state not manifesting a
$\nu=1/2$ FQHE, it is merely absent in the reasonable parameter space
of the Coulomb interaction for realistic experimental systems.  It is 
conceivable that a clever tuning of the interaction Hamiltonian far 
from the strict 2D Coulomb point will stabilize a LLL $\nu=1/2$ FQHE 
corresponding to the MR Pfaffian state.

The $\nu=1/5$ state in the SLL (i.e., $\nu=2+1/5=11/5$) is very
similar to the LLL FQH states in its dependence on the layer width.
The overlap with the Laughlin state decreases monotonically with
increasing width parameter, implying that both the LLL and SLL 1/5
FQHE would be the strongest in the strict 2D ideal limit for zero
layer thickness, similar to the situation for the $\nu=1/3$ LLL state.

Our work indicates that, in principle, wavefunction engineering should
be possible to enhance the FQHE at $\nu=5/2$ and 7/3 in the SLL by
increasing the quasi-2D width parameter $d^\prime/l$.  A trivial way
of enhancing $d^\prime/l$ is, of course, to decrease $l$ (at fixed
$d^\prime$) by increasing the magnetic field.  Since increasing the
magnetic field $B$ increases the interaction energy $\sim
e^2/l\approx\sqrt{B}$, it is obvious that this would strengthen the
FQHE trivially (the energy gap scales as $e^2/l$ thus increasing the
gap).  But increasing magnetic field while keeping $\nu$ fixed
requires a proportional increase in the carrier density which is
problematic (and may lead to the occupancy of the second subband,
reducing the sample mobility substantially).  Therefore, we suggest
that $d^\prime$ should also be increased to produce enhanced
stabilization of the SLL FQHE.  The fact that different models of
quasi-2D confinement, e.g., heterostructure (Fang-Howard), square-well,
parabolic quantum confinement, and the Zhang-Das Sarma model, all give
very similar qualitative and quantitative variation of our calculated
overlap with the width parameter indicates that our qualitative
conclusion is model independent, and applies to all quasi-2D physical
systems where the FQHE is experimentally studied.

As a direct consequence of the layer width dependence of the FQHE
found in this work, we also consider the experimentally important
situation~\cite{tilt1,tilt2} of the effect of an in-plane magnetic
field, $B_i$, applied parallel to the quasi-2D layer on the FQHE.
Assuming the system to be completely spin-polarized, as we do
throughout this work, and neglecting all disorder effects (i.e.,
ignoring for the moment that $B_i$ could cause additional
scattering~\cite{sds-hwang,prl2000} by inducing coupling between
in-plane and transverse dynamics), the only effect of $B_i$ is to
modify the quasi-2D layer width $d^\prime$ by shrinking it to
$d^\prime_i(B_i)<d^\prime_i(B_i=0)$.  This in-plane magnetic field
induced modification of the quasi-2D layer width arises from the
magneto-orbital coupling between in-plane and transverse electron
dynamics in a finite-width system, which is strictly absent in the
ideal 2D limit where, in the absence of spin (i.e., for a completely
spin-polarized system), there can be no coupling between the
in-plane and the transverse motion.  We show that in a quasi-2D
system, the in-plane field $B_i$ could destabilize the SLL 5/2
Pfaffian state by decreasing the effective layer width through the
reduction of the overlap between the exact wavefunction and the
Pfaffian wavefunction.  This follows naturally from our finding that
in the SLL, the reduction of the quasi-2D layer width always reduces
the overlap of the 5/2 state and since $B_i$ reduces the effective
value of the layer width it would naturally suppress the overlap at
$\nu=5/2$.  The same is also true for $\nu=7/3$, but \textit{not} for
$\nu=11/5$ in the SLL or for $\nu=$1/3 or 1/5 in the LLL since the
latter three FQH states are the most robust (i.e., maximum overlap) at
the smallest value of $d^\prime$.  We therefore predict that, to the
extent the 1/2, 1/3, and 1/5 FQH states are completely spin-polarized,
the application of a parallel magnetic field is likely to weaken the
$\nu=$5/2 and 7/3 FQH states and strengthen the $\nu=$11/5, 1/3, and
1/5 FQH states.  Of course, the physical effects neglected in our
approximation (e.g., spin, Landau level coupling, disorder) could play
important roles in actual experiments, but we suggest careful
experiments in an applied in-plane field to validate (or falsify) our
prediction of the weakening (strengthening) of the 5/2, 7/3 (11/5)
FQHE in the SLL.  We mention that the quantitative effect of the
in-plane field depends strongly on $l_i/l$ and $l_i/d^\prime$ where
$l_i\equiv (c\hbar/eB_i)^{1/2}$ and $l$ and $d^\prime$ are the 2D
magnetic length and the $B_i=0$ value of the width, respectively.
Since $l_i/l$ and $l_i/d^\prime$ both should be not too large for the
orbital influence of the in-plane field to be observable, it may not
be easy to observe the predicted effect in the LLL (or even for the
11/5 SLL) since the typical ``$l$'' is rather small for these cases.

Before concluding we summarize the large number of approximations we
have made in our theory: (1) We have assumed a spin-polarized system
throughout, and therefore if spin is playing an important dynamical
role in any of the fractional states we study, then our results would
not be particularly useful in understanding the corresponding
experimental observations.  (2) We have neglected the Landau level
coupling throughout our work, and it is possible that the LL coupling
plays a role in the SLL FQHE~\cite{fqhe-SLL-2008}.  (3) We have
neglected all disorder effects--in particular, the application of the
in-plane parallel field may introduce additional ``disorder'' by
opening new channels of scattering (e.g.,  intersubband scattering)
through the coupling of the in-plane and transverse dynamics.  This
will then serve to further weaken the 5/2 and 7/3 FQHE in the presence
of the in-plane field.  We expect such scattering effects to be very
weak in the $l_i\gtrsim d^\prime$ and $l_i\gtrsim l$ regime of
experimental interest.  (4) We have uncritically assumed that the
``strength'' of the incompressibility at a given filling factor (or
equivalently, the robustness of the FQHE at a given fraction) is
determined by the calculated overlap between the exactly diagonalized
numerical wavefunction for small system sizes and the corresponding
candidate theoretically postulated FQH state, i.e., the Laughlin
wavefunction for $\nu=1/5,1/3$ (in \textit{all} LLs) and the MR
Pfaffian wavefunction for $\nu=1/2$ (in \textit{all} LLs).  (5)
Related to the last point, our work will completely fail if the actual
state describing the FQHE at a particular fraction is qualitatively
different from our assumed candidate wavefunctions.  For example, one
cannot, as a matter of principle, rule out the possibility, unlikely
as it may seem, that the observed $\nu=5/2$ FQH state is the MR
Pfaffian state for finite $d^\prime/l\sim$4-5, as we find, which
continuously and adiabatically goes over to some other unknown
incompressible state at lower values of $d^\prime/l$.  Such a
continuous crossover implies that the observed 5/2 FQHE will remain
strong and robust for all values of $d^\prime$, but our calculated
overlap with just the Pfaffian decreases as $d^\prime/l$ decreases.
We believe such a scenario to be extremely unlikely, particularly
since there are no other proposed wavefunctions for 5/2, but we cannot
rule this out on purely theoretical grounds.  (The same also applies
for the 7/3 state where, if anything, such a scenario of two distinct
states, Laughlin for layer $d^\prime/l$ and ``something else'' for
smaller $d^\prime/l$, seems even more unlikely.)  (6) We have used
approximate models for considering the finite layer width in the
quasi-2D system.  However, the fact that four distinct models of
quasi-2D systems produce essentially identical results and conclusions
indicate that our results should have qualitative and
semi-quantitative validity in real samples.  Also, earlier theoretical
work~\cite{ortalano-zhang-sds, park-jain,morf} indicates that more
sophisticated models of quasi-2D confinement do not lead to
appreciable changes in the FQHE numerical results compared with the
relatively simple models used in our work.  (7) We have utilized the
standard spherical geometry for our finite size diagonalization using
rather modest system sizes (number of electrons between 5 and 10
depending on the filling factor).  We believe that the small system
size of our exact diagonalization study is not a problem since we are
not attempting any quantitative estimates of the excitation gap (or
other experimental quantities), but are interested in the qualitative
dependence of the overlap as a function of the layer width in
different LLs.  One reason for our use of relatively modest system
size is, of course, computational ease (since we are producing a very
large amount of numerical data: four different models of confinement,
and three different LLs, for three different FQH states, i.e., 36 sets
of diagonalization done as a function of the width parameter
$d^\prime$!).  Yet a second important reason for our choice of system
sizes is to avoid the well-known ``aliasing'' problem on the sphere
where two distinct fractions occur together exactly at the same
parameter values.  We have only chosen system sizes where the aliasing
issue does not arise.  The fact that we study three different orbital
LLs on an equal footing to compare the relative qualitative stability
of the 1/3, 1/5 and 1/2 states gives us confidence in our numerically
obtained trends (as a function of $d^\prime$) and conclusions, if not
the precise numbers.

Finally, we emphasize an important qualitative finding of our work.  We 
find that, in contrast to the Laughlin states (e.g., $\nu=1/3$) in the LLL 
where increasing the first pseudopotential $V_1$ compared with the 
Coulomb value necessarily strengthens the FQHE in a theoretically and 
physically meaningful manner, no such simple pseudopotential adjustments 
(either a single pseudopotential, e.g., $V_1$, $V_3$, $V_5$ or arbitrary 
combinations of them, e.g., $(V_1,V_3)$, $(V_1,V_5)$, $(V_3,V_5)$) make sense 
for studying the FQHE stability in higher (SLL or TLL) orbital Landau 
levels.  This is because of the theoretical fact that the $1/q$ Laughlin 
state becomes and \textit{exact} LLL eigenstate of the effective two-body 
Hamiltonian $\hat{H}_L^{(q)}$ (see Eq.~\ref{laugh-H}) where all pseudopotentials 
$m\geq q$ are taken to be zero, e.g., the Laughlin 1/3 state is just an 
exact LLL eigenstate of the Coulomb interaction if $V_1\rightarrow\infty$.  
This mathematical simplicity provides an adiabatic connection between 
the Laughlin state and the exact $\nu=1/3$ FQH state even in the realistic 
system as long as $V_1$ is \textit{not} too ``small''.  This mathematical 
simplicity is, however, completely lost in higher orbital LLs where the 
exact state cannot simply be written down in this manner.  In particular, 
the $\nu=5/2$ even-denominator state in the SLL, if it is indeed the Moore-Read 
Pfaffian state, is \textit{not} an eigenstate of any (even a completely 
unrealistic) two-body Hamiltonian.  As such, trying to understand the 
nature of the 5/2 state by varying the few lower pseudopotentials (e.g., 
$V_1$, $V_3$, etc.) around the Coulomb point is not 
mathematically (or physically) well-motivated since no two-body Hamiltonian 
exists with the MR state as its exact ground state.  We have therefore 
adopted the physically motivated approach in this work by working 
directly with the realistic Hamiltonian including the quasi-2D finite 
thickness effects (instead of using unrealistic arbitrary variations 
in the pseudopotentials).  Varying the finite quasi-2D layer thickness 
involves changes in \textit{all} the pseudopotentials which cannot be mimicked 
by varying any one or two pseudopotentials only.  We find that this physical 
approach leads to a rather unexpected finding: While in the LLL, finite 
quasi-2D thickness always weakens the FQHE, in the SLL, the finite 
thickness may actually stabilize the FQHE, for example, at $\nu=5/2$ 
and 7/3 filling factor.

In summary, we have theoretically considered the effect of orbital
dynamics on the stability of the primary FQH states at 1/2, 1/3, and
1/5 filling factors by calculating the wavefunction overlap (and 
the topological degeneracy expected for the MR Pfaffian state) as a
function of the quasi-2D layer width ($d^\prime$) in $n=0$ (LLL), 1
(SLL), and 2 (TLL) orbital LLs, finding that the FQHE does not occur
in the TLL (for any value of the quasi-2D width parameter), is the
most robust at $d^\prime=0$ in the LLL and for the 11/5 state in the
SLL, and is the most robust at $d^\prime/l\sim$4-5 in the SLL for the
5/2 and 7/3 state.  We also do not find any $\nu=1/2$ FQHE in the LLL
for any value of the layer width.

MRP and SDS acknowledge support from the Microsoft Q Project.  TJ
acknowledges support from IFRAF.  Numerical calculations have involved
computer time allocation IDRIS-072124.

%\bibliography{references}

\begin{thebibliography}{62}
\expandafter\ifx\csname natexlab\endcsname\relax\def\natexlab#1{#1}\fi
\expandafter\ifx\csname bibnamefont\endcsname\relax
  \def\bibnamefont#1{#1}\fi
\expandafter\ifx\csname bibfnamefont\endcsname\relax
  \def\bibfnamefont#1{#1}\fi
\expandafter\ifx\csname citenamefont\endcsname\relax
  \def\citenamefont#1{#1}\fi
\expandafter\ifx\csname url\endcsname\relax
  \def\url#1{\texttt{#1}}\fi
\expandafter\ifx\csname urlprefix\endcsname\relax\def\urlprefix{URL }\fi
\providecommand{\bibinfo}[2]{#2}
\providecommand{\eprint}[2][]{\url{#2}}





\bibitem[{\citenamefont{Tsui et~al.}(1982)\citenamefont{Tsui, Stormer, and
  Gossard}}]{fqhe}
\bibinfo{author}{\bibfnamefont{D.~C.} \bibnamefont{Tsui}},
  \bibinfo{author}{\bibfnamefont{H.~L.} \bibnamefont{Stormer}},
  \bibnamefont{and} \bibinfo{author}{\bibfnamefont{A.~C.}
  \bibnamefont{Gossard}}, \bibinfo{journal}{Phys. Rev. Lett.}
  \textbf{\bibinfo{volume}{48}}, \bibinfo{pages}{1559} (\bibinfo{year}{1982}).

\bibitem[{\citenamefont{Laughlin}(1983)}]{laughlin}
\bibinfo{author}{\bibfnamefont{R.~B.} \bibnamefont{Laughlin}},
  \bibinfo{journal}{Phys. Rev. Lett.} \textbf{\bibinfo{volume}{50}},
  \bibinfo{pages}{1395} (\bibinfo{year}{1983}).

\bibitem[{\citenamefont{Jain}(1989)}]{cftheory}
\bibinfo{author}{\bibfnamefont{J.~K.} \bibnamefont{Jain}},
  \bibinfo{journal}{Phys. Rev. Lett.} \textbf{\bibinfo{volume}{63}},
  \bibinfo{pages}{199} (\bibinfo{year}{1989}).

\bibitem[{\citenamefont{Halperin}(1984)}]{fracstats}
\bibinfo{author}{\bibfnamefont{B.~I.} \bibnamefont{Halperin}},
  \bibinfo{journal}{Phys. Rev. Lett.} \textbf{\bibinfo{volume}{52}},
  \bibinfo{pages}{1583} (\bibinfo{year}{1984}).

\bibitem[{\citenamefont{Leinaas and Myrheim}(1977)}]{fracstats1}
\bibinfo{author}{\bibfnamefont{J.~M.} \bibnamefont{Leinaas}} \bibnamefont{and}
  \bibinfo{author}{\bibfnamefont{J.}~\bibnamefont{Myrheim}},
  \bibinfo{journal}{Nuovo Cim.} \textbf{\bibinfo{volume}{B37}},
  \bibinfo{pages}{1} (\bibinfo{year}{1977}).

\bibitem[{\citenamefont{Wilczek}(1982)}]{fracstats2}
\bibinfo{author}{\bibfnamefont{F.}~\bibnamefont{Wilczek}},
  \bibinfo{journal}{Phys. Rev. Lett.} \textbf{\bibinfo{volume}{49}},
  \bibinfo{pages}{957} (\bibinfo{year}{1982}).

\bibitem[{qhe({\natexlab{a}})}]{qhe-persp}
\bibinfo{note}{{\em Perspectives in Quantum Hall Effects}, edited by S. Das
  Sarma and A. Pinczuk (Wiley, New York, 1997).}

\bibitem[{qhe({\natexlab{b}})}]{qhe-girvin}
\bibinfo{note}{{\em The Quantum Hall Effect}, edited by R. E. Prange and S. M.
  Girvin (Springer-Verlag, New York, 1990).}

\bibitem[{\citenamefont{Jain}(2007)}]{cfbook}
\bibinfo{author}{\bibfnamefont{J.~K.} \bibnamefont{Jain}},
  \emph{\bibinfo{title}{Composite Fermions}} (\bibinfo{publisher}{{Cambridge
  University Press, Cambridge}}, \bibinfo{year}{2007}).

\bibitem[{\citenamefont{Willett et~al.}(1987)\citenamefont{Willett, Eisenstein,
  St\"ormer, Tsui, Gossard, and English}}]{willett}
\bibinfo{author}{\bibfnamefont{R.}~\bibnamefont{Willett}},
  \bibinfo{author}{\bibfnamefont{J.~P.} \bibnamefont{Eisenstein}},
  \bibinfo{author}{\bibfnamefont{H.~L.} \bibnamefont{St\"ormer}},
  \bibinfo{author}{\bibfnamefont{D.~C.} \bibnamefont{Tsui}},
  \bibinfo{author}{\bibfnamefont{A.~C.} \bibnamefont{Gossard}},
  \bibnamefont{and} \bibinfo{author}{\bibfnamefont{J.~H.}
  \bibnamefont{English}}, \bibinfo{journal}{Phys. Rev. Lett.}
  \textbf{\bibinfo{volume}{59}}, \bibinfo{pages}{1776} (\bibinfo{year}{1987}).


\bibitem[{\citenamefont{Pan et~al.}(1999)\citenamefont{Pan, Xia, Shvarts,
  Adams, St\"ormer, Tsui, Pfeiffer, Baldwin, and West}}]{pan-52}
\bibinfo{author}{\bibfnamefont{W.}~\bibnamefont{Pan}},
  \bibinfo{author}{\bibfnamefont{J.-S.} \bibnamefont{Xia}},
  \bibinfo{author}{\bibfnamefont{V.}~\bibnamefont{Shvarts}},
  \bibinfo{author}{\bibfnamefont{D.~E.} \bibnamefont{Adams}},
  \bibinfo{author}{\bibfnamefont{H.~L.} \bibnamefont{St\"ormer}},
  \bibinfo{author}{\bibfnamefont{D.~C.} \bibnamefont{Tsui}},
  \bibinfo{author}{\bibfnamefont{L.~N.} \bibnamefont{Pfeiffer}},
  \bibinfo{author}{\bibfnamefont{K.~W.} \bibnamefont{Baldwin}},
  \bibnamefont{and} \bibinfo{author}{\bibfnamefont{K.~W.} \bibnamefont{West}},
  \bibinfo{journal}{Phys. Rev. Lett.} \textbf{\bibinfo{volume}{83}},
  \bibinfo{pages}{3530} (\bibinfo{year}{1999}).

\bibitem[{\citenamefont{Eisenstein et~al.}(2002)\citenamefont{Eisenstein,
  Cooper, Pfeiffer, and West}}]{fqhe-SLL}
\bibinfo{author}{\bibfnamefont{J.~P.} \bibnamefont{Eisenstein}},
  \bibinfo{author}{\bibfnamefont{K.~B.} \bibnamefont{Cooper}},
  \bibinfo{author}{\bibfnamefont{L.~N.} \bibnamefont{Pfeiffer}},
  \bibnamefont{and} \bibinfo{author}{\bibfnamefont{K.~W.} \bibnamefont{West}},
  \bibinfo{journal}{Phys. Rev. Lett.} \textbf{\bibinfo{volume}{88}},
  \bibinfo{pages}{076801} (\bibinfo{year}{2002}).

\bibitem[{\citenamefont{Xia et~al.}(2004)\citenamefont{Xia, Pan, Vicente,
  Adams, Sullivan, St\"ormer, Tsui, Pfeiffer, Baldwin, and West}}]{fqhe-SLL-1}
\bibinfo{author}{\bibfnamefont{J.~S.} \bibnamefont{Xia}},
  \bibinfo{author}{\bibfnamefont{W.}~\bibnamefont{Pan}},
  \bibinfo{author}{\bibfnamefont{C.~L.} \bibnamefont{Vicente}},
  \bibinfo{author}{\bibfnamefont{E.~D.} \bibnamefont{Adams}},
  \bibinfo{author}{\bibfnamefont{N.~S.} \bibnamefont{Sullivan}},
  \bibinfo{author}{\bibfnamefont{H.~L.} \bibnamefont{St\"ormer}},
  \bibinfo{author}{\bibfnamefont{D.~C.} \bibnamefont{Tsui}},
  \bibinfo{author}{\bibfnamefont{L.~N.} \bibnamefont{Pfeiffer}},
  \bibinfo{author}{\bibfnamefont{K.~W.} \bibnamefont{Baldwin}},
  \bibnamefont{and} \bibinfo{author}{\bibfnamefont{K.~W.} \bibnamefont{West}},
  \bibinfo{journal}{Phys. Rev. Lett.} \textbf{\bibinfo{volume}{93}},
  \bibinfo{pages}{176809} (\bibinfo{year}{2004}).

\bibitem[{\citenamefont{Csathy et~al.}(2005)\citenamefont{Csathy, Xia, Vicente,
  Adams, Sullivan, St\"ormer, Tsui, Pfeiffer, and West}}]{fqhe-SLL-2}
\bibinfo{author}{\bibfnamefont{G.~A.} \bibnamefont{Csathy}},
  \bibinfo{author}{\bibfnamefont{J.~S.} \bibnamefont{Xia}},
  \bibinfo{author}{\bibfnamefont{C.~L.} \bibnamefont{Vicente}},
  \bibinfo{author}{\bibfnamefont{E.~D.} \bibnamefont{Adams}},
  \bibinfo{author}{\bibfnamefont{N.~S.} \bibnamefont{Sullivan}},
  \bibinfo{author}{\bibfnamefont{H.~L.} \bibnamefont{St\"ormer}},
  \bibinfo{author}{\bibfnamefont{D.~C.} \bibnamefont{Tsui}},
  \bibinfo{author}{\bibfnamefont{L.~N.} \bibnamefont{Pfeiffer}},
  \bibnamefont{and} \bibinfo{author}{\bibfnamefont{K.~W.} \bibnamefont{West}},
  \bibinfo{journal}{Phys. Rev. Lett.} \textbf{\bibinfo{volume}{94}},
  \bibinfo{eid}{146801} (\bibinfo{year}{2005}).

\bibitem[{\citenamefont{Choi et~al.}(2007)\citenamefont{Choi, Kang, Das~Sarma,
  Pfeiffer, and West}}]{choi}
\bibinfo{author}{\bibfnamefont{H.~C.} \bibnamefont{Choi}},
  \bibinfo{author}{\bibfnamefont{W.}~\bibnamefont{Kang}},
  \bibinfo{author}{\bibfnamefont{S.}~\bibnamefont{Das~Sarma}},
  \bibinfo{author}{\bibfnamefont{L.~N.} \bibnamefont{Pfeiffer}},
  \bibnamefont{and} \bibinfo{author}{\bibfnamefont{K.~W.} \bibnamefont{West}},
  \bibinfo{journal}{Phys. Rev. B} \textbf{\bibinfo{volume}{77}}, 
  \bibinfo{pages}{081301}  (\bibinfo{year}{2007}).

\bibitem[{\citenamefont{Pan et~al.}(2007)\citenamefont{Pan, Xia, St\"ormer,
  Tsui, Vicente, Adams, Sullivan, Pfeiffer, Baldwin, and West}}]{pan}
\bibinfo{author}{\bibfnamefont{W.}~\bibnamefont{Pan}},
  \bibinfo{author}{\bibfnamefont{J.~S.} \bibnamefont{Xia}},
  \bibinfo{author}{\bibfnamefont{H.~L.} \bibnamefont{St\"ormer}},
  \bibinfo{author}{\bibfnamefont{D.~C.} \bibnamefont{Tsui}},
  \bibinfo{author}{\bibfnamefont{C.}~\bibnamefont{Vicente}},
  \bibinfo{author}{\bibfnamefont{E.~D.} \bibnamefont{Adams}},
  \bibinfo{author}{\bibfnamefont{N.~S.} \bibnamefont{Sullivan}},
  \bibinfo{author}{\bibfnamefont{L.~N.} \bibnamefont{Pfeiffer}},
  \bibinfo{author}{\bibfnamefont{K.~W.} \bibnamefont{Baldwin}},
  \bibnamefont{and} \bibinfo{author}{\bibfnamefont{K.~W.} \bibnamefont{West}},
  \bibinfo{journal}{Phys. Rev. B} \textbf{\bibinfo{volume}{77}},
  \bibinfo{pages}{075307}  (\bibinfo{year}{2007}).

\bibitem[{\citenamefont{Gervais et~al.}(2004)\citenamefont{Gervais, Engel,
  St\"ormer, Tsui, Baldwin, West, and Pfeiffer}}]{fqhe-TLL}
\bibinfo{author}{\bibfnamefont{G.}~\bibnamefont{Gervais}},
  \bibinfo{author}{\bibfnamefont{L.~W.} \bibnamefont{Engel}},
  \bibinfo{author}{\bibfnamefont{H.~L.} \bibnamefont{St\"ormer}},
  \bibinfo{author}{\bibfnamefont{D.~C.} \bibnamefont{Tsui}},
  \bibinfo{author}{\bibfnamefont{K.~W.} \bibnamefont{Baldwin}},
  \bibinfo{author}{\bibfnamefont{K.~W.} \bibnamefont{West}}, \bibnamefont{and}
  \bibinfo{author}{\bibfnamefont{L.~N.} \bibnamefont{Pfeiffer}},
  \bibinfo{journal}{Phys. Rev. Lett.} \textbf{\bibinfo{volume}{93}},
  \bibinfo{pages}{266804} (\bibinfo{year}{2004}).

\bibitem[{\citenamefont{MacDonald and Aers}(1984)}]{macdonald}
\bibinfo{author}{\bibfnamefont{A.~H.} \bibnamefont{MacDonald}}
  \bibnamefont{and} \bibinfo{author}{\bibfnamefont{G.~C.} \bibnamefont{Aers}},
  \bibinfo{journal}{Phys. Rev. B} \textbf{\bibinfo{volume}{29}},
  \bibinfo{pages}{5976} (\bibinfo{year}{1984}).

\bibitem[{\citenamefont{He et~al.}(1990)\citenamefont{He, Zhang, Xie, and
  Das~Sarma}}]{he}
\bibinfo{author}{\bibfnamefont{S.}~\bibnamefont{He}},
  \bibinfo{author}{\bibfnamefont{F.~C.} \bibnamefont{Zhang}},
  \bibinfo{author}{\bibfnamefont{X.~C.} \bibnamefont{Xie}}, \bibnamefont{and}
  \bibinfo{author}{\bibfnamefont{S.}~\bibnamefont{Das~Sarma}},
  \bibinfo{journal}{Phys. Rev. B} \textbf{\bibinfo{volume}{42}},
  \bibinfo{pages}{11376} (\bibinfo{year}{1990}).

\bibitem[{\citenamefont{Zhang and Das~Sarma}(1986)}]{zds}
\bibinfo{author}{\bibfnamefont{F.~C.} \bibnamefont{Zhang}} \bibnamefont{and}
  \bibinfo{author}{\bibfnamefont{S.}~\bibnamefont{Das~Sarma}},
  \bibinfo{journal}{Phys. Rev. B} \textbf{\bibinfo{volume}{33}},
  \bibinfo{pages}{2903} (\bibinfo{year}{1986}).

\bibitem[{\citenamefont{Ortalano et~al.}(1997)\citenamefont{Ortalano, He, and
  Das~Sarma}}]{ortalano-zhang-sds}
\bibinfo{author}{\bibfnamefont{M.~W.} \bibnamefont{Ortalano}},
  \bibinfo{author}{\bibfnamefont{S.}~\bibnamefont{He}}, \bibnamefont{and}
  \bibinfo{author}{\bibfnamefont{S.}~\bibnamefont{Das~Sarma}},
  \bibinfo{journal}{Phys. Rev. B} \textbf{\bibinfo{volume}{55}},
  \bibinfo{pages}{7702} (\bibinfo{year}{1997}).

\bibitem[{\citenamefont{Park and Jain}(1998)}]{park-jain}
\bibinfo{author}{\bibfnamefont{K.}~\bibnamefont{Park}} \bibnamefont{and}
  \bibinfo{author}{\bibfnamefont{J.~K.} \bibnamefont{Jain}},
  \bibinfo{journal}{Phys. Rev. Lett.} \textbf{\bibinfo{volume}{81}},
  \bibinfo{pages}{4200} (\bibinfo{year}{1998}).

\bibitem[{\citenamefont{Park et~al.}(1999)\citenamefont{Park, Meskini, and
  Jain}}]{park-meskini-jain}
\bibinfo{author}{\bibfnamefont{K.}~\bibnamefont{Park}},
  \bibinfo{author}{\bibfnamefont{N.}~\bibnamefont{Meskini}}, \bibnamefont{and}
  \bibinfo{author}{\bibfnamefont{J.~K.} \bibnamefont{Jain}},
  \bibinfo{journal}{Journal of Physics: Condensed Matter}
  \textbf{\bibinfo{volume}{11}}, \bibinfo{pages}{7283} (\bibinfo{year}{1999}).

\bibitem[{\citenamefont{Gammel et~al.}(1988)\citenamefont{Gammel, Bishop,
  Eisenstein, English, Gossard, Ruel, and St\"ormer}}]{gammel-52}
\bibinfo{author}{\bibfnamefont{P.~L.} \bibnamefont{Gammel}},
  \bibinfo{author}{\bibfnamefont{D.~J.} \bibnamefont{Bishop}},
  \bibinfo{author}{\bibfnamefont{J.~P.} \bibnamefont{Eisenstein}},
  \bibinfo{author}{\bibfnamefont{J.~H.} \bibnamefont{English}},
  \bibinfo{author}{\bibfnamefont{A.~C.} \bibnamefont{Gossard}},
  \bibinfo{author}{\bibfnamefont{R.}~\bibnamefont{Ruel}}, \bibnamefont{and}
  \bibinfo{author}{\bibfnamefont{H.~L.} \bibnamefont{St\"ormer}},
  \bibinfo{journal}{Phys. Rev. B} \textbf{\bibinfo{volume}{38}},
  \bibinfo{pages}{10128} (\bibinfo{year}{1988}).

\bibitem[{\citenamefont{Eisenstein
  et~al.}(1988{\natexlab{a}})\citenamefont{Eisenstein, Willett, St\"ormer,
  Tsui, Gossard, and English}}]{eisenstein-52}
\bibinfo{author}{\bibfnamefont{J.~P.} \bibnamefont{Eisenstein}},
  \bibinfo{author}{\bibfnamefont{R.}~\bibnamefont{Willett}},
  \bibinfo{author}{\bibfnamefont{H.~L.} \bibnamefont{St\"ormer}},
  \bibinfo{author}{\bibfnamefont{D.~C.} \bibnamefont{Tsui}},
  \bibinfo{author}{\bibfnamefont{A.~C.} \bibnamefont{Gossard}},
  \bibnamefont{and} \bibinfo{author}{\bibfnamefont{J.~H.}
  \bibnamefont{English}}, \bibinfo{journal}{Phys. Rev. Lett.}
  \textbf{\bibinfo{volume}{61}}, \bibinfo{pages}{997}
  (\bibinfo{year}{1988}{\natexlab{a}}).

\bibitem[{\citenamefont{Klitzing et~al.}(1980)\citenamefont{Klitzing, Dorda,
  and Pepper}}]{iqhe}
\bibinfo{author}{\bibfnamefont{K.~v.} \bibnamefont{Klitzing}},
  \bibinfo{author}{\bibfnamefont{G.}~\bibnamefont{Dorda}}, \bibnamefont{and}
  \bibinfo{author}{\bibfnamefont{M.}~\bibnamefont{Pepper}},
  \bibinfo{journal}{Phys. Rev. Lett.} \textbf{\bibinfo{volume}{45}},
  \bibinfo{pages}{494} (\bibinfo{year}{1980}).

\bibitem{shayegan-1992}
Y. W. Suen, L. W. Engel, M. B. Santos, M. Shayegan, and D. C. Tsui,
Phys. Rev. Lett. \textbf{68}, 1379 (1992).

\bibitem{eisenstein-1992}
J. P. Eisenstein, G. S. Boebinger, L. N. Pfeiffer, K. W. West, and Song He,
Phys. Rev. Lett. \textbf{68}, 1383 (1992).

\bibitem{eisenstein-persp}
J. P. Eisenstein in Ref.~\onlinecite{qhe-persp}.

\bibitem{macdonald-girvin-persp}
S. M. Girvin and A. H. MacDonald in Ref.~\onlinecite{qhe-persp}.

\bibitem{he-xie-sds-prb-1993}
Song He, S. Das Sarma, and X. C. Xie, 
Phys. Rev. B \textbf{47}, 4394 (1993). 

\bibitem{halp-331}
B. I. Halperin,
Helv. Phys. Acta \textbf{56}, 783 (1983).

\bibitem[{\citenamefont{Moore and Read}(1991)}]{pfaff}
\bibinfo{author}{\bibfnamefont{G.}~\bibnamefont{Moore}} \bibnamefont{and}
  \bibinfo{author}{\bibfnamefont{N.}~\bibnamefont{Read}},
  \bibinfo{journal}{Nucl. Phys. B} \textbf{\bibinfo{volume}{360}},
  \bibinfo{pages}{362} (\bibinfo{year}{1991}).

\bibitem[{\citenamefont{Morf}(1998)}]{morf-overlap}
\bibinfo{author}{\bibfnamefont{R.~H.} \bibnamefont{Morf}},
  \bibinfo{journal}{Phys. Rev. Lett.} \textbf{\bibinfo{volume}{80}},
  \bibinfo{pages}{1505} (\bibinfo{year}{1998}).

\bibitem[{\citenamefont{Rezayi and Haldane}(2000)}]{rez-hald}
\bibinfo{author}{\bibfnamefont{E.~H.} \bibnamefont{Rezayi}} \bibnamefont{and}
  \bibinfo{author}{\bibfnamefont{F.~D.~M.} \bibnamefont{Haldane}},
  \bibinfo{journal}{Phys. Rev. Lett.} \textbf{\bibinfo{volume}{84}},
  \bibinfo{pages}{4685} (\bibinfo{year}{2000}).

\bibitem[{\citenamefont{Read and Rezayi}(1996)}]{rr}
\bibinfo{author}{\bibfnamefont{N.}~\bibnamefont{Read}} \bibnamefont{and}
  \bibinfo{author}{\bibfnamefont{E.}~\bibnamefont{Rezayi}},
  \bibinfo{journal}{Phys. Rev. B} \textbf{\bibinfo{volume}{54}},
  \bibinfo{pages}{16864} (\bibinfo{year}{1996}).

\bibitem[{\citenamefont{Morf et~al.}(2002)\citenamefont{Morf,
  d\char39{}Ambrumenil, and Das~Sarma}}]{morf}
\bibinfo{author}{\bibfnamefont{R.~H.} \bibnamefont{Morf}},
  \bibinfo{author}{\bibfnamefont{N.}~\bibnamefont{d\char39{}Ambrumenil}},
  \bibnamefont{and}
  \bibinfo{author}{\bibfnamefont{S.}~\bibnamefont{Das~Sarma}},
  \bibinfo{journal}{Phys. Rev. B} \textbf{\bibinfo{volume}{66}},
  \bibinfo{pages}{075408} (\bibinfo{year}{2002}).

\bibitem[{\citenamefont{Scarola et~al.}(2000)\citenamefont{Scarola, Park, and
  Jain}}]{scarola1}
\bibinfo{author}{\bibfnamefont{V.~W.} \bibnamefont{Scarola}},
  \bibinfo{author}{\bibfnamefont{K.}~\bibnamefont{Park}}, \bibnamefont{and}
  \bibinfo{author}{\bibfnamefont{J.~K.} \bibnamefont{Jain}},
  \bibinfo{journal}{Nature} \textbf{\bibinfo{volume}{406}},
  \bibinfo{pages}{863} (\bibinfo{year}{2000}).

\bibitem[{\citenamefont{Scarola et~al.}(2002)\citenamefont{Scarola, Jain, and
  Rezayi}}]{scarola}
\bibinfo{author}{\bibfnamefont{V.~W.} \bibnamefont{Scarola}},
  \bibinfo{author}{\bibfnamefont{J.~K.} \bibnamefont{Jain}}, \bibnamefont{and}
  \bibinfo{author}{\bibfnamefont{E.~H.} \bibnamefont{Rezayi}},
  \bibinfo{journal}{Phys. Rev. Lett.} \textbf{\bibinfo{volume}{88}},
  \bibinfo{pages}{216804} (\bibinfo{year}{2002}).

\bibitem[{\citenamefont{Toke and Jain}(2006)}]{toke1}
\bibinfo{author}{\bibfnamefont{C.}~\bibnamefont{Toke}} \bibnamefont{and}
  \bibinfo{author}{\bibfnamefont{J.~K.} \bibnamefont{Jain}},
  \bibinfo{journal}{Phys. Rev. Lett.} \textbf{\bibinfo{volume}{96}},
  \bibinfo{eid}{246805} (\bibinfo{year}{2006}).

\bibitem[{\citenamefont{Toke et~al.}(2007)\citenamefont{Toke, Regnault, and
  Jain}}]{toke2}
\bibinfo{author}{\bibfnamefont{C.}~\bibnamefont{Toke}},
  \bibinfo{author}{\bibfnamefont{N.}~\bibnamefont{Regnault}}, \bibnamefont{and}
  \bibinfo{author}{\bibfnamefont{J.~K.} \bibnamefont{Jain}},
  \bibinfo{journal}{Phys. Rev. Lett.} \textbf{\bibinfo{volume}{98}},
  \bibinfo{eid}{036806} (\bibinfo{year}{2007}).

\bibitem[{\citenamefont{Wojs and Quinn}(2006)}]{wojs1}
\bibinfo{author}{\bibfnamefont{A.}~\bibnamefont{Wojs}} \bibnamefont{and}
  \bibinfo{author}{\bibfnamefont{J.~J.} \bibnamefont{Quinn}},
  \bibinfo{journal}{Phys. Rev. B} \textbf{\bibinfo{volume}{74}},
  \bibinfo{eid}{235319} (\bibinfo{year}{2006}).

\bibitem{moller}
G. M\"{o}ller and S. H. Simon, Phys. Ref. B \textbf{77}, 075319 (2008).
	
\bibitem[{\citenamefont{Feiguin et~al.}(2007)\citenamefont{Feiguin, Rezayi,
  Nayak, and Das Sarma}}]{feiguin}
\bibinfo{author}{\bibfnamefont{A.~E.} \bibnamefont{Feiguin}},
  \bibinfo{author}{\bibfnamefont{E.}~\bibnamefont{Rezayi}},
  \bibinfo{author}{\bibfnamefont{C.}~\bibnamefont{Nayak}}, \bibnamefont{and}
  \bibinfo{author}{\bibfnamefont{S.} \bibnamefont{Das Sarma}},
  \bibinfo{journal}{Phys. Rev. Lett.}\textbf{\bibinfo{volume}{100}}, 
\bibinfo{pages}{166803} (\bibinfo{year}{2008}).

\bibitem[{\citenamefont{Read}(2001)}]{mrref}
\bibinfo{author}{\bibfnamefont{N.}~\bibnamefont{Read}},
  \bibinfo{journal}{Physica B: Condensed Matter}
  \textbf{\bibinfo{volume}{298}}, \bibinfo{pages}{121} (\bibinfo{year}{2001}).

\bibitem{hald-rez-ss}
F. D. M. Haldane and E. H. Rezayi, Phys. Rev. Lett. 60, 956 (1988)

\bibitem[{\citenamefont{MacDonald}(1984)}]{macdonald-SLL}
\bibinfo{author}{\bibfnamefont{A.~H.} \bibnamefont{MacDonald}},
  \bibinfo{journal}{Phys. Rev. B} \textbf{\bibinfo{volume}{30}},
  \bibinfo{pages}{3550} (\bibinfo{year}{1984}).

\bibitem[{\citenamefont{Toke et~al.}(2005)\citenamefont{Toke, Peterson, Jeon,
  and Jain}}]{toke}
\bibinfo{author}{\bibfnamefont{C.}~\bibnamefont{Toke}},
  \bibinfo{author}{\bibfnamefont{M.~R.} \bibnamefont{Peterson}},
  \bibinfo{author}{\bibfnamefont{G.~S.} \bibnamefont{Jeon}}, \bibnamefont{and}
  \bibinfo{author}{\bibfnamefont{J.~K.} \bibnamefont{Jain}},
  \bibinfo{journal}{Phys. Rev. B} \textbf{\bibinfo{volume}{72}},
  \bibinfo{eid}{125315} (\bibinfo{year}{2005}).

\bibitem[{\citenamefont{d'Ambrumenil and Reynolds}(1988)}]{reynolds}
\bibinfo{author}{\bibfnamefont{N.}~\bibnamefont{d'Ambrumenil}}
  \bibnamefont{and} \bibinfo{author}{\bibfnamefont{A.~M.}
  \bibnamefont{Reynolds}}, \bibinfo{journal}{J. Phys. C}
  \textbf{\bibinfo{volume}{21}}, \bibinfo{pages}{119} (\bibinfo{year}{1988}).

\bibitem{scarola-hll}
V. W. Scarola, Kwon Park, and J. K. Jain, 
Phys. Rev. B \textbf{62}, R16259 (2000).

\bibitem[{\citenamefont{Sarma et~al.}(2005)\citenamefont{Sarma, Freedman, and
  Nayak}}]{tqc-1}
\bibinfo{author}{\bibfnamefont{S.} \bibnamefont{Das Sarma}},
  \bibinfo{author}{\bibfnamefont{M.}~\bibnamefont{Freedman}}, \bibnamefont{and}
  \bibinfo{author}{\bibfnamefont{C.}~\bibnamefont{Nayak}},
  \bibinfo{journal}{Phys. Rev. Lett.} \textbf{\bibinfo{volume}{94}},
  \bibinfo{eid}{166802} (\bibinfo{year}{2005}).

\bibitem{tqc-2}
C. Nayak, S. H. Simon, A. Stern, M. Freedman, and S. Das Sarma, 
arXiv:0707.1889v1 [cond-mat.str-el] (2007).

\bibitem{willett-2008}
R. L. Willett, M. J. Manfra, L. N. Pfeiffer, and K. W. West,
arXiv:0807.0221v1 [cond-mat.mes-hall] (2008).

\bibitem{dolev-2008}
M. Dolev, M. Heiblum, V. Umansky, Ady Stern,  and  D. Mahalu, 
Nature \textbf{452}, 829 (2008).

\bibitem{radu-2008}
I. P. Radu, J. B. Miller, C. M. Marcus, M. A. Kastner, L. N. Pfeiffer, and 
K. W. West, Science \textbf{320}, 899 (2008).

\bibitem{Peterson08}
M. R. Peterson, Th. Jolicoeur, and S. Das Sarma,
Phys. Rev. Lett. \textbf{101}, 016807 (2008).

\bibitem[{\citenamefont{Haldane}(1983)}]{exact-laughlin}
\bibinfo{author}{\bibfnamefont{F.~D.~M.} \bibnamefont{Haldane}},
  \bibinfo{journal}{Phys. Rev. Lett.} \textbf{\bibinfo{volume}{51}},
  \bibinfo{pages}{605} (\bibinfo{year}{1983}).

\bibitem[{\citenamefont{Haldane}()}]{haldane-qhe}
\bibinfo{author}{\bibfnamefont{F.~D.~M.} \bibnamefont{Haldane}}, \eprint{in
  Ref.~\onlinecite{qhe-girvin}.}

\bibitem[{\citenamefont{Greiter et~al.}(1991)\citenamefont{Greiter, Wen, and
  Wilczek}}]{pfaff-exact}
\bibinfo{author}{\bibfnamefont{M.}~\bibnamefont{Greiter}},
  \bibinfo{author}{\bibfnamefont{X.-G.} \bibnamefont{Wen}}, \bibnamefont{and}
  \bibinfo{author}{\bibfnamefont{F.}~\bibnamefont{Wilczek}},
  \bibinfo{journal}{Phys. Rev. Lett.} \textbf{\bibinfo{volume}{66}},
  \bibinfo{pages}{3205} (\bibinfo{year}{1991}).

\bibitem[{\citenamefont{Sarma and Mason}(1985)}]{sq-pot}
\bibinfo{author}{\bibfnamefont{S.} \bibnamefont{Das Sarma}} \bibnamefont{and}
  \bibinfo{author}{\bibfnamefont{B.~A.} \bibnamefont{Mason}},
  \bibinfo{journal}{Ann. Phys.} \textbf{\bibinfo{volume}{163}},
  \bibinfo{pages}{78} (\bibinfo{year}{1985}).

\bibitem[{\citenamefont{Ando et~al.}(1982)\citenamefont{Ando, Fowler, and
  Stern}}]{fh-pot1}
\bibinfo{author}{\bibfnamefont{T.}~\bibnamefont{Ando}},
  \bibinfo{author}{\bibfnamefont{A.~B.} \bibnamefont{Fowler}},
  \bibnamefont{and} \bibinfo{author}{\bibfnamefont{F.}~\bibnamefont{Stern}},
  \bibinfo{journal}{Rev. Mod. Phys.} \textbf{\bibinfo{volume}{54}},
  \bibinfo{pages}{437} (\bibinfo{year}{1982}).

\bibitem[{\citenamefont{Stern and Sarma}(1984)}]{fh-pot2}
\bibinfo{author}{\bibfnamefont{F.}~\bibnamefont{Stern}} \bibnamefont{and}
  \bibinfo{author}{\bibfnamefont{S.} \bibnamefont{Das Sarma}},
  \bibinfo{journal}{Phys. Rev. B} \textbf{\bibinfo{volume}{30}},
  \bibinfo{pages}{840} (\bibinfo{year}{1984}).

\bibitem[{\citenamefont{Belkhir and Jain}(1995)}]{belkhir-jain}
\bibinfo{author}{\bibfnamefont{L.}~\bibnamefont{Belkhir}} \bibnamefont{and}
  \bibinfo{author}{\bibfnamefont{J.~K.} \bibnamefont{Jain}},
  \bibinfo{journal}{Solid State Commun.} \textbf{\bibinfo{volume}{94}},
  \bibinfo{pages}{107} (\bibinfo{year}{1995}).

\bibitem[{\citenamefont{Wu and Yang}(1976)}]{yang-wu}
\bibinfo{author}{\bibfnamefont{T.~T.} \bibnamefont{Wu}} \bibnamefont{and}
  \bibinfo{author}{\bibfnamefont{C.~N.} \bibnamefont{Yang}},
  \bibinfo{journal}{Nucl. Phys. B} \textbf{\bibinfo{volume}{107}},
  \bibinfo{pages}{365} (\bibinfo{year}{1976}).

\bibitem[{\citenamefont{Fano et~al.}(1986)\citenamefont{Fano, Ortolani, and
  Colombo}}]{fano}
\bibinfo{author}{\bibfnamefont{G.}~\bibnamefont{Fano}},
  \bibinfo{author}{\bibfnamefont{F.}~\bibnamefont{Ortolani}}, \bibnamefont{and}
  \bibinfo{author}{\bibfnamefont{E.}~\bibnamefont{Colombo}},
  \bibinfo{journal}{Phys. Rev. B} \textbf{\bibinfo{volume}{34}},
  \bibinfo{pages}{2670} (\bibinfo{year}{1986}).

\bibitem[{foo({\natexlab{a}})}]{foot-sphere}
\bibinfo{note}{The single particle angular momentum $l$ is related to the
  monopole strength $Q$ through $l=Q+n$ in the spherical geometry.}

\bibitem[{foo({\natexlab{b}})}]{footnote-n10}
\bibinfo{note}{Furthermore, we are not trying to break any records by
  diagonalizing the largest systems. The larger systems will be left to other
  works, i.e., Ref.~\onlinecite{Peterson08}}

\bibitem[{foo({\natexlab{c}})}]{foot}
\bibinfo{note}{With increasing $d^\prime$ the ZDS potential gives an overlap that
  drops off significantly towards zero, cf. Ref.~\onlinecite{he}.}

\bibitem[{\citenamefont{Sbeouelji and Meskini}(2001)}]{meskini}
\bibinfo{author}{\bibfnamefont{T.}~\bibnamefont{Sbeouelji}} \bibnamefont{and}
  \bibinfo{author}{\bibfnamefont{N.}~\bibnamefont{Meskini}},
  \bibinfo{journal}{Phys. Rev. B} \textbf{\bibinfo{volume}{64}},
  \bibinfo{pages}{193305} (\bibinfo{year}{2001}).

\bibitem{cffs}
B. I. Halperin, and P. A. Lee and N. Read, 
Phys. Rev. B \textbf{47}, 7312 (1993).

\bibitem{Haldane85}
F. D. M. Haldane,
Phys. Rev. Lett. \textbf{55}, 2095 (1985).

\bibitem{Rezayi99}
E. H. Rezayi, F. D. M. Haldane, and K. Yang,
Phys. Rev. Lett. \textbf{83}, 1219 (1999).

\bibitem{Haldane00}
F. D. M. Haldane, E. H. Rezayi, and K. Yang,
Phys. Rev. Lett. \textbf{85}, 5396 (2000).

\bibitem{Yang01}
K. Yang, F. D. M. Haldane, and E. H. Rezayi,
Phys. Rev. B\textbf{64}, 081301(R) (2001).

\bibitem{HR85}
F. D. M. Haldane and E. H. Rezayi,
Phys. Rev. B \textbf{31}, 2529 (1985).

\bibitem{Greiter92}
M. Greiter, X. G. Wen, and F. Wilczek,
Nucl. Phys. B \textbf{374}, 567 (1992).

\bibitem{Chung07}
S. B. Chung and M. Stone,
J. Phys. A \textbf{40}, 4923 (2007).

\bibitem{Levin07}
M. Levin, B. I. Halperin, and B. Rosenow,
Phys. Rev. Lett. \textbf{99}, 236806 (2007).

\bibitem{Lee07}
S.-S. Lee, S. Ryu, C. Nayak, and M. P. A. Fisher,
Phys. Rev. Lett. \textbf{99}, 236807 (2007).

\bibitem{mrp-kp-sds}
M. R. Peterson, Kwon Park, and S. Das Sarma, 
arXiv:0807.0638v1 [cond-mat.mes-hall] (2008).

\bibitem[{\citenamefont{Eisenstein
  et~al.}(1988{\natexlab{b}})\citenamefont{Eisenstein, Willett, St\"ormer,
  Tsui, Gossard, and English}}]{tilt1}
\bibinfo{author}{\bibfnamefont{J.~P.} \bibnamefont{Eisenstein}},
  \bibinfo{author}{\bibfnamefont{R.}~\bibnamefont{Willett}},
  \bibinfo{author}{\bibfnamefont{H.~L.} \bibnamefont{St\"ormer}},
  \bibinfo{author}{\bibfnamefont{D.~C.} \bibnamefont{Tsui}},
  \bibinfo{author}{\bibfnamefont{A.~C.} \bibnamefont{Gossard}},
  \bibnamefont{and} \bibinfo{author}{\bibfnamefont{J.~H.}
  \bibnamefont{English}}, \bibinfo{journal}{Phys. Rev. Lett.}
  \textbf{\bibinfo{volume}{61}}, \bibinfo{pages}{997}
  (\bibinfo{year}{1988}{\natexlab{b}}).

\bibitem[{\citenamefont{Du et~al.}(1995)\citenamefont{Du, Yeh, St\"ormer, Tsui,
  Pfeiffer, and West}}]{tilt2}
\bibinfo{author}{\bibfnamefont{R.~R.} \bibnamefont{Du}},
  \bibinfo{author}{\bibfnamefont{A.~S.} \bibnamefont{Yeh}},
  \bibinfo{author}{\bibfnamefont{H.~L.} \bibnamefont{St\"ormer}},
  \bibinfo{author}{\bibfnamefont{D.~C.} \bibnamefont{Tsui}},
  \bibinfo{author}{\bibfnamefont{L.~N.} \bibnamefont{Pfeiffer}},
  \bibnamefont{and} \bibinfo{author}{\bibfnamefont{K.~W.} \bibnamefont{West}},
  \bibinfo{journal}{Phys. Rev. Lett.} \textbf{\bibinfo{volume}{75}},
  \bibinfo{pages}{3926} (\bibinfo{year}{1995}).

\bibitem[{\citenamefont{Hwang and Sarma}(2006)}]{sds-hwang}
\bibinfo{author}{\bibfnamefont{E.~H.} \bibnamefont{Hwang}} \bibnamefont{and}
  \bibinfo{author}{\bibfnamefont{S.} \bibnamefont{Das Sarma}},
  \bibinfo{journal}{Phys. Rev. B} \textbf{\bibinfo{volume}{73}},
  \bibinfo{eid}{121309} (\bibinfo{year}{2006}).

\bibitem[{\citenamefont{Das~Sarma and Hwang}(2000)}]{prl2000}
\bibinfo{author}{\bibfnamefont{S.}~\bibnamefont{Das~Sarma}} \bibnamefont{and}
  \bibinfo{author}{\bibfnamefont{E.~H.} \bibnamefont{Hwang}},
  \bibinfo{journal}{Phys. Rev. Lett.} \textbf{\bibinfo{volume}{84}},
  \bibinfo{pages}{5596} (\bibinfo{year}{2000}).

\bibitem{fqhe-SLL-2008}
C. R. Dean, B. A. Piot, P. Hayden, S. Das Sarma, G. Gervais, L. N. 
Pfeiffer, and K. W. West, Phys. Rev. Lett. \textbf{100}, 146803 (2008); 
\textit{ibid}, arXiv:0805.3349 [cond-mat.mes-hall] (2008).


%%%%%%%%%%%%%%%%%%%%%%%%%%%%%%%%%%%%%%%%%%%%%%%%%%%





\end{thebibliography}

\end{document}